\documentclass[twocolumn,english]{svjour3}
\usepackage[T1]{fontenc}
\usepackage[latin9]{inputenc}
\usepackage{color}
\definecolor{shadecolor}{rgb}{0.921875, 0.960938, 1}
\usepackage{array}
\usepackage{float}
\usepackage{booktabs}
\usepackage{framed}
\usepackage{bbding}
\usepackage{mathtools}
\usepackage{url}
\usepackage{amstext}
\usepackage{amssymb}
\usepackage{graphicx}
\usepackage[numbers]{natbib}

\makeatletter

\newcommand*\LyXZeroWidthSpace{\hspace{0pt}}
\providecommand{\tabularnewline}{\\}

\newtheorem{constraint}{\protect\constraintname}
\newtheorem{primarygoal}{\protect\primarygoalname}

\RequirePackage{fix-cm}

\smartqed  
\mathchardef\mhyphen="2D 
\makeatother

\usepackage{babel}
\providecommand{\constraintname}{Coordination Constraint}

\providecommand{\primarygoalname}{Goal}

\begin{document}
\title{Enacting Coordination Processes\thanks{This work is part of the ZAFH Intralogistik, funded by the European
Regional Development Fund and the Ministry of Science, Research and
the Arts of Baden-Württemberg, Germany (F.No. 32-7545.24-17/3/1)}}
\titlerunning{Coordination Process Execution}
\author{Sebastian Steinau \and  Kevin Andrews \and  Manfred Reichert}
\authorrunning{Steinau et al.}
\institute{Ulm University\\
 Institute of Databases and Information Systems\\
 Building O27 Level 5, James-Franck-Ring\\
 89081 Ulm, Germany\\
\email{firstname.lastname@uni-ulm.de}\\
}
\date{Preprint published on arXiv.org}
\maketitle
\begin{abstract}
With the rise of data-centric process management paradigms, interdependent
processes, such as artifacts or object lifecycles, form a business
process through their interactions. Coordination processes may be
used to coordinate these interactions, guiding the overall business
process towards a meaningful goal. A coordination process model specifies
coordination constraints between the interdependent processes in terms
of semantic relationships. At run-time, these coordination constraints
must be enforced by a coordination process instance. As the coordination
of multiple interdependent processes is a complex endeavor, several
challenges need to be fulfilled to achieve optimal process coordination.
For example, processes must be allowed to run asynchronously and concurrently,
taking their complex relations into account. This paper contributes
the operational semantics of coordination processes, which enforces
the coordination constraints at run-time. Coordination processes form
complex structures to adequately represent processes and their relations,
specifically supporting many-to-many relationships. Based on these
complex structures, markings and process rules allow for the flexible
enactment of the interdependent processes while fulfilling all challenges.
Coordination processes represent a sophisticated solution to the complex
problem of coordinating interdependent, concurrently running processes.
\end{abstract}


\section{Introduction}

\noindent Business process management, for the most part, is concerned
with large, monolithic models of business processes \citep{vanderAalst.2000b}.
Interactions between different business processes predominantly take
place only if each business process belongs to a different organization
\citep{ObjectManagementGroup.2011,vanderAalst.2000}. In practice,
these cross-organizational business processes have been mostly limited
to bidirectional one-to-one message exchanges, and interactions between
business processes in a one-to-many relationship have been an afterthought
at best. Consequently, concepts to model and describe the interactions
between processes can support one-to-many interactions conceptually,
but practically, they have been limited mostly to message exchanges
between two processes. With the advent of data-centric process management,
the postulate of only two interacting processes is no longer viable.
In general, data-centric process management elevates data to a first-class
citizen of a business process model and organizes data into objects
with attributes. Each business process may comprise multiple different
objects, and each of these objects possesses a lifecycle process that
governs the way the object is processed over time, i.e., a lifecycle
process describes object behavior. Therefore, the concepts of lifecycle
process and object are mutually intertwined. The key idea of this
object-centric variant of data-centric process management is that
a \emph{business process} is not prescribed by a monolithic model,
but instead \emph{emerges from the interactions of objects and their
lifecycle processes.} In other words, a business process is constituted
of interacting lifecycle processes in data-centric approaches using
objects or artifacts \citep{Nigam.2003,Kunzle.2011}. Generally, more
than two types of processes have interactions, and the various process
types may be in \emph{one-to-one}, \emph{one-to-many} as well as \emph{many-to-many
relationships}. 

The emergence of the overall business process from interacting lifecycle
processes possesses a fundamental property: It allows for a \emph{high
degree of flexibility} regarding the overall enactment of the business
process \citep{Kunzle.2011,Hull.2011}. The different objects and
their lifecycle processes may be created and executed in very diverse
ways, i.e., without too may constraints placed upon them. This is
possible because data-centric approaches do not presume prescribed,
imperative process choreographies that govern all object interactions
in detail (e.g., \citep{Decker.2011}). Despite the absence of imperative
process choreographies, one cannot necessarily assume that interacting
lifecycle processes converge towards a meaningful overall business
process on their own. As a consequence, the interactions between the
lifecycle processes still require the enforcement of certain constraints
at different points in time, which is denoted as \emph{process coordination}
in the following. Process coordination ensures that the processes
and their interactions are guided towards a purposeful business process. 

\subsection{Problem Statement}

For the convergence towards an overall business process, the interacting
lifecycle processes require a coordination approach that preserves
the inherent enactment flexibility of lifecycle processes. Regarding
process coordination, several challenges were encountered during the
design and conception of the object-aware approach to business process
management \citep{Kunzle.2011}. These challenges possess a high practical
relevance, as they impact a successful enactment of an object-aware
business process. These challenges represent issues that occur predominantly
when one-to-many and many-to-many relationships between processes
and the concurrent enactment of processes are concerned. Concurrent
enactment and many-to-many-relationships are essential to the object-aware
approach. The findings were supported by a systematic literature review
on data-centric approaches to BPM \citep{Steinau.2018c}, and by expert
assessments \citep{Marrella.2015}. When only one-to-one interactions
of lifecycle processes are considered, these challenges exist as well,
but are far less complicated to solve. Regarding the interacting lifecycle
processes in one-to-many and many-to-many relationships, five main
challenges may be identified: \emph{asynchronous concurrency}, \emph{complex
process relationships}, \emph{local contexts}, \emph{immediate consistency},
and \emph{manageable complexity. }These challenges are designated
as main challenges for their relevance and complexity.

Each of these challenges states an optimum of what an approach for
coordinating the interactions between lifecycle processes might support.
Addressing all challenges together requires high sophistication and
the development of new ways to deal with the coordination of processes.
These challenges not only encompass the design-time, where lifecycle
process and coordination models are specified, but in particular the
run-time, where multiple instances of these models are concurrently
executed. Much of the complexity of coordinating the lifecycle process
interactions resides with the run-time.

\subsection{Contribution}

\noindent As one data-centric approach, \emph{object-aware process
management} \citep{Kunzle.2011} has encountered these challenges.
Subsequently, a sophisticated solution to these challenges has taken
shape over recent years: \emph{coordination processes} \citep{Steinau.2018}.
A coordination process is a high-level concept to specify and enforce
the dependencies that exist between multiple interacting lifecycle
processes. These dependencies between processes are called \emph{coordination
constraints}. In other words, a coordination process provides coordination
for processes, i.e., such as lifecycle processes. The object and lifecycle
process concept is not mandatory for coordination processes; in the
following, a general notion of process is used that includes, but
is not limited to, lifecycle processes. This notion is simply named
a \emph{process}, and is distinct from the notions of coordination
process or business process.

\begin{figure}
\begin{centering}
\includegraphics[width=1\columnwidth]{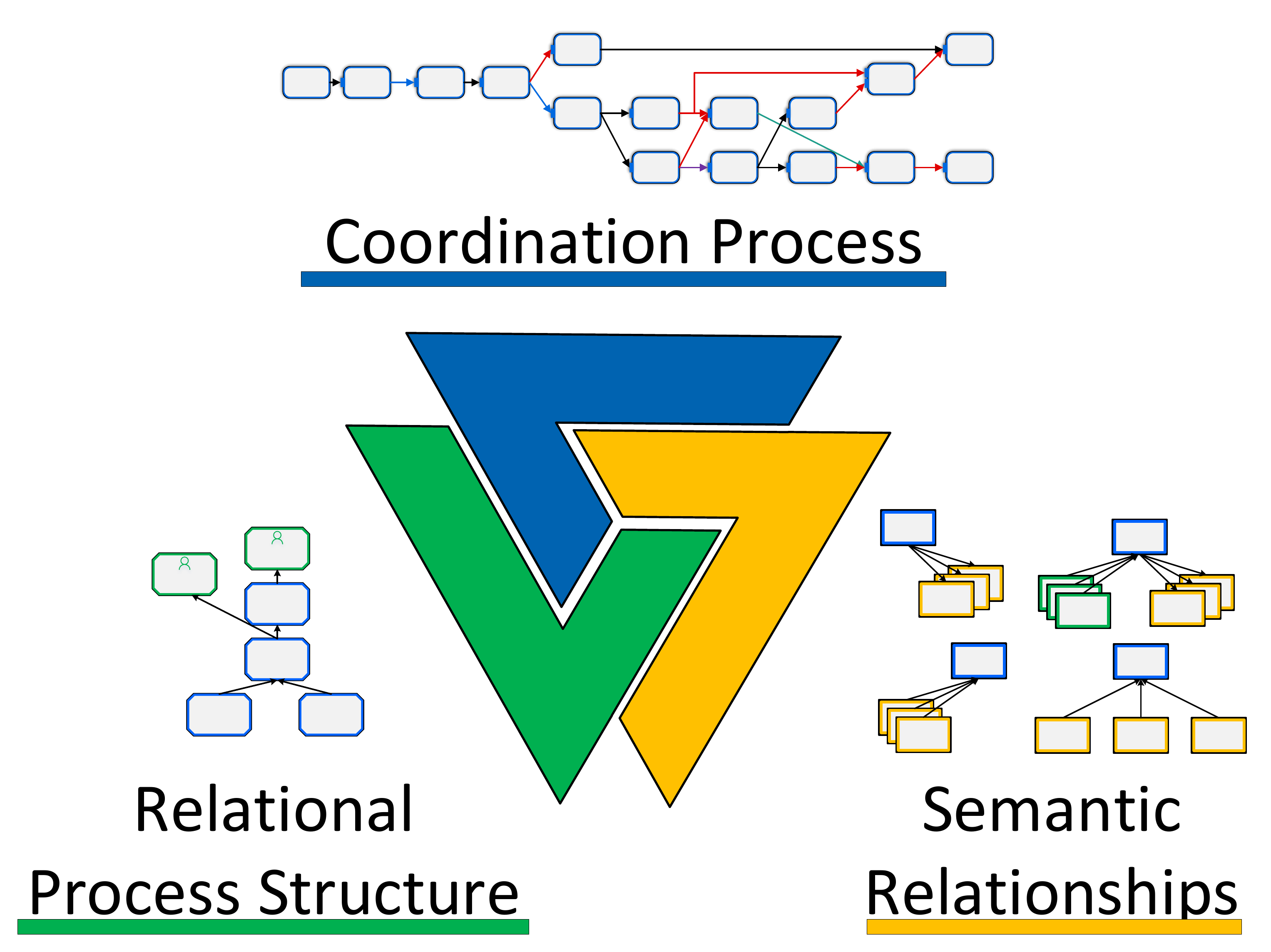}
\par\end{centering}
\begin{centering}
\caption{\label{fig:Coordination-Processes-Trinity}Coordination Process and
its foundations}
\par\end{centering}
\vspace{-0.3cm}
\end{figure}

Coordination processes alone are not able to resolve every challenge
themselves. Instead, the concept of coordination processes is the
keystone binding together two other fundamental concepts that revolve
around the aforementioned challenges: the \emph{relational process
structure} and \emph{semantic relationships} . A relational process
structure captures process types and their relations, which may be
one-to-many or many-to-many relations, at both design- and run-time.
Interactions between processes only occur between related processes,
i.e., a relation is a prerequisite for interactions between processes.
Every process instance and its relations are tracked at run-time,
enabling a complete overview over processes and their relations. Semantic
relationships describe patterns inherent in the interactions between
processes in a one-to-many relationship. Based on semantic relationships,
sophisticated coordination constraints between multiple interacting
processes can be expressed. Specifically, this paper builds upon
\begin{itemize}
\item semantics relationships at design-time and state-based views \citep{Steinau.2017}
\item the design-time and run-time of relational process structures \citep{Steinau.2018b}
\item the design-time aspects of coordination processes \citep{Steinau.2018b}
\end{itemize}
\noindent This paper subsequently presents all run-time aspects of
coordination processes and semantic relationships, concluding the
work started with \citep{Steinau.2017}. A schematic view of coordination
processes, the relational process structure, and semantic relationships
is shown in Figure \ref{fig:Coordination-Processes-Trinity}, indicating
that all three concepts are inextricably linked.

One aspect of these efforts is that coordination processes are designed
to be executable directly after instantiation from a respective coordination
process type. In consequence, object-aware process management has
achieved full design-time and run-time support for a data-centric
approach, which has been accomplished only by very few other data-centric
approaches to BPM \citep{Steinau.2018c}. The implementation of the
object-aware approach is named PHILharmonicFlows \citep{Kunzle.2011}.
PHILharmonicFlows has recently seen drastic changes due to moving
to a microservice-based architecture \citep{Andrews.2018}.\emph{
}Paramount for full run-time support is the \emph{operational semantics}
that describes how coordination processes are enacted. The operational
semantics completely account for the new microservice-based architecture
and the required adaptations in the concepts that constitute the basis
for PHILharmonicFlows.

More specifically, the first contribution in enabling the enactment
of a coordination process is the representation of process instances.
A coordination process requires an internal representation of processes
and their complex relations, i.e, one-to-many and many-to-many relationships,
which is derived from the relational process structure \citep{Steinau.2018b}.
Coordination processes combine the derived representation of processes
with components that describe the coordination constraints that restrict
the interactions between processes. This combined representation enables
the coordination process to actually enforce these coordination constraints
at run-time, i.e., the enactment of the coordination process.  The
challenge here is that this combined representation is not fixed at
run-time, but dynamically adapts to new relations and newly instantiated
processes.

The second contribution are \emph{process rules. }The operational
semantics that describes how a coordination process is enacted is
defined using a variant of event-condition-action (ECA) rules, denoted
as \emph{process rules}. Process rules read and write \emph{markings}
of elements in a coordination process instance, which indicate status
information of process instances and the coordination process instance
itself. The use of process rules and markings enables a highly flexible
enactment of coordination processes, allowing to accommodate copious
amounts of possible interactions, while prohibiting forbidden behavior
in the interactions between process instances. This paper provides
the set of process rules necessary to enact a coordination process.
Furthermore, it describes the principles of how the process rules
govern the enactment of a coordination process.

While the concept of coordination processes was devised in context
of the object-aware approach, it is by no means limited to be employed
only with object-aware processes. The design of coordination processes
includes the possibility to coordinate any kind of processes, independently
from the paradigm in which they are specified. Therefore, it is possible
to coordinate interdependent activity-centric processes with a coordination
process as well \citep{Steinau.2017,Steinau.2018b}. 

The remainder of the paper is organized as follows: Section \ref{sec:Challenges-and-Problem}
describes the problem of process coordination and presents a detailed
description of the challenges that coordination processes aim to solve.
Section \ref{sec:Background} recaps the prerequisite concepts of
the relational process structure and semantic relationships. Furthermore,
the necessary background on coordination processes at design-time
is presented. Section \ref{sec:Enacting-Coordination-Processes} gives
a general overview over the enactment of coordination processes. The
details of enacting a coordination process are presented in two distinct
parts: Part one is shown in Section \ref{sec:Representing-Process-Instances}
and describes how process instances from the relational process structure
and their semantic relationships are represented in a coordination
process. Sections \ref{sec:Enacting-Coordination-Processes} and \ref{sec:Representing-Process-Instances}
contain the first contribution. Part two is presented in Section \ref{sec:Operational-Semantics}
and describes how process rules and markings are employed to enforce
coordination constraints, constituting the second contribution. Sections
\ref{sec:Enacting-Coordination-Processes}-\ref{sec:Operational-Semantics}
discuss the run-time of coordination processes and cover the contribution
of this paper. The technical implementation of the operational semantics
and the concept of coordination processes are presented in Section
\ref{sec:Technical-Implementation}. Section \ref{sec:Related-Work}
covers related work and discusses other approaches to process coordination.
Section \ref{sec:Conclusion} concludes the paper with a summary and
an outlook.

\section{\label{sec:Challenges-and-Problem}Challenges}

The principal purpose of process coordination is to manage the complex
interactions between two or more processes. In previous works, the
interactions between processes in a one-to-one relationship, and typically
in an inter-organizational setting, have been investigated \citep{vanderAalst.2000}.
As such, concepts for process coordination have been tailored to this
setting. Extensions of this setting, e.g., the inclusion of one-to-many
or many-to-many relationships or the interactions between more than
two different types of processes, have been mostly neglected \citep{zurMuehlen.2013}.

With data-centric process management, e.g., objects with lifecycle
processes, the necessity has arisen to consider one-to-many and many-to-many
relationships between multiple lifecycle processes of different types.
With the advent of cloud computing and the Internet of Things (IoT),
it may be further conjectured that small, interdependent processes
will become increasingly important, for reasons of scalability in
cloud environments and the complexity limitations of IoT devices.

Several new challenges have arisen due to the numerous requirements
of data-centric approaches \citep{Kunzle.2011}, and these additional
challenges demand more sophisticated solutions. The principal promise
of data-centric process management is the ability to support both
unstructured and semi-structured business processes adequately. Previously,
this kind of business processes could not be correctly represented
with traditional process management approaches, e.g., BPMN 2.0. This
is due to the extraordinary requirements regarding the flexibility
of these business processes \citep{Kunzle.2009}. One contribution
for supporting this flexibility is the elimination of large, monolithic
process models that encompass everything in favor of interdependent
processes constituting the overall business process. These interdependent
processes then need to interact and, therefore, require coordination.
In consequence, much of the effort for realizing this flexibility
is placed on the coordination approach that governs these process
interactions. In particular, this effort is compounded by the requirements
to support complex process relationships and multiple types of processes.
As such, the flexibility promises of data-centric paradigms can be
fulfilled if a coordination concept delivers a solution that meets
the flexibility requirements. For a more systematic approach, five
primary challenges have been identified that a coordination approach
should fulfill in order to enable flexible interactions between different
kinds of processes with complex relationships.

For illustrating these challenges as well as the concept of coordination
processes and its operational semantics, a running example is used
throughout the paper (cf. Example \ref{exa:Running-Example}). It
describes a recruitment business process in the human resource domain.
Much of the complex behavior and interactions between different lifecycle
processes may be observed in this rather simple setting.

\noindent\begin{minipage}[t]{1\columnwidth}%
\begin{shaded}%
\begin{example}[\label{Running-Example:-Recruitment}Recruitment Business Process]
\label{exa:Running-Example}

A company has an open position for which it wants to hire a suitable
candidate. For this purpose, a company employee creates a $\mathit{Job\:Offer}$
and publishes it (e.g., on the company website). For this $\mathit{Job\:Offer}$,
interested persons may create $\mathit{Applications}$. $\mathit{Applications}$
may be created as long as the $\mathit{Job\:Offer}$ is not closed,
i.e, $\mathit{Applications}$ may arrive at different points in time
at the company. For each $\mathit{Application}$ that is sent to the
company, an evaluation is started. Company experts must create $\mathit{Reviews}$
for the application and give a recommendation on whether to invite
the applicant for an $\mathit{Interview}$ or reject him outright.
The overall recommendation requires at least three $\mathit{Reviews}$
and a majority of 50\% or more in favor of the applicant for an invite
recommendation. Depending on the availability of the company experts
and the arrival date of the respective application, $\mathit{Reviews}$
may be created and completed at different points in time. If the overall
recommendation favors the rejection of an applicant, the corresponding
$\mathit{Application}$ will be rejected. If the $\mathit{Reviews}$
are in favor of the applicant, the applicant must be invited to at
least one $\mathit{Interview}$ to further substantiate the suitability
of the applicant for the open $\mathit{Job\:Offer}$. If the majority
of $\mathit{Interviews}$ recommend hiring the applicant, the $\mathit{Application}$
may be accepted, otherwise the $\mathit{Application}$ will be rejected.
Ties are resolved in favor of acceptance. At least one $\mathit{Interview}$
must be performed. However, only one $\mathit{Application}$ may be
accepted for each $\mathit{Job\:Offer}$. Should an applicant have
been hired, the $\mathit{Job\:Offer}$ is closed and given the final
status ``position filled'' indicating success. Other applicants
must consequently be rejected. The $\mathit{Job\:Offer}$ may be closed
at any time as long as at least one $\mathit{Application}$ has been
sent to the company. If, after a reasonable amount of time, no suitable
applicant can be found, the $\mathit{Job\:Offer}$ is closed, and
its final status is set to ``position vacant''.\vspace*{-0.3cm}
\end{example}
\end{shaded}%
\end{minipage}

In the following, five challenges in the context of advanced process
coordination are characterized in detail. Coordination processes aim
to fulfill all challenges.

\subsection{\label{subsec:First-Challenge:-Asynchronous}Challenge 1: Asynchronous
concurrency}

\begin{figure}

\begin{centering}
\includegraphics[width=1\columnwidth]{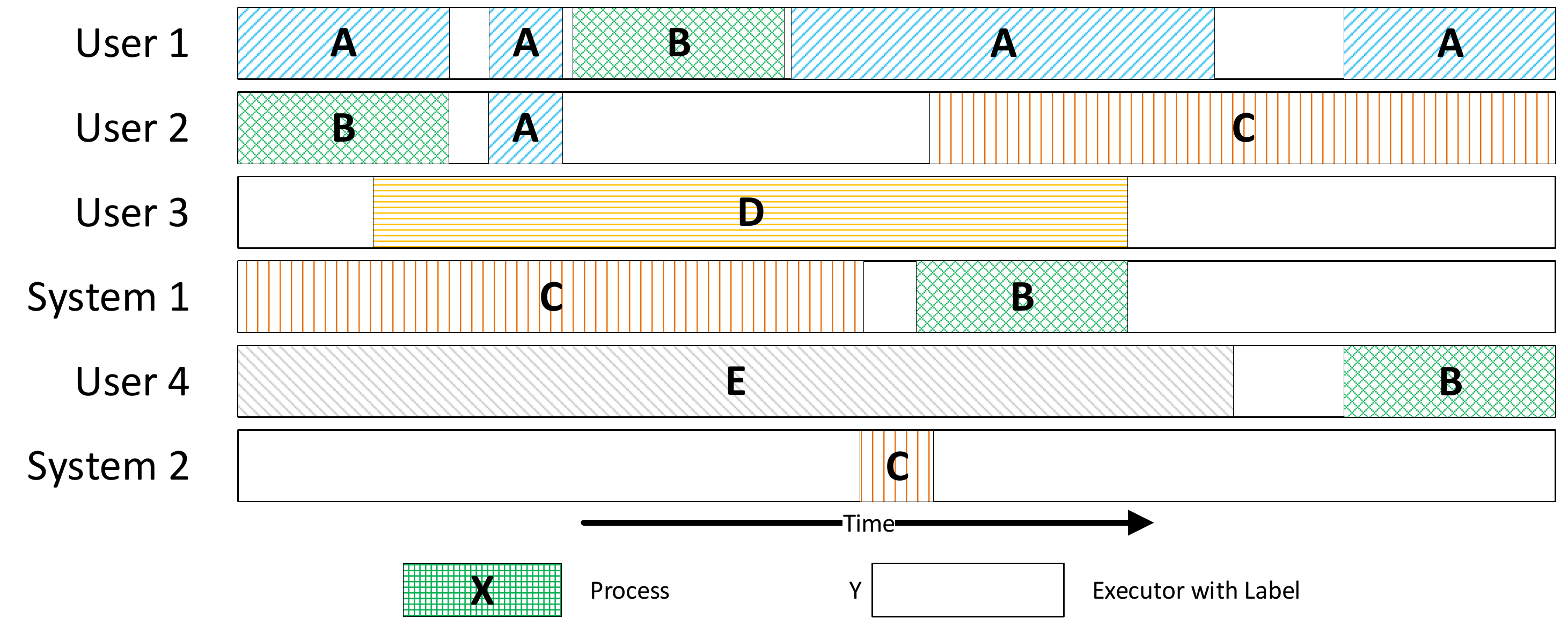}
\par\end{centering}
\caption{\label{fig:Asynchronous-concurrency-of}Asynchronous concurrency of
different processes enacted by different actors}

\vspace{-1em}
\end{figure}

The first challenge relates to the concurrent and asynchronous enactment
of processes. Processes may run concurrently to each other. Their
concurrent enactment, however, is not required to be synchronized
constantly, i.e., not every instance of a process is at the same stage
at the same point in time. Instead, the enactment of a process is
in principle fully asynchronous. At the one extreme, the enactment
of multiple processes is sequential without any parallelism, i.e.,
a process is strictly enacted one after the other. At the other extreme,
the enactment of the processes is fully parallelized, meaning every
process enacts the same step at the same time. In between these extremes,
processes may be enacted partially in parallel and partially sequential,
where any combination is feasible. For example, one group of processes
may be enacted in parallel, but sequentially after another group.
Furthermore, the enactment of a process may be suspended and resumed
at a later time, possibly by different actors. Actors may be users
or systems in this context. Figure \ref{fig:Asynchronous-concurrency-of}
shows a schematic view of asynchronously and concurrently enacted
processes by different users or systems.

In practical scenarios, there may be dependencies between processes
that restrict their concurrent enactment in some way. Of particular
importance is that such coordination constraints do not restrict process
enactment unnecessarily. Consider a process $A$ and another process
$B$, where $B$ must wait for a specific state of $A$. When $A$
has reached that state, then $B$ can continue and subsequently must
deliver a result back to $A$, which may then proceed. In a naive
implementation, process $A$ must wait for $B$ after $B$ has started
until $B$ delivers the result. However, in principle, the result
from $B$ might not be needed until a later point in the enactment
of process $A$. Therefore, it should be possible that $A$ continues
running, while process $B$ is enacted in parallel, i.e., $A$ and
$B$ run asynchronously and concurrently. This offers much more flexibility
and performance, as unneeded waiting times do not occur. Only when
the result from $B$ is absolutely needed for process $A$ to continue,
$A$ should be forced to wait for $B$. 

In summary, as processes need to interact in order to form the overall
business process, asynchronous concurrency of the processes opens
up a myriad of possibilities for different processes to interact.
A coordination approach should support these interactions as best
as possible. As concurrent and asynchronous enactment is beneficial
to the overall performance and flexibility of any interaction-focused
paradigm, such as object-aware process management, an approach that
coordinates processes should not unnecessarily restrict concurrent
enactment. Consequently, the support of asynchronous concurrency in
process enactment is the first challenge of process coordination.
Note that this challenge is not solely associated with the conceptual
level, but is also crucial for the implementation of a process management
system. A practical occurrence of the first challenge can be illustrated
with the running example (cf. Example \ref{exa:async_concurrency}).

\noindent\begin{minipage}[t]{1\columnwidth}%
\begin{shaded}%
\begin{example}[Asynchronous Concurrency]
\label{exa:async_concurrency}

In Example \ref{Running-Example:-Recruitment}, $\mathit{Applications}$
may be enacted in parallel to each other as well as to other types
of processes, e.g., $\mathit{Reviews}$ or $\mathit{Interviews}$.
Depending on when an $\mathit{Application}$ arrives, it may still
be enacted in parallel, but with a time offset (cf. processes C and
D of Figure \ref{fig:Asynchronous-concurrency-of}). However, $\mathit{Reviews}$
may only be started after the $\mathit{Application}$ has been sent
to the company. $\mathit{Interviews}$ rely on a positive preliminary
result from the $\mathit{Reviews}$ associated with the $\mathit{Application}$.
Note that the results of the $\mathit{Reviews}$ may arrive at different
points in time, i.e., asynchronously. Additionally, another $\mathit{Application}$
may be processed in parallel to conducting an $\mathit{Interview}$
for the first $\mathit{Application}$. As can be seen, processes may
run concurrently, but not without constraints placed upon them by
the process coordination.\vspace{-1em}
\end{example}
\end{shaded}%
\end{minipage}

Challenge 1 \emph{Asynchronous Concurrency} has been chosen for its
importance for business processes in general and its impact on the
performance of business process execution.

\subsection{\label{subsec:Second-Challenge:-Complex}Challenge 2: Complex Process
Relations}

\begin{figure}
\includegraphics[width=1\columnwidth]{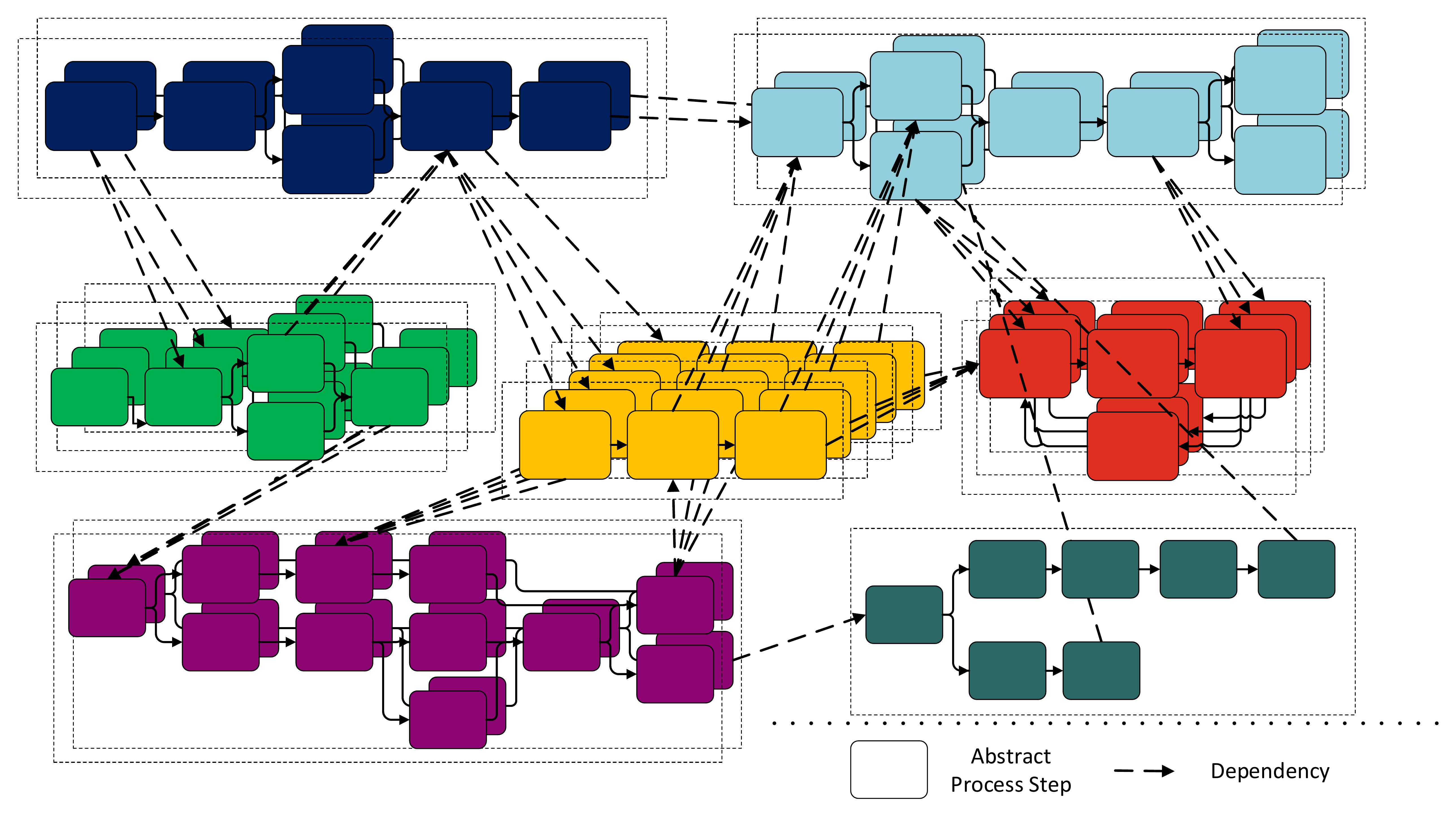}

\caption{\label{fig:Interdependent-process-instances}Interdependent process
instances with direct and transitive relations}
\end{figure}

When representing an overall business process using interacting processes,
it is likely that not just one instance of each process type is needed
at run-time. Thus, multiple instances of each process type in the
business process model may be created. These process instances do
not exist independently from each other, but have \emph{interrelations}
(cf. Figure \ref{fig:Interdependent-process-instances}). For example,
process instances may depend on multiple other process instances,
which constitutes a one-to-many relationship. Moreover, process instances
may be part of a many-to-many relationship with other process instances,
introducing a significant increase in complexity compared to one-to-many
relationships. Furthermore, processes may not only be directly related
to other processes, but relations may form \emph{paths} across different
process types. This is called a \emph{transitive relationship}, which,
together with one-to-many and many-to-many relationships, leads to
the creation of vast \emph{process structures} of interrelated and,
thus, interdependent processes. This structure of interrelated processes
sets data-centric process management apart from the predominant one-on-one
interactions of activity-centric processes\textemdash complex relationships
between processes are part of the premise.

Managing and identifying these complex relationships at design-time
and, more importantly, handling the emerging structure of process
instances and their relations at run-time, constitutes the second
challenge regarding process interactions that needs to be resolved.
The challenge is compounded by the fact that processes and their relations
are created and deleted over time. This results in a dynamic environment
of a continually evolving process structure. In summary, process structures
and their dynamic evolution must be taken into account when coordinating
process interactions.

\noindent\begin{minipage}[t]{1\linewidth}%
\begin{shaded}%
\begin{example}[Complex Process Relations]
\label{exa:complex_process_relations}

In Example \ref{Running-Example:-Recruitment}, $\mathit{Job\:Offers}$
may be related to many $\mathit{Applications}$, constituting a one-to-many
relationship. At run-time, a $\mathit{Job\:Offer}$ must know which
$\mathit{Applications}$ belong to it. In turn, each $\mathit{Application}$
may be related to several $\mathit{Reviews}$ and eventually some
$\mathit{Interviews}$, which must be tracked and related accordingly
as well. At any point in time, as long as the $\mathit{Job\:Offer}$
is not closed, $\mathit{Applications}$ may be newly created for the
$\mathit{Job\:Offer}$,\vspace{-1em}
\end{example}
\end{shaded}%
\end{minipage}

\noindent\begin{minipage}[t]{1\linewidth}%
\begin{shaded}%
or may be withdrawn or deleted, showing the dynamics of the process
structure. $\mathit{Job\:Offers}$ are transitively related to $\mathit{Reviews}$
and $\mathit{Interviews}$, e.g., it is possible to determine how
many $\mathit{Reviews}$ and $\mathit{Interviews}$ have been performed
in the context of a particular $\mathit{\mathit{Job\:Offer}}$, as
well as their status, at any point during run-time.\end{shaded}%
\end{minipage}

Challenge 2 \emph{Complex Process Relations} has been chosen as business
processes composed of smaller processes would be severely limited
in their expressiveness with only simple one-to-one relations. Considering
one-to-many and many-to-many relations for interacting processes has
been well established in literature \citep{Kunzle.2009,Fahland.2011,Meyer.2013}

\subsection{\label{subsec:Third-Challenge:-Local_contexts}Challenge 3: Local
Contexts}

As multiple different objects exist that have different kinds of
relationships, e.g., one-to-many, many-to-many, as well as transitive
relationships, several implications can be observed at run-time. As
shown before, process instances of different types form process structures
because of their relations. In consequence, one process instance may
be handled differently from another process instance of the same type
simply because of the relations it has formed to other process instances.
Thereby, some interconnected process instances form substructures
within the overall process structure, denoted as \emph{arrangements.
}Arrangements are defined by the involved process instances as well
as their exact relations \citep{Steinau.2017}. An $\mathit{Application}$
instance related to four $\mathit{Review}$ instances is an example
of an arrangement. An arrangement may therefore be subject to a \emph{local
context}. A local context represents individual constraints for the
coordination of process interactions that must be handled individually,
as part of an overall coordination effort. As opposed to this individualized
approach, process choreography approaches in activity-centric process
management always operate in the same context, as only one-to-one
relations and no transitive relations have been considered. With many-to-many
relations and transitive relations, the individual handling of emerging
local contexts creates an entirely new challenge. 

\begin{figure}

\centering{}\includegraphics[width=1\columnwidth]{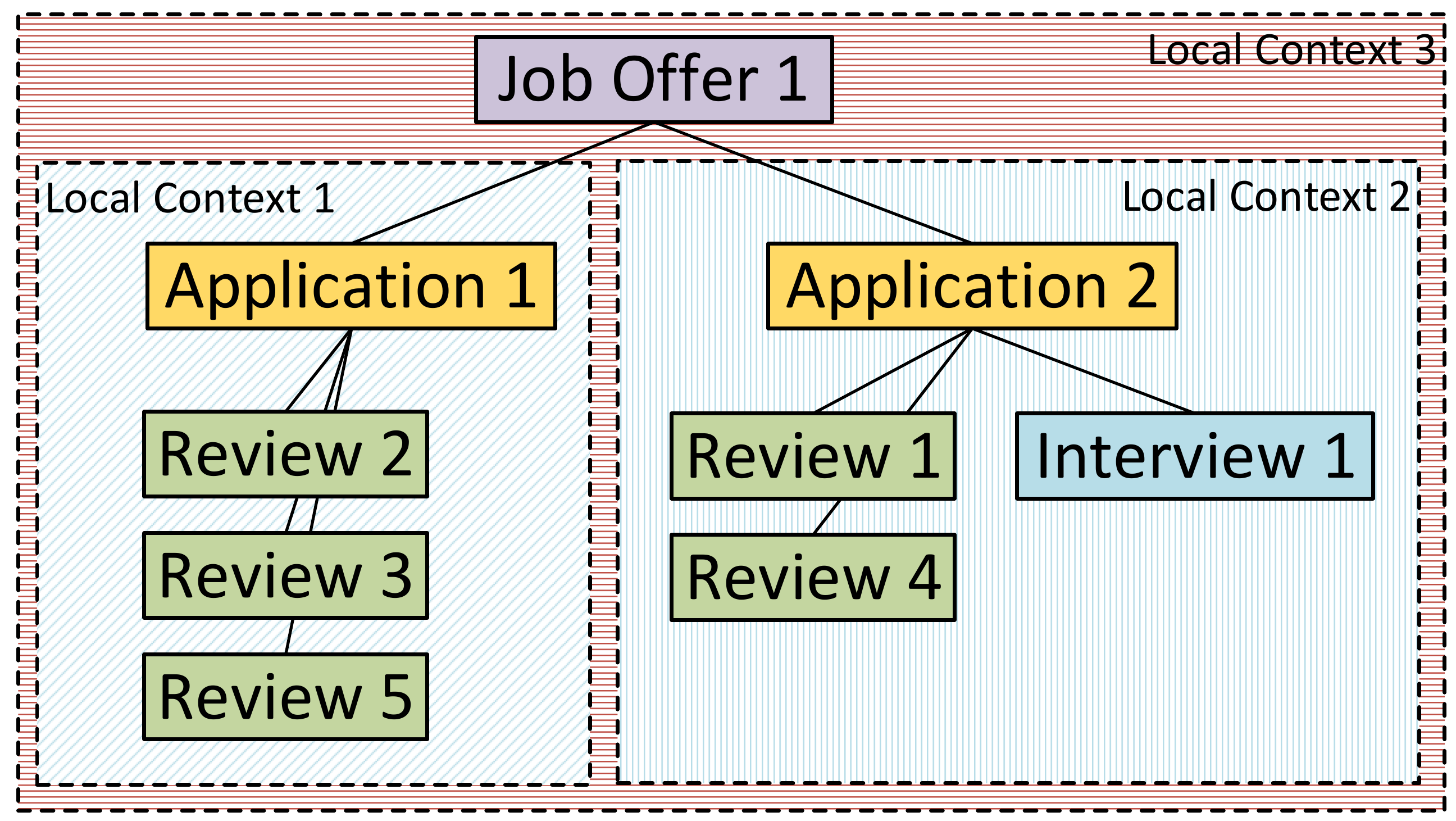}\caption{\label{fig:Schematic-view-of}Schematic view of local context regarding
the running example}
\vspace{-1em}
\end{figure}

Numerous different local contexts may exist within the same process
structure at the same time. More specifically, process instances and
their arrangements may belong to different local contexts at the same
time, i.e., local contexts may overlap or contain each other (cf.
Figure \ref{fig:Schematic-view-of}). The arrangement consisting of
$\mathit{Application\:1}$ as well as $\mathit{Reviews}$ $2$, $3$,
and $5$ is subject to its own local context (Local Context 1) and
is included in the local context of $\mathit{Job\:Offer\:1}$ (Local
Context 3). Local contexts change over time due to changes in process
relations. Altogether, the correct consideration of local contexts
in the overall process coordination is crucial and therefore designated
as the third challenge.

\noindent\begin{minipage}[t]{1\columnwidth}%
\begin{shaded}%
\begin{example}[Local Contexts]
\label{exa:local_contexts}Each $\mathit{Application}$ creates a
local context together with its $\mathit{Reviews}$ and $\mathit{Interviews}$.
While each $\mathit{Application}$ is embedded in the overall coordination
of the local context of a $\mathit{Job\:Offer}$ (cf. Figure \ref{fig:Schematic-view-of}),
each $\mathit{Application}$ must also be handled individually. For
one $\mathit{Application}$, $\mathit{Reviews}$ may still be in progress
and have not reached consensus yet (cf. Local Context 1, Figure \ref{fig:Schematic-view-of}),
whereas for another $\mathit{Application}$, $\mathit{Interviews}$
\vspace{-2em}
\end{example}
\end{shaded}%
\end{minipage}

\noindent\begin{minipage}[t]{1\columnwidth}%
\begin{shaded}%
are being conducted (cf. Local Context 2, Figure \ref{fig:Schematic-view-of}).
While $\mathit{Applications}$ must obey the same coordination constraints
in general, due to their local context, different subsets of these
constraints are relevant at different points in time. In other words,
$\mathit{Applications}$ for which an interview invite is proposed
are subject to other constraints than $\mathit{Appli\mhyphen}$\\
$\mathit{cations}$ for which a reject is proposed. These different
local contexts must be recognized and the respective appropriate subset
of constraints enforced.\end{shaded}%
\end{minipage}

The emergence of local contexts is a direct consequence of considering
processes in one-to-many and many-to-many as well as transitive relations.
Therefore, Challenge 3 \emph{Local Contexts} has been included for
consideration in coordination processes. As transitive one-to-many
and many-to-many relationships have only been considered in relational
process structures \citep{Steinau.2018b} and object-aware process
management \citep{Kunzle.2011}, the phenomenon of local contexts
in process interactions has not been established otherwise in business
process management literature.

\subsection{\label{subsec:Immediate-Consistency}Challenge 4: Immediate Consistency}

As mentioned in Section \ref{subsec:Second-Challenge:-Complex},
process structures are highly dynamic, as processes and relations
may be created at any point in time and, later on, deleted. Furthermore,
this dynamic is amplified by the actual progress of processes, which
change enactment status according to their process model over time.
Concerning process coordination, any of these changes might have enormous
implications for other processes within the same process structure
(cf. Example \ref{Process-Coordination-Consistency1}).

\noindent\begin{minipage}[t]{1\columnwidth}%
\begin{shaded}%
\begin{example}[\label{Process-Coordination-Consistency1}Process Coordination Consistency
I]

Advancing of the enactment status of a process instance may allow
other process instances to advance as well. When $\mathit{Reviews}$
reach either state $\mathit{Invite\;Proposed}$ or $\mathit{Reject\;Proposed}$,
the related $\mathit{Application}$ may enter state $\mathit{Checked}$.
\vspace{-1em}
\end{example}
\end{shaded}%
\end{minipage}

In general, such an event may trigger a \emph{cascade of changes}
that propagates through a process structure. A cascade is triggered
by an initial event that causes many subsequent changes to occur,
which again might cause additional changes. An event is the emergence
of a new process or a new enactment status of a process being enacted.
A change means that coordination constraints are re-evaluated, which
may cause processes, which previously have been suspended, to continue.
This, in turn, might trigger new re-evaluations of other coordination
constraints. The cascade may affect one or more processes that are
only distantly related to the process where the initial event occurred.
The cascade is highly relevant for process coordination, as previously
disallowed actions, e.g., creating a new process instance, may become
available or vice versa, depending on the given coordination constraints.
In order to ensure correctness, subsequent actions may only be performed
if it is ensured that such an action is indeed allowed by the respective
coordination constraints. Performing the action triggers another event,
leading to another cascade.

This requires that the process structure has reached an \emph{overall
consistent state}, i.e, all follow-up changes from an event have been
processed until no more changes occur and the cascade has ended. Only
then can it be determined with certainty whether or not coordination
constraints allow performing a given action. Consequently, consistency
is crucial when changes occur within a process structure. Only then
can the process coordination approach ensure a correct evaluation
of all coordination constraints on subsequent actions. Otherwise,
subsequent actions may be performed based on an overall inconsistent
state, i.e., on outdated information, leading to incorrect processing
and possibly violated coordination constraints. Example \ref{Process-Coordination-Consistency2}
shows this based on an online shop scenario.\vspace{-1em}

\noindent\begin{minipage}[t]{1\columnwidth}%
\begin{shaded}%
\begin{example}[\label{Process-Coordination-Consistency2}Process Coordination Consistency
II]

When a user has placed an order in an online shop, the ordered products
must be commissioned, packaged, and loaded onto a delivery truck.
The user may modify the \vspace*{-2em}
\end{example}
\end{shaded}%
\end{minipage}

\noindent\begin{minipage}[t]{1\columnwidth}%
\begin{shaded}%
order as long as commissioning has not started. The commissioning
and packaging starts once respective workers are free from other duties.
Thus, timing and consistency are crucial for the possibility to modify
the order. The respective coordination between order and delivery
must ensure that this constraint actually holds. Furthermore, once
the last package has been loaded onto the delivery truck, the delivery
starts, allowing the order status to display ``on the way''. The
correct processing of the order is only possible when constraints
are upheld, which requires a consistent process structure of order,
commissioning, packaging, and delivery processes.\end{shaded}%
\end{minipage}

Keeping consistency is not only a question of the \emph{correctness
of process coordination}. It also concerns the \emph{performance of
the system} that implements the coordination approach, i.e., how fast
actions are performed by the system. The required time to reach an
overall consistent state is, among others, dependent on the system's
performance. Moreover, the required time for propagating changes is
proportional to the size of the process structure and the number of
changes, i.e., the size of the changeset. Considering this, ``immediate''
consistency is certainly unobtainable, as changes need time to propagate
through a process structure. Moreover, the system cannot process the
changes arbitrarily fast. There is always some time required to process
the changes leading to an overall consistent state. The term ``immediate''
should be understood as an \emph{idealized goal}.

However, the required time to process the cascading changes leading
to a consistent overall state should be minimal, i.e., as close to
``immediate'' as possible. Overly long processing times cause a
variety of issues, including the decreased acceptance by an end user.
An obvious threshold for judging if the performance of the system
is adequate is whether an end user experiences a delay when issuing
two consecutive actions in the system. Note that remaining below this
threshold for all possible sizes of the follow-up changeset is unreasonable,
as large changesets require more time to be processed. As a general
rule, processing time for cascading changes should be below this threshold
for small- to medium-sized sets of follow-up changes, for large sizes
of the changeset it is arguably understandable if the user notices
a reasonable delay. 

In summary, it is required that changes are properly propagated within
a process structure to reach an overall consistent state, so that
coordination constraints are not evaluated based on outdated information.
This requires performance considerations on the part of the coordination
approach, as better performance leads to reaching an overall consistent
state more quickly, benefiting the end user experience. Keeping consistency
and as-best-as-possible performance in regard to the coordinated processes
is paramount and therefore designated as the fourth challenge. Note
again that this challenge is not only concerned with concepts, but
primarily concerns the implementation of a process management system.

Consistency is a desirable property for the correct execution of vast
structures of processes. Furthermore, establishing consistency rapidly
is relevant for the performance of the overall system. Challenge 4
\emph{Immediate Consistency} logically follows from the consideration
of vast process structures enabled by Challenge 2 \emph{Complex Process
Relations}

\subsection{\label{subsec:Manageable-Complexity}Challenge 5: Manageable Complexity}

Any of the four previously described challenges introduces an enormous
complexity. Obviously, an approach to process coordination that aims
to fulfill these challenges is bound to have a high complexity as
well. Especially in regard to coordination constraint modeling, any
solution that is overly complex should be avoided. 

The fifth challenge consists of managing this complexity in such a
manner that the complexity of the solution does not outweigh the complexity
of the challenges or overall benefits of the approach; at least, the
complexity of the concepts should be on an appropriate level for the
challenges. In other terms, the Challenge 5 is concerned with finding
suitable abstractions, simplifications, and ideas to make the overall
complexity manageable. Ideally, the intricacies of the solutions for
the challenges can be abstracted and simplified in such a way that
the overall complexity is lower than the complexity of the challenges.
A coordination approach that solves the other challenges, but is hard
to use due to overbearing complexity, may not find acceptance with
the users.

The consideration of this challenge can be subdivided into the complexity
for end users executing a business process, the complexity for process
modelers creating the business process models, and the complexity
for developers implementing a corresponding process engine for running
these business processes. Furthermore, given the nature of this challenge,
the assessment of whether this challenge is fulfilled in the absolute
sense is challenging in itself. However, it is also feasible to compare
the complexity of an approach relative to established standard approach,
e.g., an industry standard. As this approach has become an established
(industry) standard, by definition its complexity must be manageable
for the majority of end users, process modelers, and developers. 

Challenge 5 \emph{Manageable Complexity }is motivated by research
into simplicity and understandability of BPM and business process
models \citep{Fahland.2009,VomBrocke.2014,Reijers.2011,Houy.2010}.
One common factor that influences simplicity and understandability
is complexity. Complexity may therefore be seen as the root problem.
Keeping a check on complexity has a myriad of benefits and is therefore
designated as challenge, as a groundbreaker for more specific issues
like understandability

\medskip{}

For all five challenges, it is important to note that for achieving
full support, it is\emph{ not only required to conceptually fully
support a particular challenge}. It is furthermore equally important,
if not more so, that the challenge is also \emph{fully supported at
run-time} by an appropriate process engine or general implementation.
Especially as some challenges require a working process engine to
be evaluated properly, e.g., Challenge 1 \emph{Asynchronous Concurrency}
and Challenge 4 \emph{Immediate Consistency}.

Specifically not an issue for process coordination is data exchange
between processes. Logically, data exchange occurs when processes
have been coordinated in order to send or receive data so that an
successful data exchange can take place. While message-based approaches
to process coordination, e.g., BPMN \citep{ObjectManagementGroup.2011},
tend to couple process coordination and data exchange, this is not
a necessary consequence. Technically, process coordination and data
exchange are separate issues. Though there are connections, e.g.,
data exchange builds upon process coordination. The separation of
process coordination and data exchange is further reinforced as data
exchange can occur by means other than messages, e.g., by writing
into shared memory or databases. Coordination processes therefore
focus on the pure coordination of processes and consequently do not
presume or prescribe the means of data exchange between processes.

\begin{figure*}[t]
\includegraphics[width=1\textwidth]{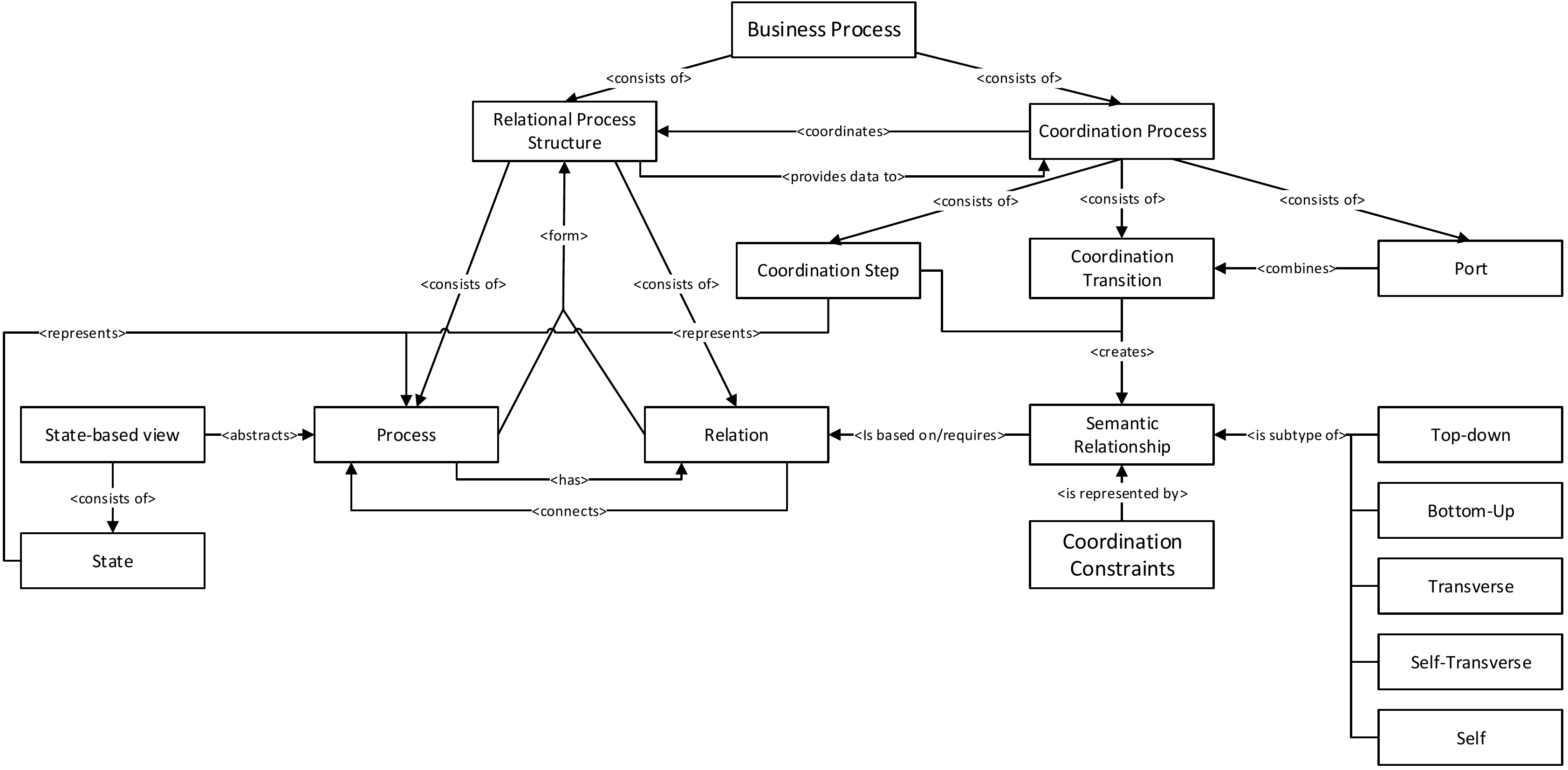}\caption{\label{fig:Essential-Object-aware-Process-Meta-Model}Essential Object-aware
Process Management Meta-Model}
\end{figure*}

It is noteworthy that for most individual challenges, attempts at
solutions or solutions exist. However, jointly considering all five
challenges requires a sophisticated approach as well as the proper
consideration of possible trade-offs. Coordination processes aim at
fulfilling exactly all five challenges, using the relational process
structure and semantic relationships (cf. Figure \ref{fig:Coordination-Processes-Trinity}).
Both concepts and the modeling parts of coordination processes are
discussed in Section \ref{sec:Background}.

\section{\label{sec:Background}Background}

The following section presents terminology and concepts required for
the definition of the operational semantics of coordination processes.

Object-aware process management is an comprehensive approach for managing
data-centric processes \citep{Kunzle.2011}. The core of object-aware
process management is presented as a meta-model in Figure \ref{fig:Essential-Object-aware-Process-Meta-Model}.
Object-aware process management describes \emph{business processes}
in terms of \emph{interacting processes}, e.g., object lifecycles,
with the goal of providing better support for data and better flexibility.
The business process only emerges through interactions between processes,
and this requires \emph{coordination} for guiding the business process
towards a meaningful goal. Note that the meta model is intended as
an overview and only represents a simplified view and does not convey
all details of either the design-time or the run-time.

The coordination approach of object-aware process management consists
of three concepts: \emph{relational process structures} capture and
track process types and their relations. \emph{Semantic Relationships
}use these relations for describing constraints for coordinating the
interactions between processes. \emph{Coordination processes} are
used for concretely specifying semantic relationships and enforcing
these constraints at run-time, relying on the information provided
by the relational process structure. Moreover, for obtaining only
the relevant information coordinating the processes, these processes
are abstracted using a state-based view. 

\textcolor{black}{The concepts that constitute and support a coordination
process are inextricably linked to each other, which necessitates
mutual references and forward references in the formal definitions
for completeness. The formal definitions mirror the implementation
of the concepts and do not contain cyclic dependencies, but simply
mutual references for navigating the resulting graph. Consequently,
formal definitions may mention concepts and entities that will only
be defined later in this section. Still, the introduction of concepts
and entities follows a logical top-down manner despite the forward
references. The intention is to keep this background section as concise
as possible while still conveying the essential information. The (mutual)
references are implicitly resolved using a globally unique identifier
(GUID) for each entity, omitted in all definitions for conciseness.
Furthermore, as this article is part of a larger body of work in context
of the PHILharmonicFlows project, the formal definitions are kept
consistent in every article.}

\begin{figure*}

\begin{centering}
\includegraphics[width=1\textwidth]{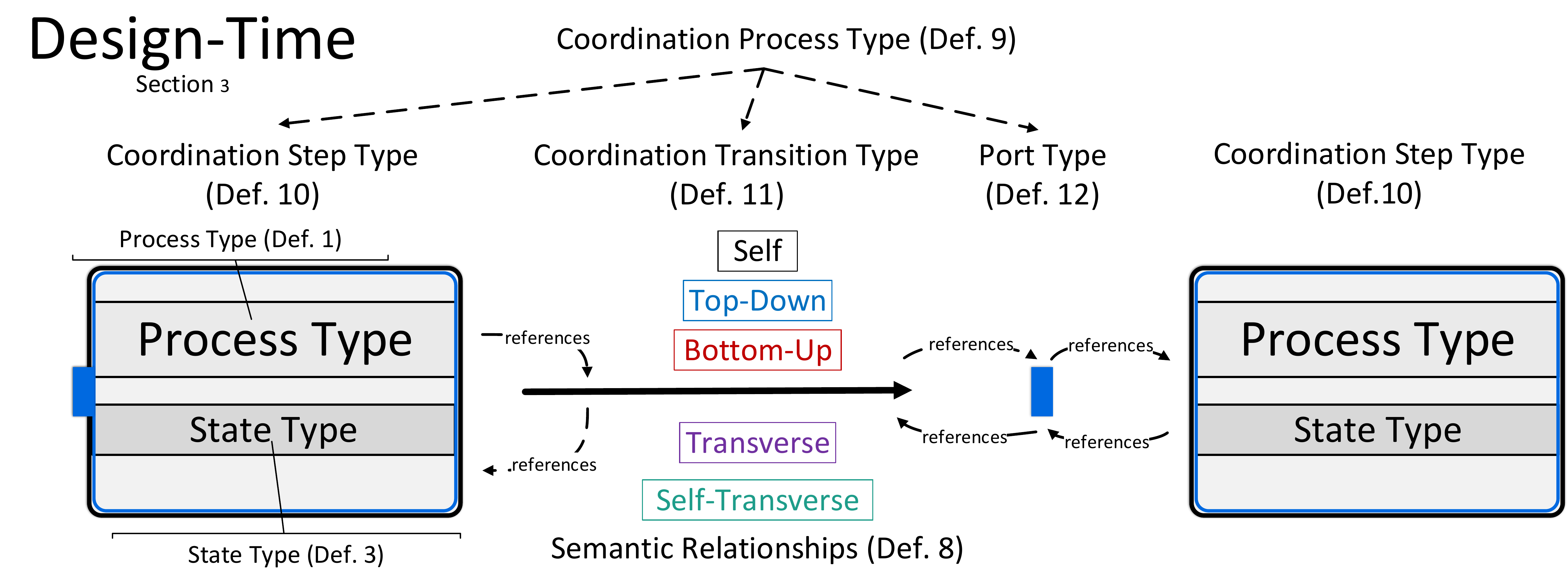}\caption{\label{fig:Design-Time-Entities-Overview}Design-Time Entities Overview}
\par\end{centering}
\end{figure*}

This section is concerned exclusively with the design-time of coordination
processes and otherwise the design-and run-time of other concepts
as far as required for understanding Sections \ref{sec:Enacting-Coordination-Processes}-\ref{sec:Operational-Semantics}.
Figure \ref{fig:Design-Time-Entities-Overview} gives a brief over
coordination process design-time entities introduced in this section
and their relationship to each other.

\subsection{Basics}

Coordination processes and their related concepts operate on a strict
distinction between design-time and run-time entities. A design-time
entity is designated as a \emph{type }(formally superscript$^{T}$),
whereas run-time entities are \emph{instances }(formally $^{I}$).
``Entity'' is used as an umbrella term comprising all types and
instances defined in the following. For the sake of brevity, when
referring to entities without a type or instance superscript or word
member, e.g., just \emph{process} instead of \emph{process instance},
this means that a statement applies to both types and instances. By
convention, instances are created by instantiating a type. The dot
(.) represents the member access operator. The symbol $<:$ signifies
the subtype relation, i.e., $x$ is a subtype of $y$ is written as
$x<:y$. Again, by convention, any set is denoted by a capital letter,
whereas an element of the set is denoted with the same lowercase letter.
The concepts that constitute and support a coordination process are
inextricably linked to each other, which necessitates mutual references
and forward references in the formal definitions for completeness.
For sake of clarity, formal definitions are presented in a top-down
manner.
\begin{definition}[Process Type]
\label{Def:ObjectType-Normal} A process type $\omega^{T}$ has the
form $(d^{T},n{\normalcolor ,\theta_{priv}^{T},\theta^{T}})$ where 
\end{definition}
\begin{itemize}
\item $d^{T}$ refers to relational process structure to which this process
type belongs (cf. Definition \ref{Def:DataModel})
\item $n$ is the unique identifier (name) of the process type
\item $\theta_{priv}^{T}$ is a process model specification not publicly
visible
\item $\theta^{T}$ is a publicly visible state-based view mapped to $\theta_{priv}^{T}$(cf.
Definition \ref{Def:StateBasedView})
\end{itemize}
While objects and their lifecycle processes have provided the initial
motivation for coordination processes, objects with lifecycle processes
are not a prerequisite for coordination processes to work. Therefore,
a generalized notion of process $\theta_{priv}$ is used that may
represent, in principle, any kind of process model specification,
except for a coordination process. Object lifecycle processes are
just one example of a process $\omega$. For the purpose of coordination
processes and their operational semantics, the paradigm and modeling
language in which processes are specified is unimportant. Consequently,
a process $\theta_{priv}$ may be an object-aware process or a process
that is specified using BPMN 2.0 \citep{ObjectManagementGroup.2011}.
Consequently, no formal definition of $\theta_{priv}$ is provided.
Instead, a state-based view $\theta$ provides an abstraction level
over the actual process specification $\theta_{priv}$ \citep{Steinau.2019b},
which is used by a coordination process. 

\begin{figure}
\begin{centering}
\includegraphics[width=0.7\columnwidth]{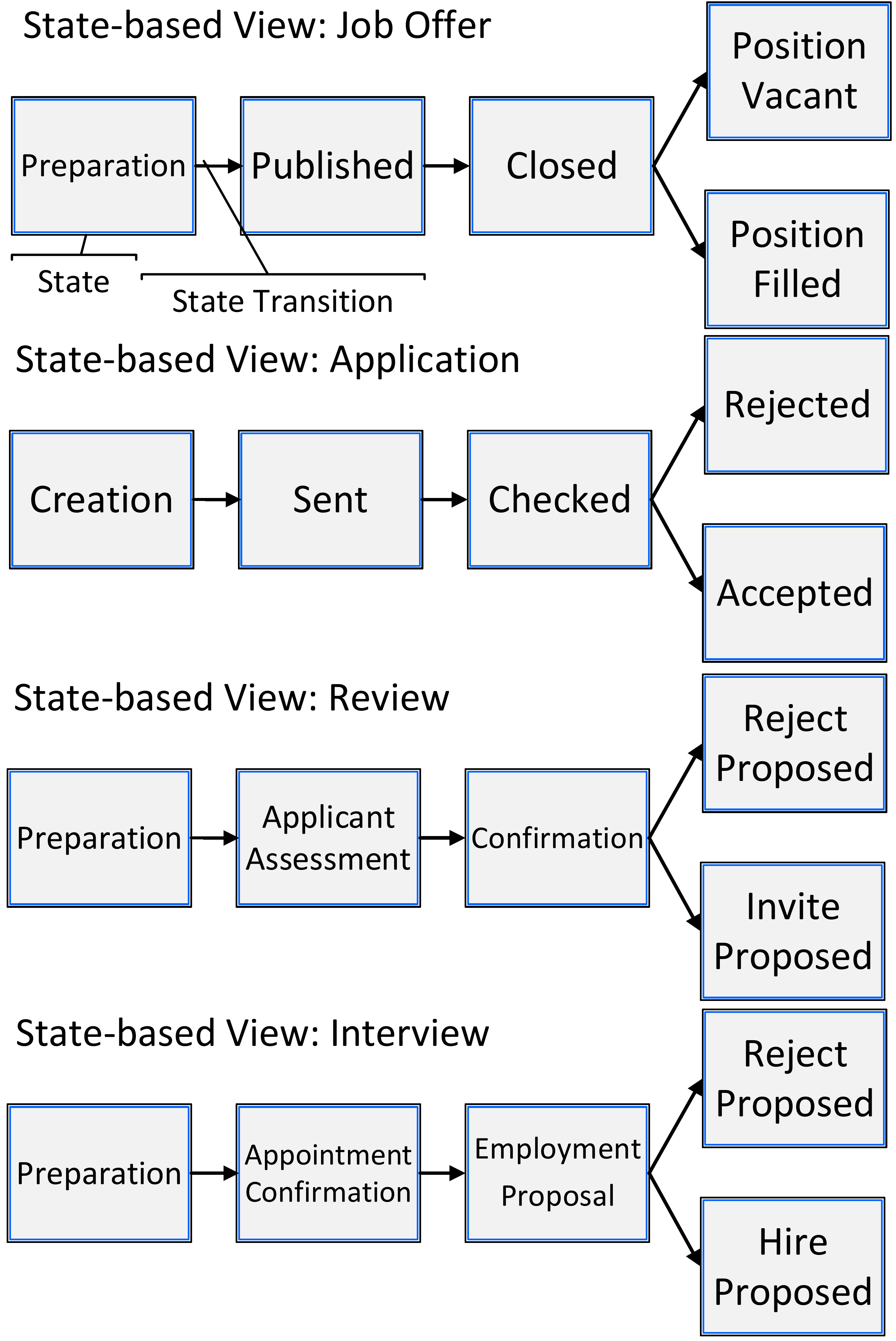}
\par\end{centering}
\caption{\label{fig:State-based-views-of}State-based Views of the Processes
from the Running Example}
\end{figure}

Thereby, every process to be coordinated $\omega$ is partitioned
into different states $\sigma$ that provide significant meaning for
process coordination. State-based views enable a coordination process
to be \emph{paradigm-agnostic}, i.e., processes from any paradigm
or even different paradigms may be coordinated. This applies to both
type and instance levels.
\begin{definition}[Process Instance]
\label{Def:ObjectInstance-Normal}A process instance $\omega^{I}$
has the form $(\omega^{T},d^{I},l,\theta_{priv}^{I},\theta^{I})$
where
\end{definition}
\begin{itemize}
\item $\omega^{T}$ refers to the process type from which $\omega^{I}$
has been instantiated (cf. Definition \ref{Def:ObjectType-Normal})
\item $d^{I}$ refers to the relational process instance structure to which
this object instance belongs (cf. Definition \ref{Def:DataStructure})
\item $l$ is the unique identifier (name) of the process instance. Default
is $\omega^{T}.n$
\item $\theta_{priv}^{I}$ is a process instance specification not publicly
visible
\item $\theta^{I}$ is a publicly visible state-based view mapped to $\theta_{priv}^{I}$(cf.
Definition \ref{Def:StateBasedView})
\end{itemize}
\emph{State-based views} partition a process specification into distinct
and non-overlapping states (cf. Definition \ref{Def:StateBasedView}).
More precisely, a state-based view $\theta$ is an abstraction over
$\theta_{priv}$, the actual process specification, mapping elements
of $\theta_{priv}$ to states of the state-based view so that each
process element (e.g., an activity) belongs to exactly one state (cf.
Figure \ref{fig:State-based-views-of})\citep{Steinau.2019b}. States
are used to indicate the progress of the underlying process $\theta_{\mathit{priv}}$.
State-based views are virtually identical for design-time and run-time,
indicated by the missing superscript. 
\begin{definition}[State-based View]
\label{Def:StateBasedView}A state-based view $\theta$ has the form
$(\omega,\Sigma,T,\Psi)$ where
\end{definition}
\begin{itemize}
\item $\omega$ refers to the process to which this state-based view belongs
(cf. Definitions \ref{Def:ObjectType-Normal} and \ref{Def:ObjectInstance-Normal})
\item $\Sigma$ is a set of states $\sigma$ 
\item $T$ is a set of transitions\textbf{ $\tau$ }
\item $\Psi$ is a set of backwards transitions $\psi$.
\end{itemize}
States $\sigma$ are connected with directed edges $\tau$ denoting
state transitions. At run-time, an \emph{active state $\sigma_{a}^{I}$}
of a process signifies its current execution status; the active state
is determined by $\theta_{priv}$, e.g., the currently executed activity
is mapped to $\sigma_{a}^{I}$. As states are an abstraction over
executable elements, e.g., activities, for the sake of abstraction,
the term \emph{executing a state} is used to refer to work being done
within the state, e.g., executing the activities in the state. Only
one state $\sigma$ may be active at a given point in time. As a consequence,
branching state transitions categorically implement an exclusive choice
semantics, i.e., states may be mutually exclusive regarding activation.
This does not prohibit parallel execution of activities, as parallelism
may still occur within a state. Note that this is in addition to the
concurrent or parallel execution of processes. As only one state may
be active, in case of mutually exclusive states, non-active states
are denoted as \emph{skipped}.\emph{ }Furthermore, state-based views
may include \emph{backwards transitions} $\psi$ that allow re-activating
a previous state $\sigma$, i.e., $\sigma$ is a predecessor of the
currently active state $\sigma_{a}$. For changing a state, transitions
$\tau$ and backwards transitions $\psi$ require an explicit commitment
per default, e.g., a user or system must explicitly commit the activation
of the transition. Figure \ref{fig:State-based-views-of} shows state-based
views of the processes referenced in Example \ref{exa:Running-Example}.

States and their transitions are, by default, the only entities that
are publicly visible to an outside observer of a process. The state
transitions $\tau^{I}$ and the active state $\sigma_{a}^{I}$ are
driven by $\theta_{priv}^{I}$. Despite the simplistic specification,
state-based views capture the essentials of a process in regard to
process coordination. In addition, if desired, state-based views may
introduce additional process properties, e.g., specific data attributes
that may subsequently be used for process coordination.

Generally, processes may be interconnected by \emph{relations}. A
relation represents a connection between two processes, indicating
one or more dependencies between them, i.e., multiple coordination
constraints can be defined over the same relation. A relation type
(cf. Definition \ref{Def:RelationType}) and relation instance (cf.
Definition \ref{Def:RelationInstance}) are defined as follows:
\begin{definition}[Relation Type]
\label{Def:RelationType}A relation type $\pi^{T}$ represents a
many-to-many relation between two processes and has the form $(\omega_{\mathit{source}}^{T},\omega_{\mathit{target}}^{T},m_{upper},m_{lower},n_{upper},n_{lower})$
where
\end{definition}
\begin{itemize}
\item $\omega_{source}^{T}$ refers to the source process type (cf. Definition
\ref{Def:ObjectType-Normal})
\item $\omega_{target}^{T}$ refers to the target process type (cf. Definition
\ref{Def:ObjectType-Normal})
\item $m_{upper}$ is an upper bound on the number of process instances
$\omega_{target}^{I}$ with which $\omega_{source}^{I}$ may be related.
Default: $m_{upper}=$$\infty$
\item $m_{lower}$ is a lower bound on the number of process instances $\omega_{target}^{I}$
with which $\omega_{source}^{I}$ may be related. Default: $m_{lower}=0$
\item $n_{upper}$ is an upper bound on the number of process instances
$\omega_{source}^{I}$ with which $\omega_{target}^{I}$ may be related.
Default: $n_{upper}=$$\infty$
\item $n_{lower}$ is a lower bound on the number of process instances $\omega_{source}^{I}$
with which $\omega_{target}^{I}$ may be related. Default: $n_{lower}=0$
\end{itemize}
As a relation type represents a many-to-many relationship, four bounds
are needed to have restrictions on both source and target sides. By
choosing appropriate bounds, a relation type may represent one-to-many
and one-to one relationships as well.
\begin{definition}[Relation Instance]
\label{Def:RelationInstance}A relation instance $\pi^{I}$ has the
form $(\pi^{T},\omega_{\mathit{source}}^{I},\omega_{\mathit{target}}^{I})$
where
\end{definition}
\begin{itemize}
\item $\pi^{T}$ refers to the relation type from which $\pi^{I}$ has been
instantiated (cf. Definition \ref{Def:RelationType})
\item $\omega_{source}^{I}$ refers to the source process instance (cf.
Definition \ref{Def:ObjectInstance-Normal})
\item $\omega_{target}^{I}$ refers to the target process instance (cf.
Definition \ref{Def:ObjectInstance-Normal})
\end{itemize}
\begin{figure}
\begin{centering}
\includegraphics[width=1\columnwidth]{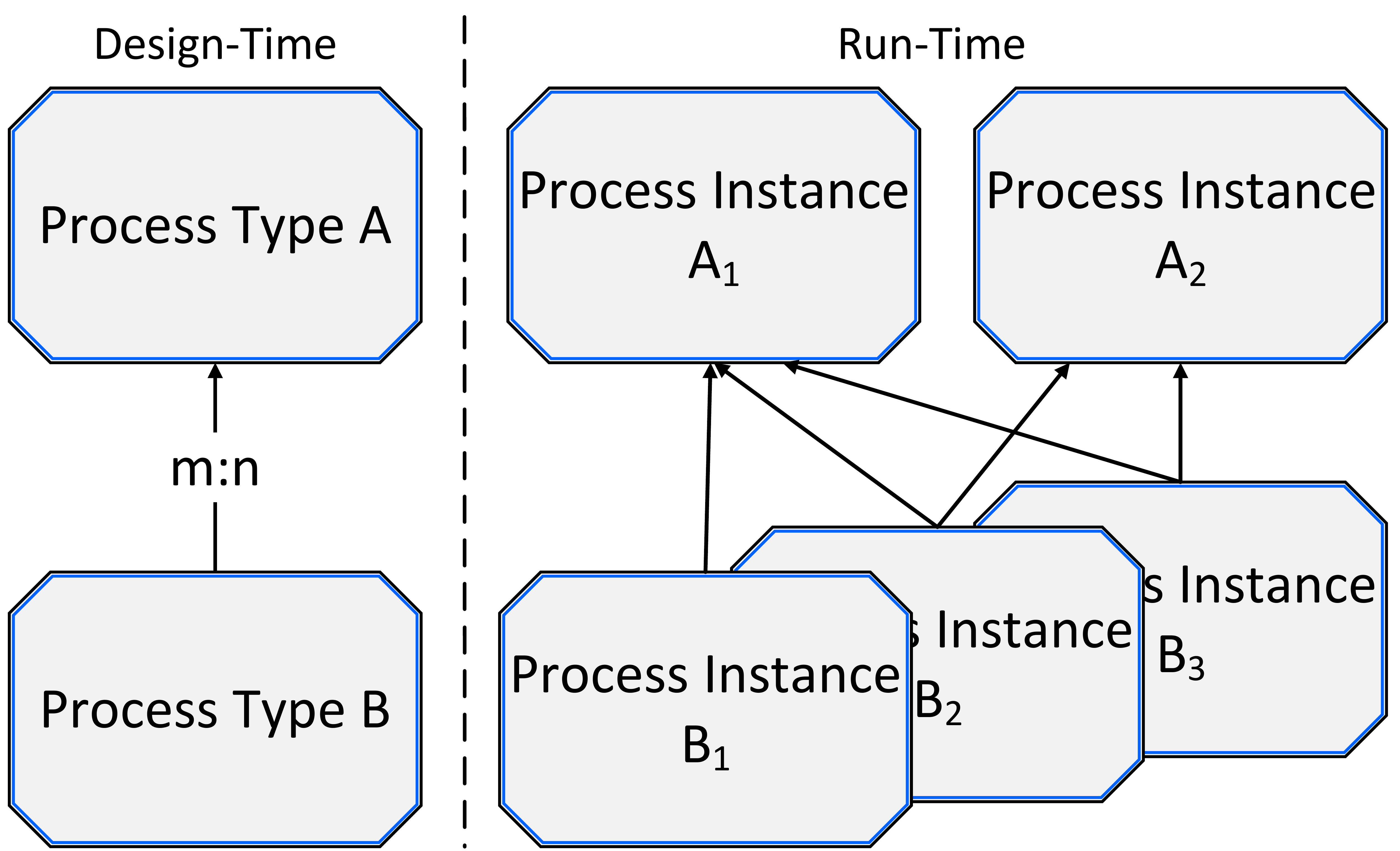}
\par\end{centering}
\caption{\label{fig:process_types_and_relations}Processes and Relations at
Design- and Run-time}
\end{figure}

Relation instances always have exactly one source and one target process
instance, as one-to-many or many-to-many relationships are comprised
of multiple relation instances $\pi^{I}$ (cf. Figure \ref{fig:process_types_and_relations}).
In particular, two processes may be related by a \emph{transitive
relation}, i.e., a path of relations exists connecting one process
with another. Contrary to, for example, Entity-Relationship-Diagrams,
relations are directed, which serves various purposes, among them
the definition of \emph{semantic relationships} (cf. Section \ref{subsec:Semantic-Relationships}).
For any process type or instance $\omega$, two sets are maintained
in regard to relations: $\Pi_{in}$ is the set of incoming relation
instances for a process instance $\omega^{I}$ , i.e., $\Pi_{in}=\{\pi\,|\,\pi.\omega_{target}=\omega^{I}\}$,
and $\Pi_{out}$, which is defined analogously for outgoing relations.
These sets allow realizing some efficiency optimizations in coordination
process execution and are therefore mentioned for accuracy. 

\subsection{\label{subsec:Relational-Process-Structures}Relational Process Structures}

Relational Process Structures provide a major building block for solving
Challenge 2 \emph{Complex Process Relations} (cf. Section \ref{subsec:Second-Challenge:-Complex}).
Moreover, they are a factor in solving Challenge 3 \emph{Local Contexts}
and Challenge 4 \emph{Immediate Consistency} (cf. Sections \ref{subsec:Third-Challenge:-Local_contexts}
and \ref{subsec:Immediate-Consistency}). At design-time, a \emph{relational
process type structure} captures all processes and their relations
(cf. Definition \ref{Def:DataModel}) \citep{Steinau.2018b}. Formally,
relational process type and relational process instance structure
(cf. Definition \ref{Def:DataStructure}) are defined as follows: 
\begin{definition}[Relational Process Type Structure]
\label{Def:DataModel}A relational process type structure $d^{T}$
has the form $(n,\Omega^{T},\Pi^{T})$ where
\end{definition}
\begin{itemize}
\item $n$ is the name of the relational process structure
\item $\Omega^{T}$ is the set of process types $\omega^{T}$ (cf. Definition
\ref{Def:ObjectType-Normal})
\item $\Pi^{T}$ is the set of relation types $\pi^{T}$ (cf. Definition
\ref{Def:RelationType})
\end{itemize}
\begin{definition}[Relational Process Instance Structure]
\label{Def:DataStructure}A relational process instance structure
$d^{I}$ has the form $(d^{T},\Omega^{I},\Pi^{I})$ where
\end{definition}
\begin{itemize}
\item $d^{T}$ refers to the relational process type structure from which
$d^{I}$ has been instantiated
\item $\Omega^{I}$ is the set of process instances $\omega^{I}$(cf. Definition
\ref{Def:ObjectInstance-Normal})
\item $\Pi^{I}$ is the set of relation instances $\pi^{I}$ (cf. Definition
\ref{Def:RelationInstance})
\end{itemize}
Relation types $\pi$ (and by extension, relation instances) that
belong to relational process structure $d$ only exist between processes
in $d.\Omega$. Creating a new relation between two process instances
is referred to as \emph{linking process instances}. The new process
instance and the new relation are then added to the respective sets
of the relational process structure the existing process instance
belongs to.

At run-time, the purpose of the relational process instance structure
is to track and capture every creation and deletion of processes and
relation instances, enabling full process relation awareness \citep{Steinau.2018b}.
Process instances may be added dynamically during run-time to an existing
relational process instance structure, each creating a new relation
between the process instance to be added and a process instance that
is already part of the relational process instance structure. 

\begin{figure}[h]
\begin{centering}
\includegraphics[width=1\columnwidth]{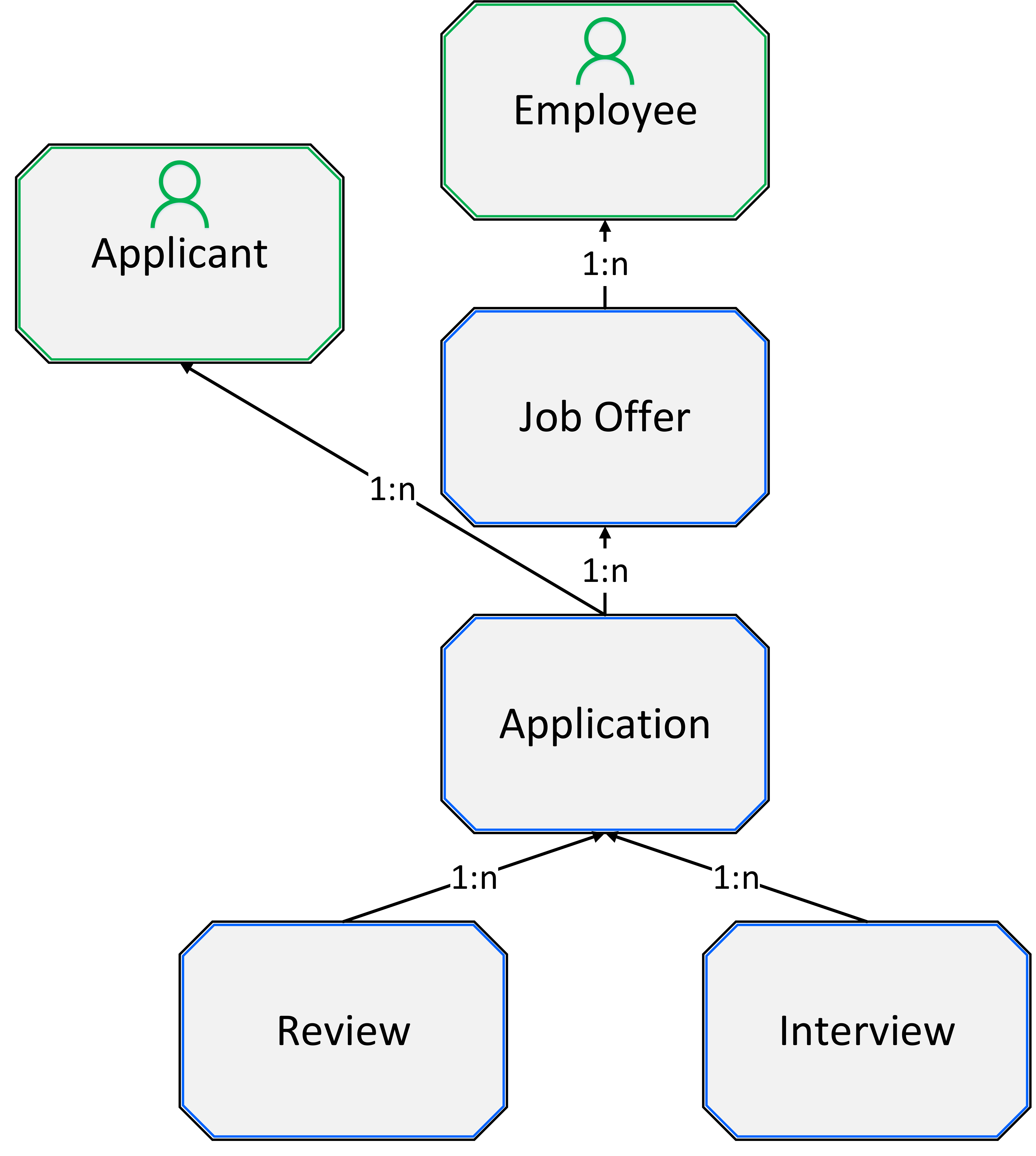}
\par\end{centering}
\caption{\label{fig:Exemplary-Relational-Process}Relational Process Type Structure
for the Running Example}
\end{figure}

A coordination process can query the relational process instance structure
to obtain up-to-date information about processes and their relations.
Figure \ref{fig:Exemplary-Relational-Process} shows an example of
a relational process type structure in context of the running example. 

A coordination process can query the relational process instance structure
to obtain up-to-date information about processes and their relations.

\noindent\begin{minipage}[t]{1\columnwidth}%
\begin{shaded}%
\begin{example}[Relational Type Structure]
Figure \ref{fig:Exemplary-Relational-Process} shows the corresponding
relational process type structure for the running example (cf. Example
\ref{exa:Running-Example}), showing various process types and their
relations. \vspace{-1em}
\end{example}
\end{shaded}%
\end{minipage}

The formal notation $\omega_{i}\twoheadrightarrow\omega_{j}$ is used
to signify a (transitive) directed relation from process $\omega_{i}$
to process $\omega_{j}$. The directed relation between processes
induce a hierarchy in a relational process structure. In this context,
the terms \emph{lower- }and\emph{ higher-level} become important\emph{.
}For illustration, $\mathit{Job\:Offer}$ is denoted as a higher-level
process in respect to process $\mathit{Application}$, as there is
a directed relation from $\mathit{Application}$ to $\mathit{Job\:Offer}$
(cf. Figure \ref{fig:Exemplary-Relational-Process}). Transitively,
$\mathit{Job\:Offer}$ is also a higher-level process to $\mathit{Review}$
and $\mathit{Interview}$. Analogously, $\mathit{Review}$ and $\mathit{Interview}$
are lower-level processes in respect to process $\mathit{Application}$.
This terminology applies to transitive relations as well. The process
types $\mathit{Applicant}$ and $\mathit{Employee}$ are user process
subtypes concerned with representing users, relevant for authorizations
and permissions in object-aware process management \citep{Andrews.2017}.

For the purpose of coordination processes, each process is required
to know all its related processes, specifically, its lower- and higher-level
processes. In order to avoid computationally expensive queries every
time lower- or higher-level instances are needed, the relational process
structure maintains two sets per process instance $\omega^{I}$: $L_{\omega^{I}}^{I}$
for all lower-level instances and $H_{\omega^{I}}^{I}$ for all higher-level
instances. These sets are kept up-to-date as the process structure
evolves, providing a performance benefit to process coordination \citep{Steinau.2018b}. 

Altogether, relational process structures allow for a coordination
approach to gain full knowledge over processes and their relations,
and thus enable fine-grained process coordination. Relational process
structures represent one foundation for coordination processes (cf.
Figure \ref{fig:Coordination-Processes-Trinity}).

\subsection{\label{subsec:Semantic-Relationships}Semantic Relationships}

Semantic relationships are means to specify \emph{coordination constraints}
at a high level of abstraction \citep{Steinau.2017}. A coordination
constraint is a formal or informal statement describing one or more
conditions or dependencies that exist between processes. For example,
the statement ``An application may only be accepted if three or more
reviews are positive'' is a coordination constraint. \emph{In essence,
process coordination is tasked with formally capturing and enforcing
coordination constraints.} Other coordination approaches, e.g., BPMN
choreographies \citep{ObjectManagementGroup.2011}, choose messages
to express the necessary interactions between the processes to be
coordinated. However, due to complex process relationships and large
numbers of process instances, defining messages in a procedural manner
is cumbersome. This is especially true for larger relational process
structures. Specifying individual messages also negatively impacts
the fulfillment of Challenge 1 \emph{Asynchronous Concurrency}. A
process modeler has to ensure asynchronous concurrency manually when
specifying messages, requiring large efforts.

\begin{figure*}[t]
\includegraphics[width=1\textwidth]{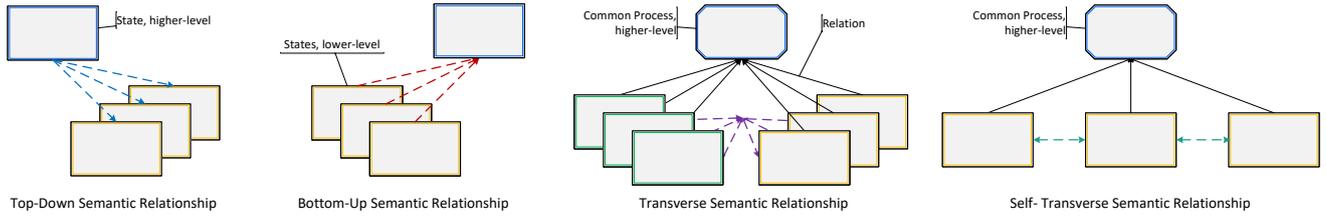}\caption{\label{fig:Semantic-Relationships}Semantic Relationships}
\end{figure*}

A coordination constraint must be expressed in terms of semantic relationships
for its use in a coordination process. \emph{A semantic relationship
describes a recurring semantic pattern inherent in the coordination
of processes in a one-to-many or many-to-many relationship} (cf. Table
\ref{tab:Overview-over-semantic}). As one example of a pattern, several
process instances may depend on the execution of one other process
instance. For a proper representation of coordination constraints,
the combination of multiple different semantic relationships might
become necessary. Moreover, a semantic relationship may only be established
between processes if a (transitive) relation within the relational
process structure, i.e., a dependency, exists between these processes.
Figure \ref{fig:Semantic-Relationships} illustrates the types of
semantic relationships between different processes in the running
example.

\begin{table}
\caption{\label{tab:Overview-over-semantic}Overview of semantic relationships}
\begin{tabular*}{1\columnwidth}{@{\extracolsep{\fill}}>{\raggedright}p{0.2\columnwidth}>{\raggedright}p{0.8\columnwidth}}
\toprule 
{\footnotesize{}Name} & {\footnotesize{}Description of the semantic relationship}\tabularnewline
\midrule
\midrule 
{\footnotesize{}Top-Down} & {\footnotesize{}}%
\parbox[t]{0.96\linewidth}{%
{\footnotesize{}The execution of one or more lower-level processes
depends on the execution status of one common higher-level process.}%
}\tabularnewline
\midrule 
{\footnotesize{}Bottom-Up} & {\footnotesize{}}%
\parbox[t]{0.96\linewidth}{%
{\footnotesize{}The execution of one higher-level process depends
on the execution status of one or more lower-level processes of the
same type.}%
}\tabularnewline
\midrule 
{\footnotesize{}Transverse} & {\footnotesize{}}%
\parbox[t]{0.96\linewidth}{%
{\footnotesize{}The execution of one or more processes is dependent
on the execution status of one or more processes of different type.
Both types of processes have a common higher-level process.}%
}\tabularnewline
\midrule 
{\footnotesize{}Self} & {\footnotesize{}}%
\parbox[t]{0.96\linewidth}{%
{\footnotesize{}The execution of a process depends upon the completion
of a previous step of the same process.}%
}\tabularnewline
\midrule 
{\footnotesize{}Self-Transverse} & {\footnotesize{}}%
\parbox[t]{0.96\linewidth}{%
{\footnotesize{}The execution of a process depends on the execution
status of other processes of the same type. All processes have a common
higher-level process.}%
}\tabularnewline
\bottomrule
\end{tabular*}\vspace{-0.3cm}
\end{table}

Semantic relationships are specified at design-time in the context
of a coordination process. Formally, a semantic relationship $s^{T}$
is defined as follows.
\begin{definition}
\label{Def:CoordinationComponentType}A semantic relationship $s^{T}$
has the form $(\mathit{\iota},\lambda,\Sigma_{valid}^{T},\omega_{ca}^{T})$
where 
\end{definition}
\begin{itemize}
\item $\mathit{\iota}$ is the identifier of the semantic relationship,
$\mathit{\iota}\in\{\mathit{top\mhyphen down},\mathit{\mathit{bottom\mhyphen up},\mathit{transverse},\mathit{self},}$\\
$\mathit{\mathit{self\mhyphen transverse}}\}$
\item $\lambda$ is an expression, configuring $s^{T}$ in case of $\iota\in\{bottom\mhyphen up,transverse,self\mhyphen transverse\}$
\item $\Sigma_{valid}^{T}$ is a set of state types in case of $\iota\in\{top\mhyphen down\}$.
\item $\omega_{ca}^{T}$ refers to the common ancestor of $s^{T}$ in case
of $\iota\in\{transverse,self\mhyphen transverse\}$
\end{itemize}
Semantic relationships are always defined between two (not necessarily
different) types of processes. Different semantic relationships, determined
by the identifier $\mathit{\iota}$, signify different basic constraints
(cf. Table \ref{tab:Overview-over-semantic}). One of the outstanding
features of semantic relationships is that the appropriate semantic
relationship between processes can be automatically inferred from
a relational process structure. This is possible as the direction
of the relations directly implies certain semantic relationships between
process types \citep{Steinau.2018b}. This is exemplified in Example
\ref{exa:SemanticRelsTopDownBottomUpProcessTypes}.

\noindent\begin{minipage}[t]{1\columnwidth}%
\begin{shaded}%
\begin{example}[\emph{Top-Down and Bottom-Up Semantic Relationships I}]
\emph{\label{exa:SemanticRelsTopDownBottomUpProcessTypes}} Consider
Figure \ref{fig:Exemplary-Relational-Process}: A top-down semantic
relationship can be established from\textit{\emph{ $\mathit{Job\:Offer}$
to an}} $\mathit{Application}$, as there is a relation from $\mathit{Application}$
to\textit{\emph{ $\mathit{Job\:Offer}$. Additionally, a bottom-up
semantic relationship can be established from}} $\mathit{Application}$
to a \textit{\emph{$\mathit{Job\:Offer}$. The direction of the connection
and the direction of the relation determine directly the type of semantic
relationship. Note also that one relation supports establishing multiple
semantic relationships on top.}}\vspace{-0.3cm}
\end{example}
\end{shaded}%
\end{minipage}

The \emph{execution status} referred to in Table \ref{tab:Overview-over-semantic}
is represented by the state-based view of the process (cf. Section
\ref{sec:Background}). At run-time, each semantic relationship has
a logical value to indicate whether or not it is satisfied; Boolean
operators may be used to express more complicated coordination logic
involving more than one semantic relationship. Semantic relationships
have been designed with Challenge 1 \emph{Asynchronous Concurrency}
in mind. The details on how Challenge 1 is fulfilled are presented
in Sections \ref{sec:Representing-Process-Instances}-\ref{sec:Technical-Implementation}.

Semantic relationships feature an expression in case of a bottom-up,
transverse, or self-transverse semantic relationship. Top-down semantic
relationships feature a state set. Self semantic relationships cannot
be configured and do not possess an expression or a state set (cf.
Definition \ref{Def:CoordinationComponentType}). Instead, they may
be addressed collectively by using the umbrella term \emph{coordination
condition}. A coordination condition modifies the basic semantics
of the semantic relationship (cf. Table \ref{tab:Overview-over-semantic}),
which is needed to customize a semantic relationship to specifically
represent a coordination constraint. Details on the coordination conditions
can be found in \citep{Steinau.2018}.

\subsection{\label{subsec:Coordination-Processes}Coordination Process Types}

\emph{Coordination processes} are a generic concept for coordinating
interdependent processes by expressing coordination constraints with
the help of semantic relationships, which are then enforced at run-time
\citep{Steinau.2018}. The concept allows specifying sophisticated
coordination constraints for vast structures of interrelated process
instances with an expressive, high-level graphical notation using
a minimum number of modeling elements.

\begin{figure}
\centering{}\includegraphics[width=1\columnwidth]{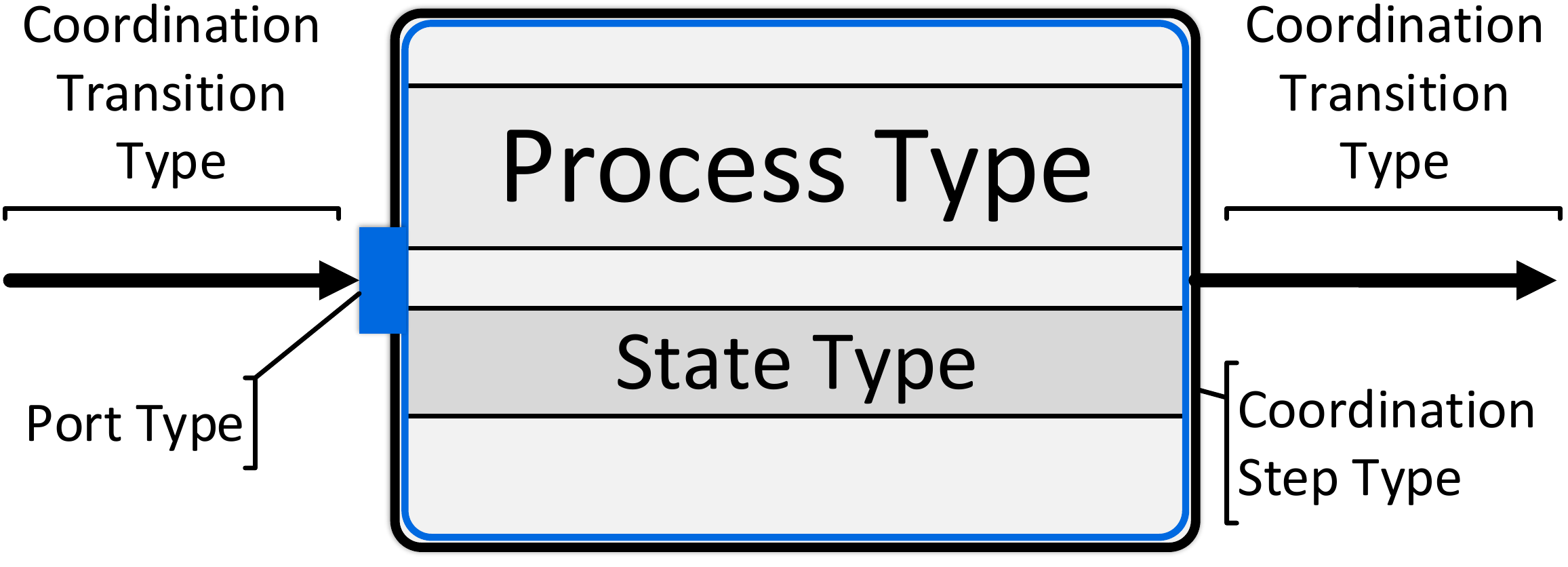}\caption{\label{fig:Coordination-Process-Modeling}Coordination Process Modeling
Elements}
\end{figure}

\begin{figure*}[t]
\begin{centering}
\includegraphics[width=1\textwidth]{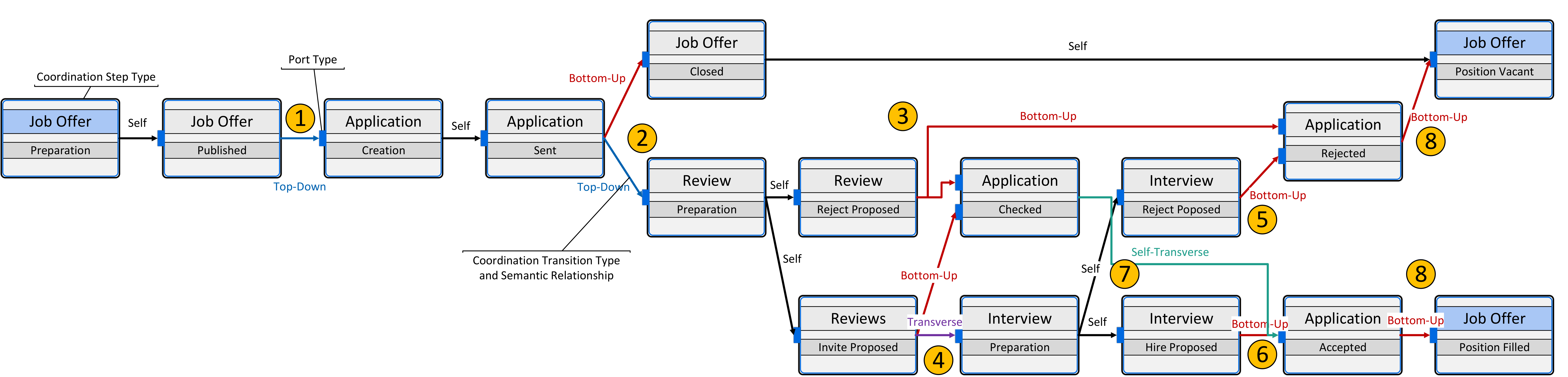}
\par\end{centering}
\caption{\label{fig:Coordination-Process-for}Coordination Process for the
Running Example}
\end{figure*}

A coordination process type is a design-time entity, which is represented
as a directed, connected, and acyclic graph that consists of \emph{coordination
step types},\emph{ coordination transition types,} and\emph{ port
types} (cf. Figure \ref{fig:Coordination-Process-Modeling}). A formal
definition of coordination process types is presented in Definition
\ref{Def:CoordinationProcessType}. Figure \ref{fig:Coordination-Process-for}
shows the coordination process type for the processes of the running
example, which ensures the correct enactment of the overall recruitment
business process.
\begin{definition}[Coordination Process Type]
\label{Def:CoordinationProcessType}A coordination process type $c^{T}$
has the form $(\omega_{coord}^{T},B^{T},$\\
$\Delta^{T},H^{T})$ where 
\end{definition}
\begin{itemize}
\item $\omega_{\mathit{coord}}^{T}$ refers to the process type to which
the coordination process type $c^{T}$ belongs
\item $B^{T}$ is a set of coordination step types $\beta^{T}$ (cf. Definition
\ref{Def:CoordinationStepType})
\item $\Delta^{T}$ is a set of coordination transition types $\delta^{T}$
(cf. Definition \ref{Def:CoordinationTransitionType})
\item $H^{T}$ is a set of port types $\eta^{T}$ (cf. Definition \ref{Def:PortType})
\end{itemize}
Coordination steps are the vertices of the graph referring to a process
type $\omega^{T}$ as well as to one of the states $\sigma^{T}$ of
its state-based view $\theta^{T}$, e.g. \textit{\emph{$\mathit{Job\:Offer}$}}
and state \textit{\emph{$\mathit{Published}$}}. For the sake of convenience,
a coordination step $\beta^{T}$ is addressed with referenced process
type and state in the form of \textit{\emph{$\mathit{ProcessType}$}}:$\mathit{State}$,
e.g. $\mathit{Job\:Offer}$:$\mathit{Published}$. A formal definition
for coordination steps is presented in Definition \ref{Def:CoordinationStepType}.
\begin{definition}[Coordination Step Type]
\label{Def:CoordinationStepType}A coordination step type $\beta^{T}$
has the form $(c^{T},\omega^{T},\sigma^{T},\Delta_{out}^{T},$\\
$H^{T})$ where
\end{definition}
\begin{itemize}
\item $c^{T}$ refers to the coordination process type (cf. Definition \ref{Def:CoordinationProcessType})
\item $\omega^{T}$ refers to a process type (cf. Definition \ref{Def:ObjectType-Normal})
\item $\sigma^{T}$ refers to a state type belonging to $\omega^{T}$, i.e.,
$\sigma^{T}\in\omega^{T}.\theta^{T}.\Sigma^{T}$ (cf. Definition \ref{Def:StateBasedView})
\item $\Delta_{out}^{T}$ is a set of outgoing coordination transition types
$\delta^{T}$ (cf. Definition \ref{Def:CoordinationTransitionType})
\item $H^{T}$ is a set of port types $\eta^{T}$ (cf. Definition \ref{Def:PortType})
\end{itemize}
A coordination transition $\delta^{T}$ is a directed edge that connects
a \emph{source coordination step type }$\beta_{\mathit{src}}^{T}$
with a \emph{target coordination step type }$\beta_{\mathit{tar}}^{T}$
(cf. Figure \ref{fig:Coordination-Process-for} and Definition \ref{Def:CoordinationTransitionType}). 
\begin{definition}[Coordination Transition Type]
\label{Def:CoordinationTransitionType}A coordination transition
type $\delta^{T}$ has the form $(\beta_{src}^{T},\eta_{tar}^{T},$\\
$s^{T})$ where
\end{definition}
\begin{itemize}
\item $\beta_{src}^{T}$ refers to the source coordination step type (cf.
Definition \ref{Def:CoordinationStepType})
\item $\eta_{tar}^{T}$ refers to the target port type (cf. Definition \ref{Def:PortType})
\item $s^{T}$ is a semantic relationship between $\beta_{src}^{T}.\omega^{T}$
and $\eta_{tar}^{T}.\beta^{T}.\omega^{T}$
\end{itemize}
More precisely, $\delta^{T}$ connects to one of multiple ports $\eta_{tar}^{T}$
that are attached to $\beta_{\mathit{tar}}^{T}$. A formal definition
of ports is shown in Definition \ref{Def:PortType}
\begin{definition}[Port Type]
\label{Def:PortType}A port type $\eta^{T}$ has the form $(\beta^{T},\Delta_{in}^{T})$
where
\end{definition}
\begin{itemize}
\item $\beta^{T}$ refers to the coordination step type to which this port
type belongs (cf. Definition \ref{Def:CoordinationStepType})
\item $\Delta_{in}^{T}$ is the set of all incoming coordination transitions
$\delta^{T}$ (cf. Definition \ref{Def:CoordinationTransitionType})
\end{itemize}
By creating a coordination transition between source step $\beta_{src}^{T}$
and target step $\beta_{tar}^{T}$, a semantic relationship $s^{T}$
is created as well. Conceptually, a semantic relationship is attached
to a coordination transition. With the relations from the relational
process structure and the definitions of semantic relationships (cf.
Table \ref{tab:Overview-over-semantic}), the identifier $\mathit{\iota}$
can be automatically derived. The identifier $\mathit{\iota}$ determines
which semantic relationship is established between the process types
referenced by the two coordination steps. 

\noindent\begin{minipage}[t]{1\columnwidth}%
\begin{shaded}%
\begin{example}[\emph{Top-Down and Bottom-Up Semantic Relationships}]
\emph{} Connecting\textit{\emph{ $\mathit{Job\:Offer}$:$\mathit{Published}$}}
with $\mathit{Application}$:$\mathit{Creation}$ constitutes a top-down
relationship (cf. Figure \ref{fig:Coordination-Process-for}). The
sequence in which the steps occur is important for determining the
type of semantic relationship. By connecting $\mathit{Application}$:$\mathit{Sent}$
with $\mathit{Job\:Offer}$:$\mathit{Closed}$, a bottom-up semantic
relationship is established instead, as $\mathit{Application}$ is
a lower-level process type of $\mathit{Job\:Offer}$.\vspace{-0.3cm}
\end{example}
\end{shaded}%
\end{minipage}

As coordination transitions represent coordination constraints with
semantic relationships, coordination constraints depend on previous
constraints for fulfillment. In Example \ref{exa:SemanticRelsTopDownBottomUpProcessTypes},
activating $\mathit{Job\:Offer}$:$\mathit{Closed}$ requires at least
one $\mathit{Application}$ in state $\mathit{Sent}$, which in turn
requires \textit{\emph{$\mathit{Job\:Offer}$:$\mathit{Published}$
to be activated. The coordination constraint between $\mathit{Job\:Offer}$:$Closed$
and }}$\mathit{Application}$:$\mathit{Sent}$\textit{\emph{ depends
on the constraint between $\mathit{Job\:Offer}$:$\mathit{Published}$
and }}$\mathit{Application}$:$\mathit{Creation}$. Therefore, coordination
process graphs must be acyclic, otherwise cyclic dependencies and,
therefore, deadlocks are possible. Consequently, the acyclicity of
coordination processes is not a restriction of expressiveness, but
a requirement for correctness.

Moreover, a coordination process is not required to coordinate all
processes at every point in time. Depending on the coordination constraints,
only the processes and states that are necessary for these constraints
need to be modeled and are therefore subject to coordination. States
and processes that do not occur in a coordination process model are
not constrained in their execution by process coordination. Consequently,
coordination process allow for a high degree of freedom in executing
processes by only providing coordination when absolutely required.

Ports allow realizing different semantics for combining semantic relationships
\citep{Steinau.2018}. Connecting multiple coordination transitions
to the same port corresponds to AND-semantics, i.e., all semantic
relationships attached to the incoming transitions must be enabled
for the port to become enabled as well. Enabling a port also enables
the coordination step, allowing the state of the coordination step
to become active. Generally, at least one port of a coordination step
must be enabled for the coordination step to become enabled as well.
Consequently, connecting transitions to different ports of the same
coordination step corresponds to OR-semantics.

A coordination process $c$ is a directed, acyclic graph which possesses
exactly one start coordination step $\beta_{\mathit{start}}$ $\in c.B$
and a finite set of end coordination step types $B_{\mathit{end}}\subset c.B$.
The notions of start and end coordination step apply equally to types
and instances. A start coordination step type has no port types and
consequently no incoming transitions, i.e., $\beta_{\mathit{start}}.H=\emptyset$.
Analogously, an end coordination step $\beta_{\mathit{end}}$ has
no outgoing transitions, i.e., $\beta_{\mathit{end}}.\Delta_{\mathit{out}}=\emptyset$.
Coordination process enactment begins at the start step $\beta_{\mathit{start}}$
and terminates after reaching an end step $\beta_{\mathit{end}}\in B_{\mathit{end}}$.

A coordination process is attached to a particular process type within
the relational process structure. This process type is denoted as
a \emph{coordinating process type} $\omega_{coord}^{T}$. Note that
$\omega_{coord}^{T}$ is a short-hand notation for a process $\omega_{i}^{T}$
being a coordinating process type, i.e, $\exists c:c.\omega^{T}=\omega_{i}^{T}$,
and does not signify one specific process. There may be many processes
$\omega_{i}^{T}$ in a relational process structure that are coordinating
processes, i.e., $|\{w_{i}^{T}\,|\,\exists c_{i}^{T}:c_{i}^{T}.\omega^{T}=\omega_{i}^{T}\}|\geq1$
, as many coordination processes $c_{i}^{T}$ may be used to coordinate
the same relational process structure $d^{T}$. The notion of coordinating
process also applies at the instance level, i.e., there may be one
or more coordinating process instances. For the running example (cf.
Example \ref{Running-Example:-Recruitment}), $\mathit{Job\,Offer}$
is designated as the coordinating process type. 

The \emph{scope} of a coordination process determines which processes
$\omega^{T}$ it needs to coordinate, in relation to the coordinating
process type $\omega_{\mathit{coord}}^{T}$. In general, the scope
is defined as all lower-level process types $L_{\omega_{coord}^{T}}^{T}$
of the coordinating process type, which includes $\omega_{\mathit{coord}}^{T}$
itself. The processes contained in $L_{\omega_{coord}^{T}}^{T}$ are
called the \emph{coordinated processes}. In the running example, the
$\mathit{Job\,Offer}$ coordination process is responsible for coordinating
$\mathit{Job}$ $\mathit{Offers}$, $\mathit{Applications}$, $\mathit{Reviews}$,
and $\mathit{Interviews}$. Coordinating process type and the concept
of scope serve for dealing with large process structures, coordinated
by multiple coordination processes. 

In the following, this paper gives a rundown of the coordination process
(cf. Figure \ref{fig:Coordination-Process-for}) of the running example
(cf. Example \ref{Running-Example:-Recruitment}) and the \emph{most
important} coordination constraints. Encircled numbers \raisebox{.5pt}{\textcircled{\raisebox{-.9pt} {n}}}
represent points of interest in Figure \ref{fig:Coordination-Process-for}.\vspace{-0.3cm}

\noindent\begin{minipage}[t]{1\columnwidth}%
\begin{shaded}%
\begin{example}[Coordination Process Rundown]
\label{exa:Rundown}

Any $\mathit{Job\:Offer}$ process begins enactment in the start state
$\mathit{Preparation}$, represented by the start coordination step
type of the coordination process. The outgoing self semantic relationship
signifies the transition of the $\mathit{Job\:Offer}$ to state $\mathit{Published}$.
Then, Coordination Constraint \ref{const:coordConstraint1} is represented
using a top-down semantic relationship \raisebox{.5pt}{\textcircled{\raisebox{-.9pt} {1}}}.
\end{example}
\begin{constraint}
\label{const:coordConstraint1}An application may only be created
as long as the corresponding job offer is published.
\end{constraint}
After coordination step type $\mathit{Application}$:$\mathit{Creation}$,
a self semantic relationship allows an $\mathit{Application}$ to
transition to state $\mathit{Sent}$. When in state $\mathit{Sent}$,
$\mathit{Reviews}$ may be created for the $\mathit{Application}$
(cf. Coordination Constraint \ref{const:coordConstraint2}), again
represented by a top-down semantic relationship \raisebox{.5pt}{\textcircled{\raisebox{-.9pt} {2}}}.
Multiple lower-level processes ($\mathit{Reviews}$) depend upon the
execution status (state $\mathit{Sent}$) of one higher-level process
(the $\mathit{Application}$) (cf. Table \ref{tab:Overview-over-semantic}).
\begin{constraint}
\label{const:coordConstraint2}An application may only be reviewed
once it has been sent to the company.
\end{constraint}
Moreover, at least one $\mathit{Application}$ in state $\mathit{Sent}$
allows a $\mathit{Job\:Offer}$ to reach state $\mathit{Closed}$
(cf. Coordination Constraint \ref{const:coordConstraint3}). For representing
the coordination constraint, a bottom-up semantic relationship is
established between coordination step types $\mathit{Application}$:$\mathit{Sent}$
and $\mathit{Job\:Offer}$:$\mathit{Closed}$ \raisebox{.5pt}{\textcircled{\raisebox{-.9pt} {2}}}.
It is a bottom-up semantic relationship as $\mathit{Job\:Offer}$
is a higher-level process of $\mathit{Application}$ (cf. Table \ref{tab:Overview-over-semantic}).
\begin{constraint}
\label{const:coordConstraint3}A job offer may be closed after at
least one application has been received.
\end{constraint}
Coordination Constraint \ref{const:coordConstraint4} determines when
$\mathit{Applica\mhyphen}$\LyXZeroWidthSpace$\mathit{tions}$ may
reach state $\mathit{Rejected}$ or when $\mathit{Interviews}$ may\vspace{-0.2em}
\end{shaded}%
\end{minipage}

\noindent\begin{minipage}[t]{1\columnwidth}%
\begin{shaded}%
be created. Rejection is handled by a bottom-up semantic relationship
between coordination step types $\mathit{Review}$:$\mathit{Reject\:Proposed}$
and $\mathit{Application}$:$\mathit{Rejected}$ \raisebox{.5pt}{\textcircled{\raisebox{-.9pt} {3}}}.
The precise semantics of the bottom-up semantic relationship is accomplished
with an expression $\lambda$ (cf. Definition \ref{Def:CoordinationComponentType}).
\begin{constraint}
\label{const:coordConstraint4}An interview with the applicant may
only be performed if at least three reviews or a simple majority of
reviews are in favor of the applicant. Applications for which this
is not the case must be rejected.
\end{constraint}
In case of favorable $\mathit{Reviews}$, a transverse semantic relationship
is established between $\mathit{Review}$:\LyXZeroWidthSpace$\mathit{Invite\:Proposed}$
and $\mathit{Interview}$:$Preparation$ \raisebox{.5pt}{\textcircled{\raisebox{-.9pt} {4}}}.
$\mathit{Interviews}$ depend on $\mathit{Reviews}$ in the context
of a particular $\mathit{Application}$ (cf. Table \ref{tab:Overview-over-semantic}).
The $\mathit{Application}$ serves as the common ancestor $\omega_{ca}$
of the transverse semantic relationship (cf. Definition \ref{Def:CoordinationComponentType}).
The precise semantics of the transverse semantic relationship are
again accomplished with an expression $\lambda$ (cf. Definition \ref{Def:CoordinationComponentType}).
In case of unfavorable reviews, the $\mathit{Application}$ must be
rejected. $\mathit{Interview}$:$\mathit{Reject\:Proposed}$ is connected
to a second port of coordination step $\mathit{Application}$:$\mathit{Rejected}$
\raisebox{.5pt}{\textcircled{\raisebox{-.9pt} {5}}}. 

This constitutes OR-Semantics, as an $\mathit{Application}$ may be
rejected because of unfavorable $\mathit{Reviews}$ or unfavorable
$\mathit{Interviews}$. After $\mathit{Interviews}$ have been created
and conducted, another assessment of the applicant is accomplished.
In case of favorable $\mathit{Interviews}$, the $\mathit{Application}$
may be $\mathit{Accepted}$ (cf. Coordination Constraint \ref{const:coordConstraint5}).
Therefore, a bottom-up semantic relationship is established between
$\mathit{Interview}$:$\mathit{Hire\:Proposed}$ and $\mathit{Application}$:$\mathit{Accepted}$
\raisebox{.5pt}{\textcircled{\raisebox{-.9pt} {6}}}.
\begin{constraint}
\label{const:coordConstraint5}At least one interview or a simple
majority of interviews must be in favor of the applicant before the
applicant can be accepted for the job offer
\end{constraint}
In addition to the bottom-up semantic relationship representing Coordination
Constraint \ref{const:coordConstraint5}, another coordination constraint
affects the acceptance of an $\mathit{Application}$ \end{shaded}%
\end{minipage}

\noindent\begin{minipage}[t]{1\columnwidth}%
\begin{shaded}%
(cf. Coordination Constraint \ref{const:coordConstraint6}).
\begin{constraint}
\label{const:coordConstraint6}Only one applicant may be accepted
for a job offer.
\end{constraint}
For Coordination Constraint \ref{const:coordConstraint6}, $\mathit{Applications}$
depend on other $\mathit{Applications}$, hence a self-transverse
semantic relationship is established \raisebox{.5pt}{\textcircled{\raisebox{-.9pt} {7}}}
(cf. Table \ref{tab:Overview-over-semantic}). The self-transverse
semantic relationship permits only one $\mathit{Application}$ to
reach state $\mathit{Accepted}$, whereas other $\mathit{Applications}$
are blocked. The self-transverse semantic relationship connects to
the same port of $\mathit{Application}$:$\mathit{Accepted}$ as the
previous bottom-up semantic relationship. This represents AND-semantics,
as both Coordination Constraints \ref{const:coordConstraint5} and
\ref{const:coordConstraint6} need to be fulfilled simultaneously.

Finally, Coordination Constraint \ref{const:coordConstraint7} determines
under which conditions a $\mathit{Job\:Offer}$:$\mathit{Closed}$
may terminate.
\begin{constraint}
\label{const:coordConstraint7}The job offer is successfully completed
when an applicant has been found. If no suitable applicant is found,
the job offer ends with status ``Position vacant''.
\end{constraint}
The representation of Coordination Constraint \ref{const:coordConstraint7}
must be split into two semantic relationships. One bottom-up semantic
relationship established between $\mathit{Application}$:$\mathit{Rejected}$
and $\mathit{Job\:Offer}$:$\mathit{Position\:Vacant}$ represents
the case where no suitable applicant could be found. A second bottom-up
semantic relationship between $\mathit{Application}$:$\mathit{Accepted}$
and $\mathit{Job\:Offer}$:$\mathit{Position\:Filled}$ represents
the opposite case, i.e., a suitable applicant could be found \raisebox{.5pt}{\textcircled{\raisebox{-.9pt} {8}}}. \end{shaded}%
\end{minipage}

How a coordination process can fulfill its role and enforce these
coordination constraints at run-time is explained in the subsequent
sections. The enactment of a coordination process involves two primary
components: The \emph{accurate representation of process instances,
their relations, and their semantic relationships} in a coordination
process at run-time, and the comprehensive \emph{operational semantics}
to enforce the constraints imposed by semantics relationships on the
coordinated process instances.

\section{\label{sec:Enacting-Coordination-Processes}Enacting Coordination
Processes}

A coordination process model represents coordination constraints between
multiple process types in terms of (multiple) semantic relationships
(cf. Section \ref{subsec:Semantic-Relationships}). The process types
to be coordinated and their relations are captured in a relational
process structure (cf. Section \ref{subsec:Relational-Process-Structures}).
At run-time, multiple process instances may be created from each of
these types, which then form relations to other instances. Compared
to the design-time, this represents an enormous additional complexity.
The specific challenges of the run-time have been discussed in Section
\ref{sec:Challenges-and-Problem}. Consequently, a coordination process
instance is required to deal with this complexity adequately if proper
process coordination shall be provided.

One of the main ideas of a coordination process is that it may be
enacted, similarly to any regular process. A coordination process
is not simply a collection of coordination constraints, but the representation
of constraints is done in a process-like fashion. An instance of a
coordination process has a start and an end, as well as steps in between
signifying important goals for process coordination. Through the enactment
of a coordination process, it enforces the correct coordination constraints
at the appropriate time. Furthermore, a coordination process is able
to react to changing circumstances, e.g., when a new process instance
emerges or existing process instances are deleted. This enactment
can be split into two parts: The first part consists of the correct
representation of process instances and their semantic relationships,
taking Challenge 2 \emph{Complex Process Relations} and Challenge
3 \emph{Local Contexts} into account. Based on this representation,
the operational semantics, constituting the second part, describes
how the coordination process is enacted, i.e., which coordination
constraint is fulfilled or not at which point in time. In the following,
the general properties of run-time coordination processes are discussed. 

\begin{figure*}[t]
\centering{}\includegraphics[width=1\textwidth]{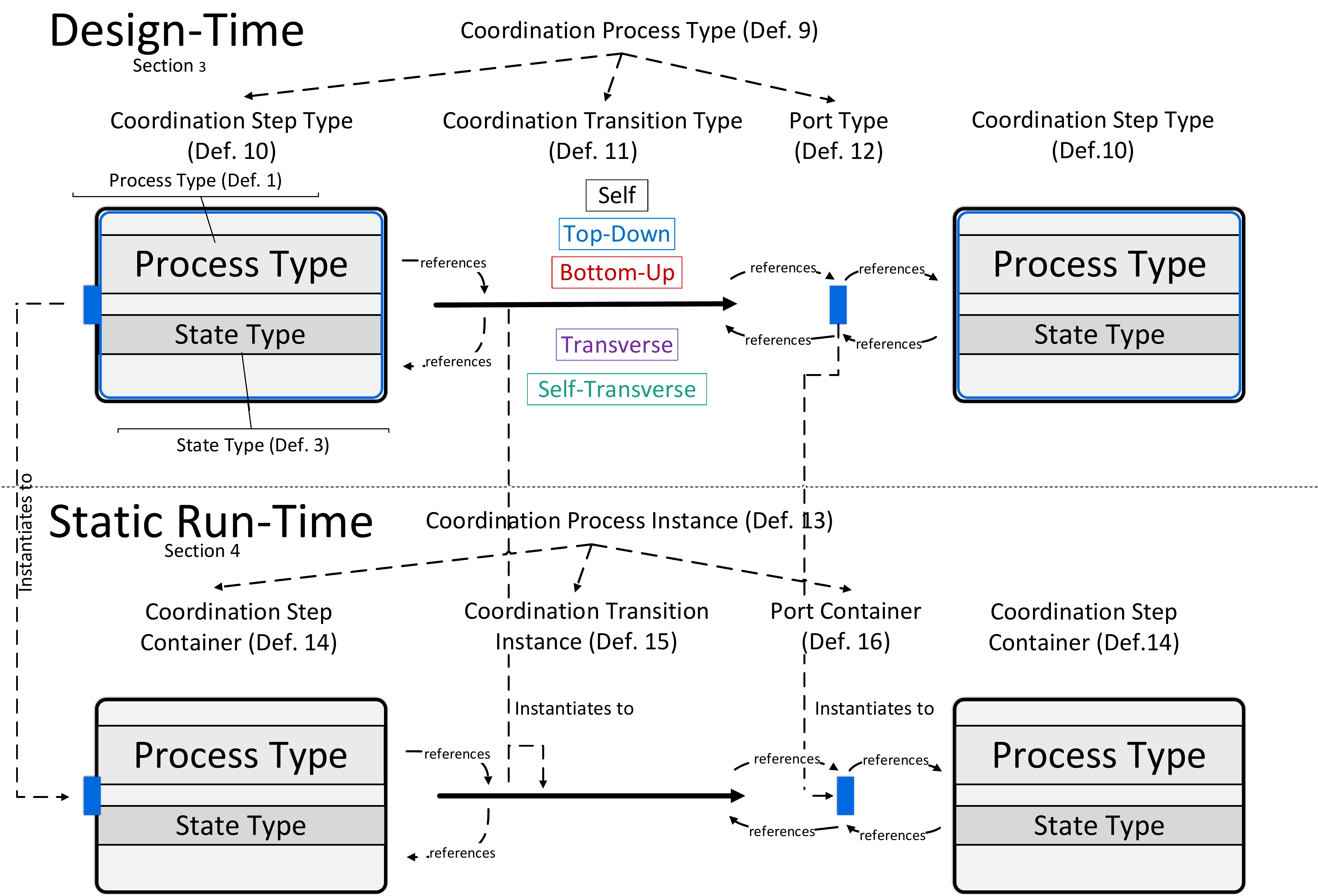}\caption{\label{fig:Static-Run-Time-Entities-Overview}Static Run-time Entities
Overview}
\end{figure*}

This section is concerned with the static parts of the run-time, called
the ``static run-time''. Static in this context means that certain
entities are instantiated once per coordination process instance and
persist over the lifetime of the coordination process. Dynamic entities
are instantiated and deleted conditionally and introduced in Section
\ref{sec:Representing-Process-Instances}. Figure \ref{fig:Static-Run-Time-Entities-Overview}
gives a brief over entities introduced in this section and their relationship
to each other. Further, it displays their connection to the design-time
entities of Section \ref{sec:Background}.

\begin{figure}
\centering{}\includegraphics[width=1\columnwidth]{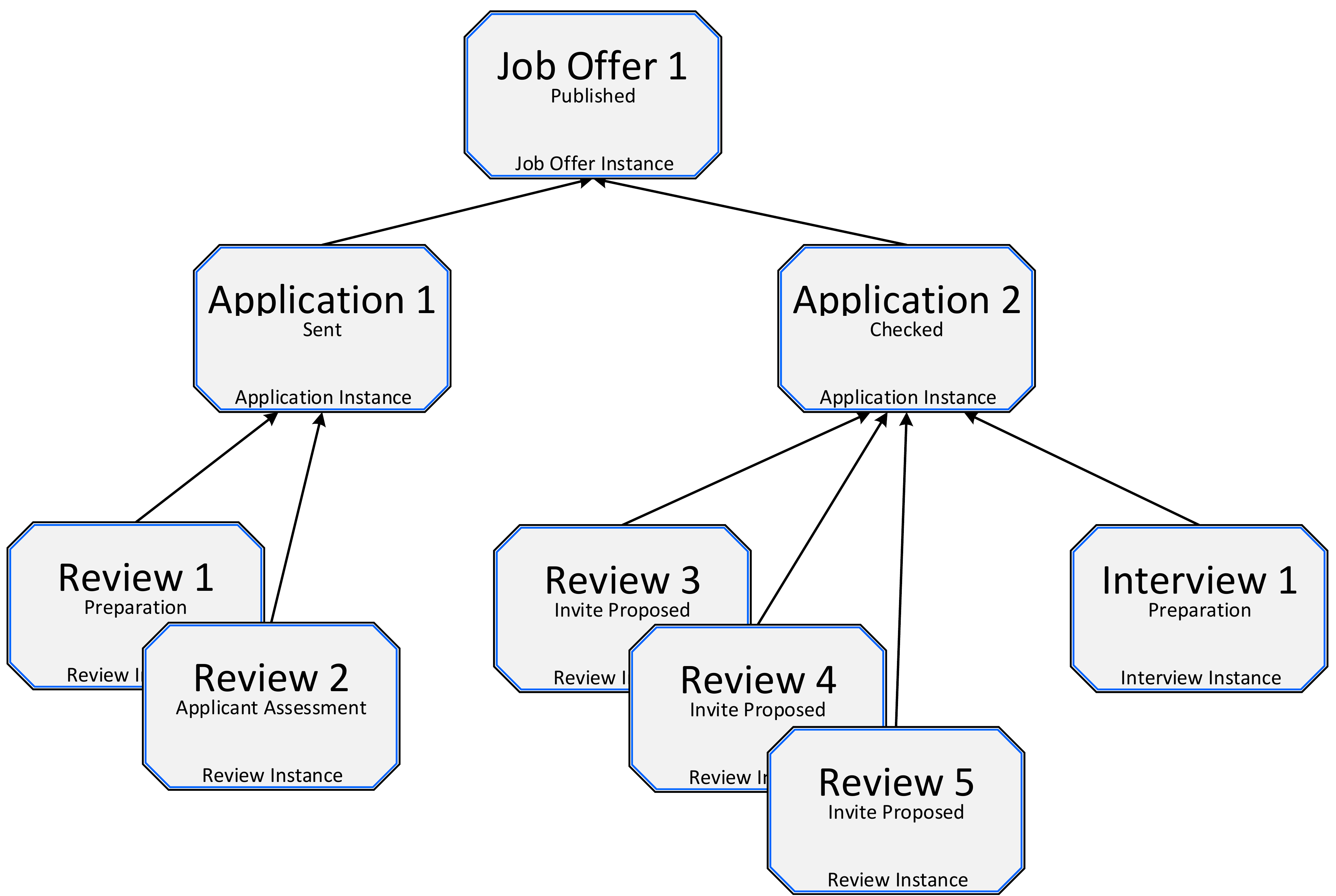}\caption{\label{fig:Relational-Process-Instance}Example of a Relational Process
Instance Structure}
\vspace*{-1em}
\end{figure}

Coordination process instances coordinate process instances captured
in a relational process instance structure. For this purpose, the
coordination process model is instantiated along with its coordinating
process type $\omega_{\mathit{coord}}^{T}$, e.g., a coordination
process instance is created along with an instance of $\mathit{Job\,Offer}$
(cf. Example \ref{Running-Example:-Recruitment}). The coordination
process instance is responsible for enforcing all coordination constraints
in the context of the $\mathit{Job\,Offer}$ process instance. Formally,
a coordination process instance is defined as follows:
\begin{definition}[Coordination Process Instance]
\label{Def:CoordinationProcessInstance}A coordination process instance
$c^{I}$ has the form $(\omega_{coord}^{I},$\\
$G^{I},\Delta^{I},Q^{I})$ where
\end{definition}
\begin{itemize}
\item $\omega_{\mathit{coord}}^{I}$ is the coordinating process instance
to which $c^{I}$ belongs (cf. Definition \ref{Def:ObjectInstance-Normal})
\item $G^{I}$ is a set of coordination step containers $g^{I}$ (cf. Definition
\ref{Def:CoordinationStepInstanceContainer})
\item $\Delta^{I}$ is a set of coordination transition instances $\delta^{I}$
(cf. Definition \ref{Def:CoordinationTransitionInstance})
\item $Q^{I}$ is a set of port containers $\mathit{q}^{I}$ (cf. Definition
\ref{Def:PortInstanceContainer})
\end{itemize}
The model of a coordination process type contains coordination step
types $\beta^{T}$, coordination transition types $\delta^{T}$, and
port types $\eta^{T}$. By principle, a coordination step type $\beta^{T}$
represents multiple process instances $\omega^{I}$ at run-time, e.g.,
multiple $\mathit{Application}$ instances are represented by an $\mathit{Application}$
coordination step type. This has consequences for the overall instantiation
of a coordination process: A coordination step type $\beta^{T}$ is
not instantiated once per coordination process instance, but instead
multiple times, depending on how many process instances of corresponding
type $\omega^{T}$ exist in the relational process structure. Figure
\ref{fig:Relational-Process-Instance} shows an example configuration
of a relational process instance structure for the running example.
Accordingly, an $\mathit{Application}$ coordination step type would
create two instances representing each $\mathit{Application}$.

Regarding a relational process structure at run-time, the term \emph{arrangement}
denotes a specific substructure of the relational process structure.
It is characterized by specific process instances and their concrete
relations. Arrangements are defined as immutable, i.e., adding a new
relation or process instance creates a new arrangement, which is consequently
not identical to the first arrangement. In other words, arrangements
can be thought of as snapshots of a part of the relational process
structure. Note that an entire relational process structure, at a
specific point in time, is likewise an arrangement.

Arrangements are connected to Challenge 3 \emph{Local Contexts}, but
are however not identical to local contexts. The main difference is
that arrangements are precisely defined by processes and relations,
and any change destroys the arrangement and creates a new one (immutability).
In contrast, local contexts persists over changes of the relational
process structure, i.e., linking a new $Review$ to an $\mathit{Application}$
instance creates a new arrangement, but the local context of the $\mathit{Application}$
remains the same.

\noindent\begin{minipage}[t]{1\columnwidth}%
\begin{shaded}%
\begin{example}[Arrangements and relational process instance structure]
 In Figure \ref{fig:Relational-Process-Instance}, the combination
of processes $\mathit{Application\:1}$, $\mathit{Review\:1}$, and
$\mathit{Review\:2}$ and their relations is one example of an arrangement.
Generally, Figure \ref{fig:Relational-Process-Instance} shows a snapshot
of the recruitment business process (cf. Example \ref{exa:Running-Example}).
One $\mathit{Job\,Offer}$ instance and related process instances
are shown to be in various stages of processing (cf. Figure \ref{fig:State-based-views-of}
for an overview of the state-based views). Specifically, two $\mathit{Applications}$
have been submitted for $\mathit{Job\,Offer\:1}$. More may still
be sent, as the $\mathit{Job\,Offer}$ is still in state $\mathit{Published}$.
The process instance $\mathit{Application\:1}$ is in the first states
of processing, as its active state is $\mathit{Sent}$ and only two
$\mathit{Reviews}$ are linked, which have not yet reached a conclusion
(states $\mathit{Preparation}$ and $\mathit{Applicant\:Assessment}$,
respectively). By contrast, $\mathit{Application\:2}$ already has
three completed $\mathit{Reviews}$ with a positive conclusion, signified
by state $\mathit{Invite:Proposed}$. Because of the positive verdict,
it is allowed conducting $\mathit{Interviews}$ with the applicant,
for whom $\mathit{Interview\:1}$ is currently in state $\mathit{Preparation}$.
This is another arrangement different from the arrangement involving
$\mathit{Application\:1}$.\vspace*{-1em}
\end{example}
\end{shaded}%
\end{minipage}

The coordination process of $\mathit{Job\,Offer\:1}$(cf. Figure \ref{fig:Coordination-Process-for})
ensures that all coordination constraints are upheld during the execution
of all process instances. For example, each coordination step type
$\beta^{T}$ that represents the $\mathit{Application}$ process type,
namely $\mathit{Application}$:$\mathit{Creation}$, $\mathit{Application}$:$\mathit{Sent}$,
$\mathit{Application}$:$\mathit{Checked}$, $\mathit{Application}$:$\mathit{Ac}$\-\LyXZeroWidthSpace$cepted$,
and $\mathit{Application}$:$\mathit{Rejected}$, references each
$\mathit{Appli}$\-\LyXZeroWidthSpace$cation$ instance. A coordination
step instance would be required to have a reference to process type,
state type, and process instance information. As depicted in Figure
\ref{fig:Relational-Process-Instance}, the relational process instance
structure has two $\mathit{Application}$ instances. In consequence,
this results in ten coordination step instances, representing two
$\mathit{Application}$ instances and each of the five coordination
step types. The state type and process type is however held redundantly
in each coordination step instance. Note that this implies that there
are groups of coordination step instances based on state type and
process type information. In other words, each coordination step type
corresponds to a group of coordination step instances at run-time. 

In order to organize this information efficiently, the instatiation
of a coordination step type follows a non-standard pattern. Instantiating
a coordination step type creates two different entities: 
\begin{enumerate}
\item A \emph{coordination step container}, which is instantiated exactly
once with the creation of the coordination process instance and contains
process and state information, representing the group.
\item The actual \emph{coordination step instance}, which is instantiated
dynamically and represents a corresponding process instance belonging
to the group.
\end{enumerate}
Formally, a coordination step container is defined as follows:
\begin{definition}[Coordination Step Container]
\label{Def:CoordinationStepInstanceContainer}A coordination step
container $g^{I}$ has the form $(c^{I},\omega^{T},\sigma^{T},$\\
$B^{I},\Delta_{out}^{I},Q^{I})$ where
\end{definition}
\begin{itemize}
\item $c^{I}$ is the coordination process instance (cf. Definition \ref{Def:CoordinationProcessInstance})
\item $\omega^{T}$ is the process type (cf. Definition \ref{Def:ObjectType-Normal})
\item $\sigma^{T}$ is a state type of $\omega^{T}.\theta$ (cf. Definition
\ref{Def:StateBasedView})
\item $B^{I}$ is a set of coordination step instances $\beta^{I}$ (cf.
Definition \ref{Def:CoordinationStepInstance})
\item $\Delta_{out}^{I}$ is a set of outgoing coordination transition instances
$\delta^{I}$(cf. Definition \ref{Def:CoordinationTransitionInstance})
\item $Q^{I}$ is a set of port instance containers $q^{I}$ (cf. Definition
\ref{Def:PortInstanceContainer})
\end{itemize}
Coordination transition types and port types follow the same non-standard
pattern. Coordination transition types create\emph{ coordination transition
instances,} functioning as containers, and \emph{coordination components},
functioning as \emph{dynamic representations of semantic relationships}
(cf. Definition \ref{Def:CoordinationTransitionInstance}). Port instances
and coordination step instances are analogous to coordination step
instances regarding containers.
\begin{definition}
\label{Def:CoordinationTransitionInstance}A coordination transition
instance $\delta^{I}$ has the form $(g_{src}^{I},q_{tar}^{I},S^{I})$
where
\end{definition}
\begin{itemize}
\item $g_{src}^{I}$ is the source coordination step container (cf. Definition
\ref{Def:CoordinationStepInstanceContainer})
\item $q_{tar}^{I}$ is the target port instance container (cf. Definition
\ref{Def:PortInstanceContainer})
\item $S^{I}$ is a set of coordination component instances $s^{I}$ (cf.
Definition \ref{Def:CoordinationComponentBaseInstance})
\end{itemize}
The instantiation of port types functions analogously to the instantiation
of coordination step types. \emph{Port instance containers} that contain
port \emph{instances} are defined as follows: 
\begin{definition}
\label{Def:PortInstanceContainer}A port instance container $q^{I}$
has the form $(c^{I},H^{I},\Delta_{in}^{I},g^{I})$ where
\end{definition}
\begin{itemize}
\item $c^{I}$ is the coordination process instance (cf. Definition \ref{Def:CoordinationProcessInstance})
\item $H^{I}$ is a set of port instances $\eta^{I}$ (cf. Definition \ref{Def:PortInstance})
\item $\Delta_{in}^{I}$ is a set of outgoing coordination transition instances
$\delta^{I}$ (cf. Definition \ref{Def:CoordinationTransitionInstance})
\item $g^{I}$ is a coordination step instance container (cf. Definition
\ref{Def:CoordinationStepInstanceContainer})
\end{itemize}

\begin{figure}
\begin{centering}
\includegraphics[width=1\columnwidth]{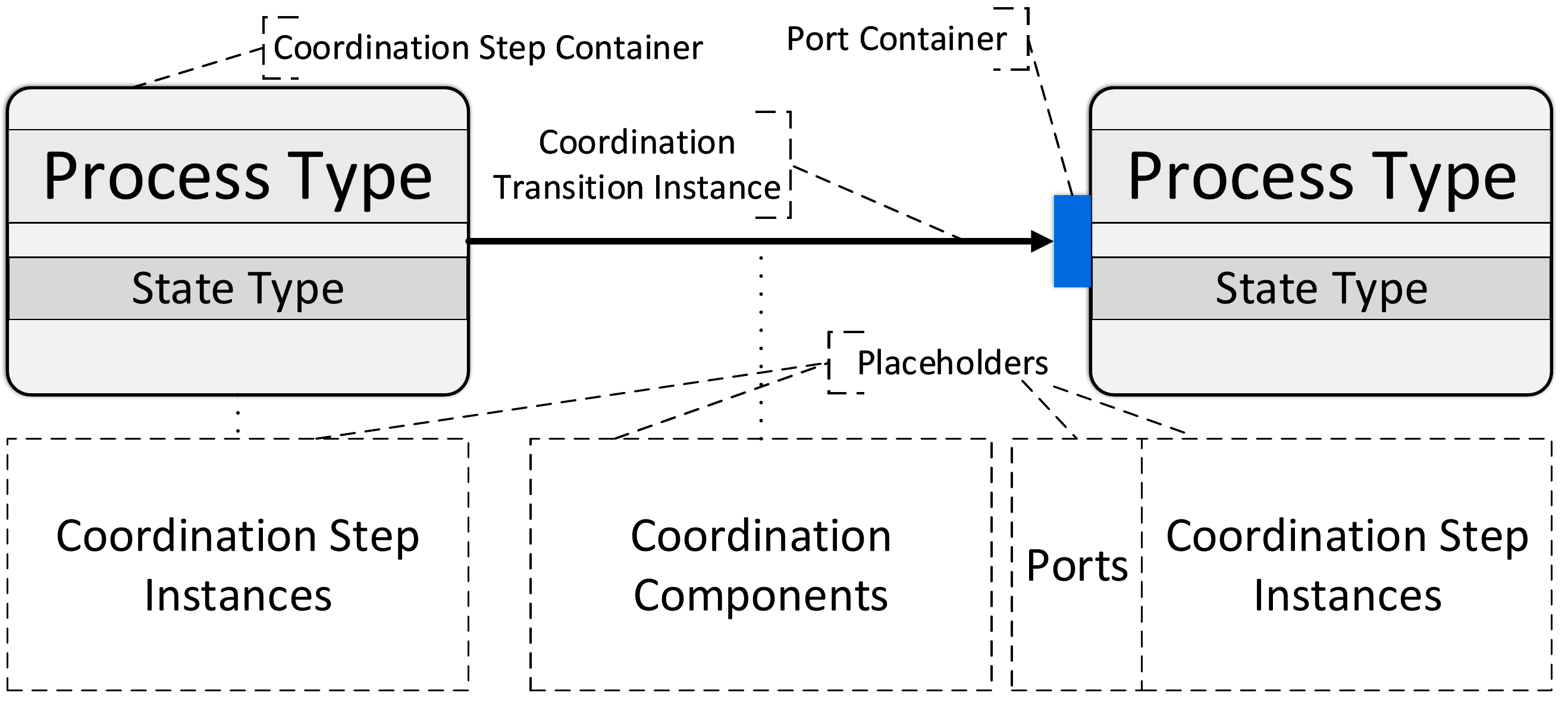}
\par\end{centering}
\caption{\label{fig:Example-containers-of}Example containers of a coordination
process instance}

\end{figure}

In essence, containers represent the scaffolding of a coordination
process instance. Containers replicate the structure of the coordination
process model at run-time and are static components, i.e., they persist
over the lifetime of the coordination process instance. Within the
containers, coordination step instances, coordination component instances,
and port instances are located. These are dynamic components, which
depend on the existence of process instances and their relations.
For example, when a new process instance emerges, a new coordination
step instance becomes instantiated as well, and is put into the appropriate
container. In the same way, coordination step instances are deleted
when the process instance is deleted. Coordination step instances,
coordination component instances, and port instances are discussed
in detail in Section \ref{sec:Representing-Process-Instances}. Figure
\ref{fig:Example-containers-of} shows a part of a coordination process
instance consisting solely of containers, with placeholders standing
for the dynamically instantiated entities.

Coordination containers were introduced for technical reasons in order
to maintain correct references and associations in an implementation.
However, coordination containers find usage in the (real-time) monitoring
of a coordination process instance, as they can be used for providing
aggregated status information over the contained coordination steps
and their process instances.

Coordination step containers, coordination transition instances, and
port containers are the static entities of a coordination process
instance. In Section \ref{sec:Representing-Process-Instances}, the
dynamic entities, namely coordination step instances, coordination
components, and port instances, are discussed in detail.

\section{\label{sec:Representing-Process-Instances}Representing Process Instances
and Semantic Relationships}

A coordination process instance $c^{I}$ coordinates process instances
that fall within its scope in the relational process structure $d^{I}$.
More precisely, it coordinates process instances that fall within
its scope and whose type is referenced by a coordination step type
$\beta^{T}$ of the coordination process model $c^{T}$. Process instances
whose process type $\omega^{T}$ is not referenced by at least one
coordination step type $\beta^{T}$ are irrelevant for the coordination
process instance $c^{I}$. Both the creation of new process instances
and deletion of old process instances affect a coordination process
instance. For the sake of brevity, the following examples use the
creation of a new process instance or existing process instances;
deletions are handled correctly as well by a coordination process
instance, but are omitted in this paper. The following elaborations
continue using the running example: one coordination process instance
$c^{I}$ is attached to the $\mathit{Job\:Offer\:1}$ process instance,
and $c^{I}$ is responsible for coordinating the entire relational
process structure $d^{I}$ (cf. Figure \ref{fig:Relational-Process-Instance}).

Whenever a new process instance $\omega^{I}$ is created and, hence,
emerges in a relational process structure, it must first be checked
whether it is to be coordinated by $c^{I}$. To efficiently accomplish
the check, a relational process instance structure $d^{I}$ maintains
the set of lower level process instances $L_{\omega^{I}}^{I}$ for
each object instance $\omega_{i}^{I}$ (cf. Section \ref{subsec:Relational-Process-Structures}).
If $\omega_{i}^{I}$ is a coordinating process type, i.e., $\exists c^{I}:c^{I}.\omega^{I}=\omega_{i}^{I}$,
and the new process instance is added to $L_{\omega_{i}^{I}}^{I}$
of $\omega_{i}^{I}$, the corresponding coordination process instance
$c^{I}$ is notified of this addition. This optimized approach avoids
the performance penalty of continual depth-first traversals of the
relational process structure in order to recognize new instances,
which contributes to fulfilling Challenge 4 \emph{Immediate Consistency}
\citep{Steinau.2018b}.

\begin{figure*}
\centering{}\includegraphics[width=1\textwidth]{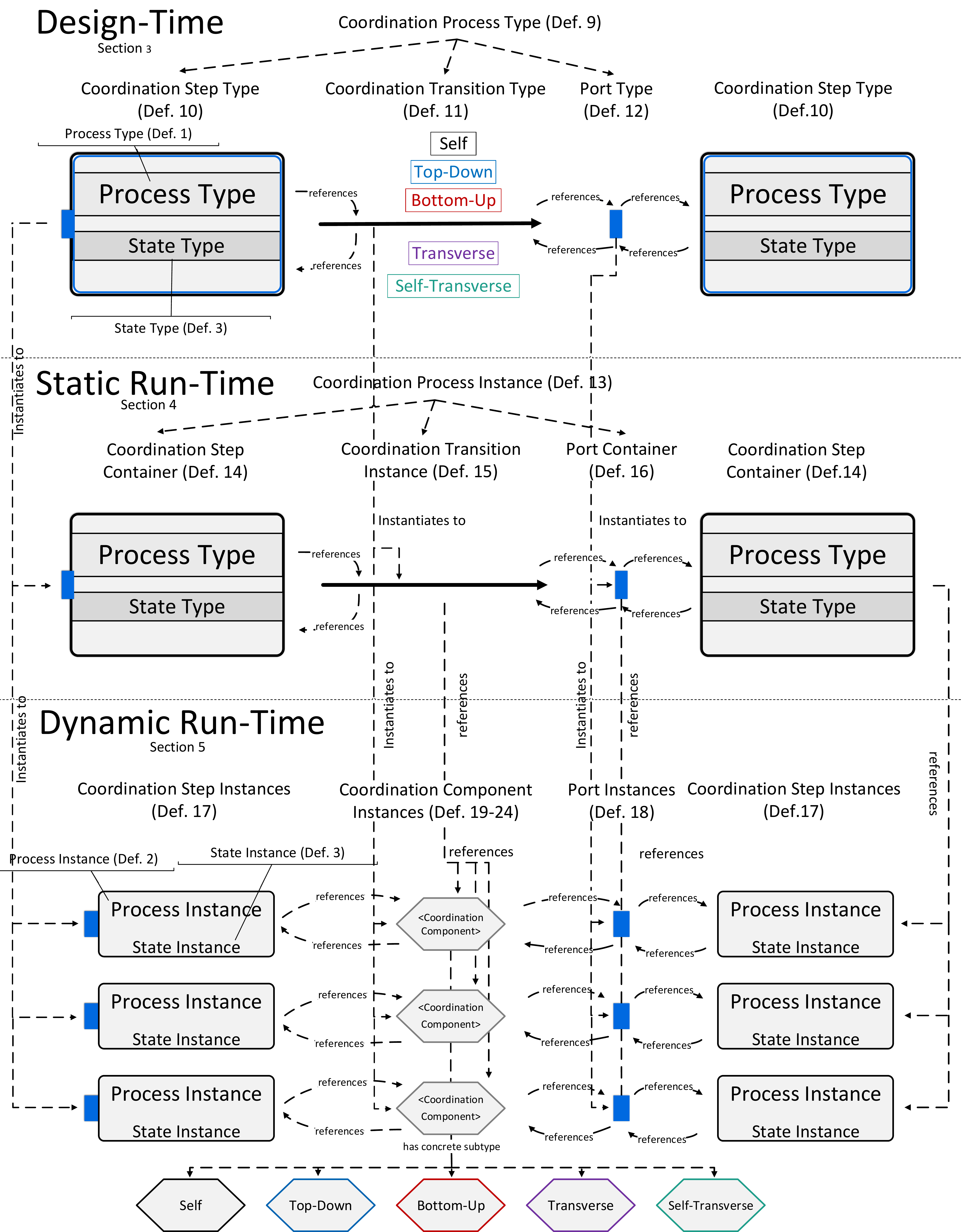}\caption{\label{fig:Dynamic-Run-Time-Entities-Overview}Dynamic Run-time Entities
Overview}
\end{figure*}

This section is concerned with the dynamic parts of the run-time.
Dynamic entities are instantiated or deleted conditionally, based
on the existence of processes and relations. Figure \ref{fig:Dynamic-Run-Time-Entities-Overview}
gives a brief over entities introduced in this section and their relationship
to each other. Further, it displays their connection to the design-time
entities of Section \ref{sec:Background}.

\subsection{Coordination Step and Port Instances}

If the new process instance needs to be coordinated by the coordination
process instance, the new process instance is represented with a \emph{coordination
step instance} in the coordination process. Formally, a coordination
step instance is defined as follows:
\begin{definition}[Coordination Step Instance]
\label{Def:CoordinationStepInstance}A coordination step instance
$\beta^{I}$ has the form $(g^{I},\omega^{I},\sigma^{I},S_{out}^{I},$\\
$H^{I})$ where
\end{definition}
\begin{itemize}
\item $g^{I}$ is a coordination step instance container (cf. Definition
\ref{Def:CoordinationProcessInstance})
\item $\omega^{I}$ is a reference to a process instance (cf. Definition
\ref{Def:ObjectInstance-Normal})
\item $\sigma^{I}$ is a reference to a state instance belonging to $\omega^{I}$,
i.e., $\sigma^{I}\in\omega^{I}.\Sigma^{I}$ (cf. Definition \ref{Def:StateBasedView})
\item $S_{out}^{I}$ is a set of outgoing coordination component instances
$s^{I}$ (cf. Definition \ref{Def:CoordinationComponentBaseInstance})
\item $H^{I}$ is a set of port instances $\eta^{I}$ (cf. Definition \ref{Def:PortInstance})
\end{itemize}
For each process instance $\omega^{I}$, there is exactly one corresponding
coordination step instance $\beta^{I}$ in \emph{each container} $g^{I}$.
Consequently, a particular process instance $\omega_{j}^{I}$ may
be represented by multiple coordination step instances $\beta^{I}$,
depending on the number of containers $g^{I}$ that reference the
corresponding process type $\omega_{j}^{T}$. The reason is that coordination
step types $\beta^{T}$ not only reference a process type $\omega^{T}$,
but also a state type $\sigma^{T}$, analogous to the coordination
step instance (cf. Definition \ref{Def:CoordinationStepInstance}).
Consequently, a coordination step instance $\beta^{I}$ is created
for each combination of process type $\omega^{T}$ and state type
$\sigma^{T}$ referenced by coordination step type $\beta^{T}$ in
each container $g^{I}$. The state type $\sigma^{T}$ is important
for the operational semantics of a coordination process.

The instantiation of a coordination process $c^{I}$ itself is a good
example of how a process is represented by multiple coordination step
instances. Coordination process types $c^{T}$ are instantiated together
with their coordinating process type $\omega_{\mathit{coord}}^{T}$.
By definition, a coordinating process type $\omega_{\mathit{coord}}^{T}$
belongs to the scope of the corresponding coordination process, i.e.,
$\omega_{\mathit{coord}}^{T}\in L_{\omega_{\mathit{coord}}^{I}}^{I}$. 

As such, coordinating process instance $\mathit{Job\:Offer\:1}$ must
be represented by coordination step instances. In case of the running
example (cf. Example \ref{Running-Example:-Recruitment}) and the
$\mathit{Job\:Offer}$ coordination process type (cf. Figure \ref{fig:Coordination-Process-for}),
the coordination process type contains five coordination step types
$\beta^{T}$ referencing the $\mathit{Job\:Offer}$ process type.
These are the states $\mathit{Job\:Offer}$:$\mathit{Preparation}$,
$\mathit{Job\:Offer}$:\LyXZeroWidthSpace$\mathit{Published}$, $\mathit{Job\:Offer}$:$\mathit{Closed}$,
$\mathit{Job\:Offer}$:$\mathit{Position\,Filled}$, and $\mathit{Job}$\LyXZeroWidthSpace{}
$\mathit{Offer}$:$\mathit{Position\,Vacant}$. The naming scheme
is chosen in this manner for simplicity. Coordination step types can
be unambiguously identified by their references to $\omega^{T}$ and
$\sigma^{T}$, as this combination is unique in a coordination process.
At run-time, coordination step types are instantiated as containers
and coordination step instances, which are referenced by using the
same naming scheme $\mathit{Process}$:$\mathit{State}$ plus additional
type information, e.g., coordination step container $\mathit{Job\:Offer}$:\LyXZeroWidthSpace$\mathit{Preparation}$.

\begin{figure}[h]
\centering{}\includegraphics[width=1\columnwidth]{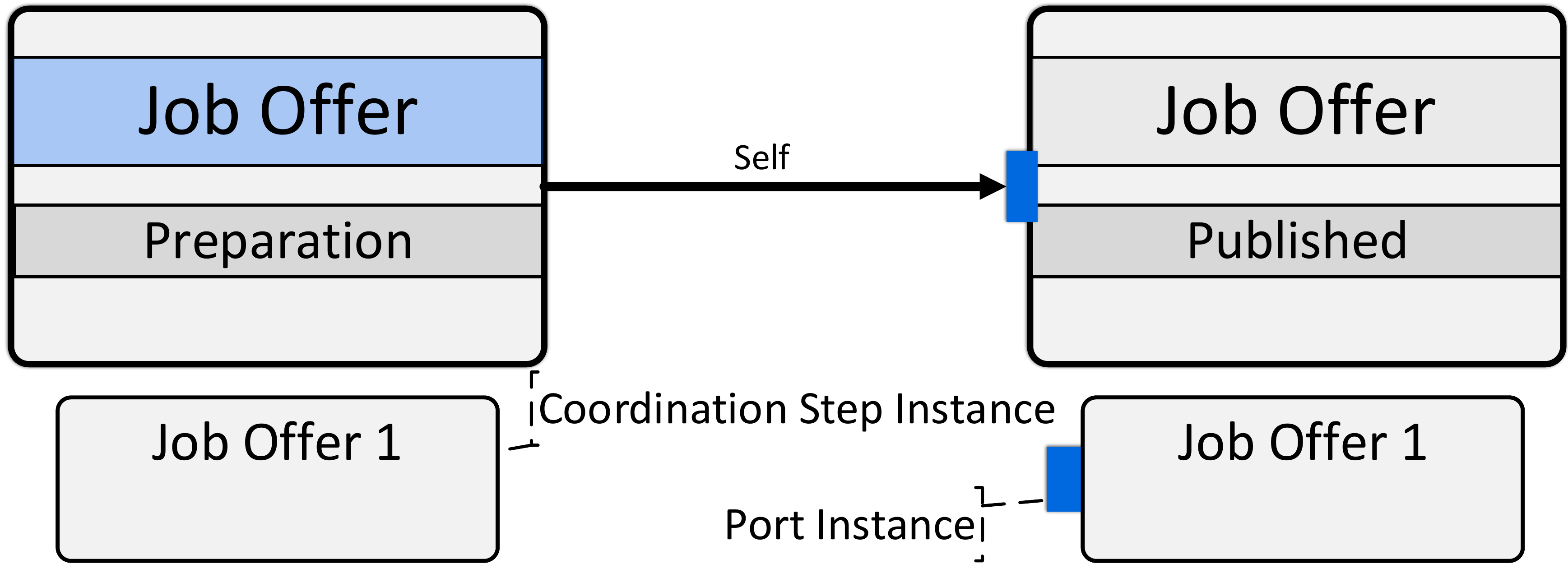}\caption{\label{fig:Coordination-Step-Instances}Coordination Step Instances
for Process Instance $\mathit{Job\:Offer\:1}$}
\end{figure}

Consequently, the $\mathit{Job\:Offer\:1}$ instance is represented
by five coordination step instances $\beta^{I}$ (cf. Figure \ref{fig:Coordination-Process-for}).
For the sake of simplicity, Figure \ref{fig:Coordination-Step-Instances}
shows an excerpt of the entire coordination process. This excerpt
exemplarily shows the first two coordination step containers $\mathit{Job\:Offer}$:$\mathit{Preparation}$
and $\mathit{Job\:Offer}$:$\mathit{Published}$ as well as the corresponding
coordination step instances for $\mathit{Job\:Offer\:1}$. In contrast
to the start coordination step instance, the coordination step instance
for $\mathit{Job\:Offer}$:\LyXZeroWidthSpace$\mathit{Published}$
possesses a \emph{port instance}. Ports are used to combine semantic
relationships for expressing and enforcing more complex coordination
constraints. When a coordination step type $\beta^{T}$ bercomes instantiated,
any port types defined by the coordination step model, i.e., $\beta^{T}.H^{T}$,
are instantiated as well. Formally, a port instance $\eta^{I}$ is
defined as follows:

\begin{definition}[Port Instance]
\label{Def:PortInstance}A port instance $\eta^{I}$ has the form
$(q^{I},\beta^{I},S_{in}^{I})$ where
\end{definition}
\begin{itemize}
\item $q^{I}$ is the corresponding port instance container (cf. Definition
\ref{Def:PortInstanceContainer})
\item $\beta^{I}$ is the coordination step instance to which this port
instance belongs (cf. Definition \ref{Def:CoordinationStepInstance})
\item $S_{in}^{I}$ is the set of all incoming coordination component instances
$s^{I}$ (cf. Definition \ref{Def:CoordinationComponentBaseInstance})
\end{itemize}
For each port type $\eta^{T}$, a corresponding port instance container
$q^{I}$ exists at run-time. When a new coordination step instance
$\beta^{I}$ is created, new port instances $\eta_{i}^{I}$ are instantiated
alongside $\beta^{I}$ as well. The newly created coordination step
instance $\beta^{I}$ gains exactly $n$ port instances $\eta_{i}^{I}$,
$i\in\{1..n\}$ for a given number $n$. The number $n$ is defined
by the number of port instance containers $q^{I}$ of the corresponding
coordination step container $g^{I}$, i.e., $n=|g^{I}.Q^{I}|$. Each
newly created port instance $\eta_{i}^{I}$ is additionally assigned
to the respective port instance container $q_{i}^{I}$. In short,
each process instance is treated individually in a coordination process,
with sets of coordination instances and respective port instances
to allow for proper process coordination. This becomes an important
building block towards fulfilling Challenge 3 \emph{Local Contexts}
and contributes to Challenge 2 \emph{Complex Process Relations}.

Process instances which belong to the scope of the coordination process
are adequately represented by coordination step instances and port
instances. However, the most crucial part of coordination processes
is still missing: the representation of the semantic relationships
between process instances. Whether or not a semantic relationship
can be established between two process instances is determined by
the relations that exist between these instances and the coordination
transitions of the coordination process type. Note that multiple different
semantic relationships may be established based on the same relation
between the process instances (cf. Example \ref{exa:multiple_semantic_relationships}). 

\noindent\begin{minipage}[t]{1\columnwidth}%
\begin{shaded}%
\begin{example}[Multiple Semantic Relationships]
\label{exa:multiple_semantic_relationships}The coordination transition
in Figure \ref{fig:Coordination-Process-for} shows a top-down semantic
relationship between $\mathit{Job\:Offer}$:$\mathit{Published}$
and $\mathit{Application}$:$\mathit{Creation}$. Conversely, there
is a bottom-up semantic relationship between $\mathit{Application}$:$\mathit{Sent}$
and $\mathit{Job\:Offer}$:$\mathit{Closed}$. Both semantic relationships
rely on the same relation between $\mathit{Application}$ and $\mathit{Job\:Offer}$
(cf. Figure \ref{fig:Exemplary-Relational-Process}).\vspace{-1em}
\end{example}
\end{shaded}%
\end{minipage}

\subsection{Semantic Relationships and Coordination Components}

Semantic relationships are means to specify coordination constraints
at a high level of abstraction. At design-time, semantic relationships
$s^{T}$ are essentially labels for a coordination transition type.
The semantic relationship $s^{T}$ indicates which \emph{coordination}
\emph{component instance} must be created. A coordination component
instance provides the necessary functionality for representing and
enforcing semantic relationships between process instances at run-time.
Each semantic relationship is represented by a coordination component
instance whose exact properties depend on the semantic relationship
$s^{T}$(cf. Table \ref{tab:Overview-over-semantic}). All coordination
component instances are derived from a common base $s^{I}$ (cf. Definition
\ref{Def:CoordinationComponentBaseInstance}), which encapsulates
properties common to all coordination component instances.
\begin{definition}[Coordination Component Instance]
\label{Def:CoordinationComponentBaseInstance}A coordination component
instance $s^{I}$ is the common base for $s_{top\mhyphen down}^{I}$,
$s_{bottom\mhyphen up}^{I}$, $s_{transverse}^{I}$, $s_{self}^{I}$,
as well as $s_{self\mhyphen transverse}^{I}$ and has the form $(s^{T},\delta^{I})$
where
\end{definition}
\begin{itemize}
\item $s^{T}$ is the coordination component type from which $s^{I}$ has
been instantiated (cf. Definition \ref{Def:CoordinationComponentType})
\item $\delta^{I}$ is the coordination transition instance to which $s^{I}$
is attached (cf. Definition \ref{Def:CoordinationTransitionInstance})
\end{itemize}
A coordination component instance $s^{I}$ corresponds to an edge
in the coordination process graph, whereas the combination of a port
instance $\eta^{I}$ and coordination step instance $\beta^{I}$ corresponds
to a vertex. Therefore, these elements are treated with the same vocabulary
as edges and vertices, for the sake of simplicity. Coordination components
have sources and targets, whereas ports and coordination steps have
incoming or outgoing coordination components.

The following coordination component definitions extend $s^{I}$ with
additional properties. In total, there are five unique coordination
components, one for each semantic relationship (cf. Table \ref{tab:Overview-over-semantic}).
All following examples are based on the relational process structure
of the running example presented in Figure \ref{fig:Relational-Process-Instance},
i.e., the relational process structure contains one $\mathit{Job\:Offer}$
instance, two $\mathit{Application}$ instances, five $\mathit{Review}$
instances, and one $\mathit{Interview}$ instance, as well as their
respective relations.

Furthermore, establishing a coordination component between process
instances is highly dynamic at run-time. Process instances may be
created and deleted arbitrarily, and their relations may change. For
establishing semantic relationships it is important under which circumstances
a coordination component becomes instantiated (or subsequently deleted).
As a semantic relationship represents coordination constraints between
two process instances, its existence at run-time is logically dependent
on the existence of these process instances and their relations.

Consequently, the \emph{creation of coordination components is coupled
to one specific entity in a coordination process}. This entity is
denoted as an\emph{ instantiator} entity. Whenever such an instantiator
entity is created, the corresponding semantic relationship is instantiated
as well, i.e. a new coordination component instance is created. Depending
on the type of semantic relationship, it is not always the same type
of entity that is designated as the instantiator. It is noted in the
formal definition of the respective coordination component instance
(cf. Defs \ref{def:SelfCoordinationComponent}-\ref{def:A-self-transverse-coordination})
which entity causes the instantiation of a coordination component,
i.e, which entity is the instantiator.

\subsubsection{Self Semantic Relationship}

The simplest semantic relationship is the self semantic relationship
$s_{self}^{I}$. It represents the normal progression of a process
between states. It is represented by the \emph{self coordination component}
(cf. Definition \ref{def:SelfCoordinationComponent})\emph{. }Its
purpose in a coordination process is to keep the graph connected,
i.e., all coordination process graph elements are connected by transitions/edges.

\begin{figure}
\begin{centering}
\includegraphics[width=1\columnwidth]{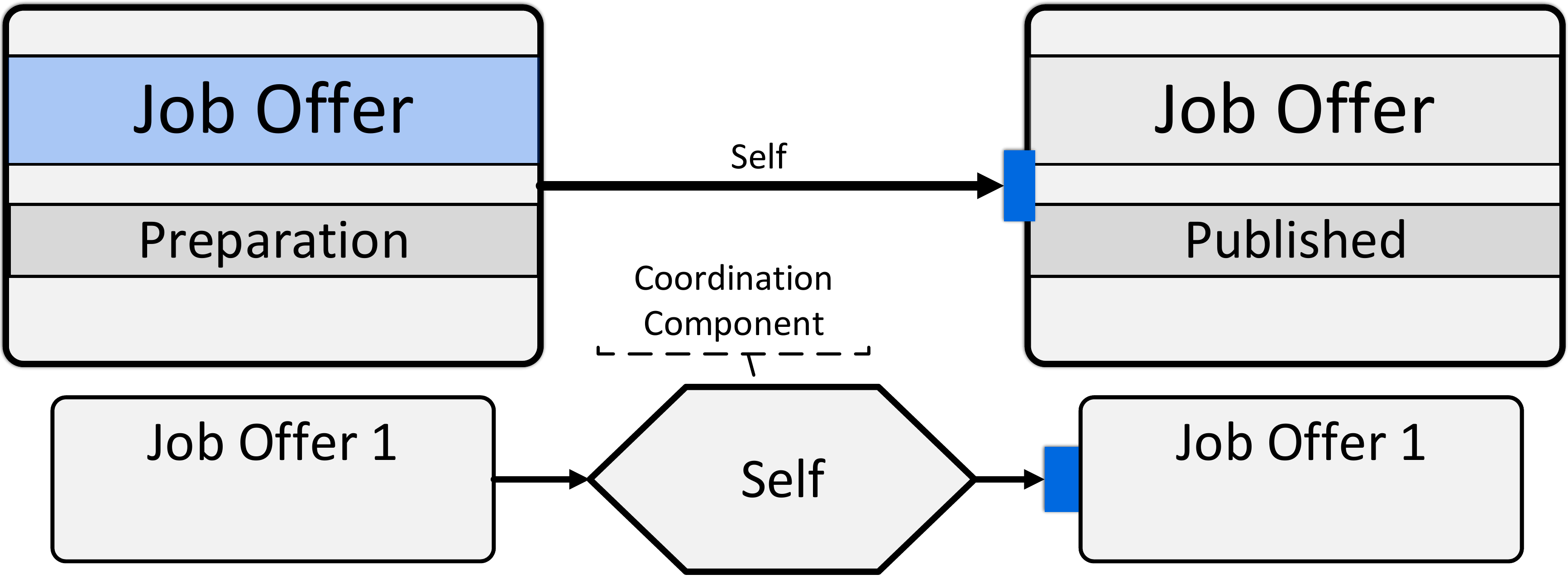}
\par\end{centering}
\caption{\label{fig:Example-Self-CC}Self Coordination Component Example}
\end{figure}

\noindent\begin{minipage}[t]{1\columnwidth}%
\begin{shaded}%
\begin{example}[Self Relationship]
The $\mathit{Job\:Offer}$ coordination process (cf Figure \ref{fig:Coordination-Process-for})
has $\mathit{Job\:Offer}$:$\mathit{Preparation}$ as a start step.
Coordination Constraint \ref{const:coordConstraint1} states that
$\mathit{Applications}$ may only be created once the $\mathit{Job\:Offer}$
has been $\mathit{Published}$.

This constraint however is not represented by a self semantic relationship
and involves coordination steps $\mathit{Job\:Offer}$:$\mathit{Published}$
and $\mathit{Application}$:$\mathit{Creation}$. Consequently, a
self coordination component $s_{self}^{I}$ is used to connect the
start coordination step $\mathit{Job\:Offer}$:$\mathit{Preparation}$
with the coordination step $\mathit{Job\:Offer}$:$\mathit{Published}$.
This mirrors normal state progression of the $\mathit{Job\:Offer}$
process (cf. Figure \ref{fig:State-based-views-of}), but is necessary
to be able to specify Coordination Constraint \ref{const:coordConstraint1}
(cf. Example \ref{exa:Rundown}). With $\mathit{Job\:Offer}$:$\mathit{Published}$,
the aforementioned coordination constraint may be specified using
semantic relationships other than the self semantic relationship.\vspace{-1em}
\end{example}
\end{shaded}%
\end{minipage}

. 
\begin{definition}[Self Coordination Component Instance]
\label{def:SelfCoordinationComponent}A self coordination component
instance $s_{\mathit{self}}^{I}$ has the form $(s^{I},\beta_{src}^{I},\eta_{tar}^{I})$
where
\end{definition}
\begin{itemize}
\item $s_{\mathit{self}}^{I}<:s^{I}$, with $s^{I}$ defined as in Definition
\ref{Def:CoordinationComponentBaseInstance}
\item $\beta_{src}^{I}$ is a coordination step instance, $\beta_{src}^{I}$
is instantiator (cf. Definition \ref{Def:CoordinationStepInstance})
\item $\eta_{tar}^{I}$ is a port instance with\\
 $\beta_{src}^{I}.\omega^{I}=\text{\ensuremath{\eta_{tar}^{I}.\beta^{I}.}}\omega^{I}$
(cf. Definition \ref{Def:PortInstance})
\end{itemize}
As the subtype relation $s_{\mathit{self}}^{I}<:s^{I}$ holds, $s_{\text{\ensuremath{\mathit{self}}}}^{I}$
obtains the properties and notions defined for $s^{I}$. A self coordination
component $s_{\mathit{self}}^{I}$ is established between exactly
one source coordination step $\beta_{src}^{I}$ and exactly one port
instance $\eta_{tar}^{I}$. The coordination step $\beta$ of the
port instance $\eta_{tar}^{I}$ must reference the same process instance
as $\beta_{src}^{I}$ (cf. Figure \ref{fig:Example-Self-CC}). As
both source and target coordination step instance reference the same
process instance $\omega^{I}$, any of the coordination steps may
be designated as instantiator. Definition \ref{def:SelfCoordinationComponent}
designates $\beta_{src}^{I}$ as instantiator, avoiding arbitraryness.

\subsubsection{Top-Down Semantic Relationship}

A top-down semantic relationship $s_{top\mhyphen down}^{I}$ is employed
whenever multiple process instances depend on the execution status
of exactly one common higher-level process. It is represented by a
\emph{top-down coordination component} (cf. Definition \ref{def:A-top-down-coordination}).
\vspace{-1em}

\begin{figure}[h]
\centering{}\includegraphics[width=1\columnwidth]{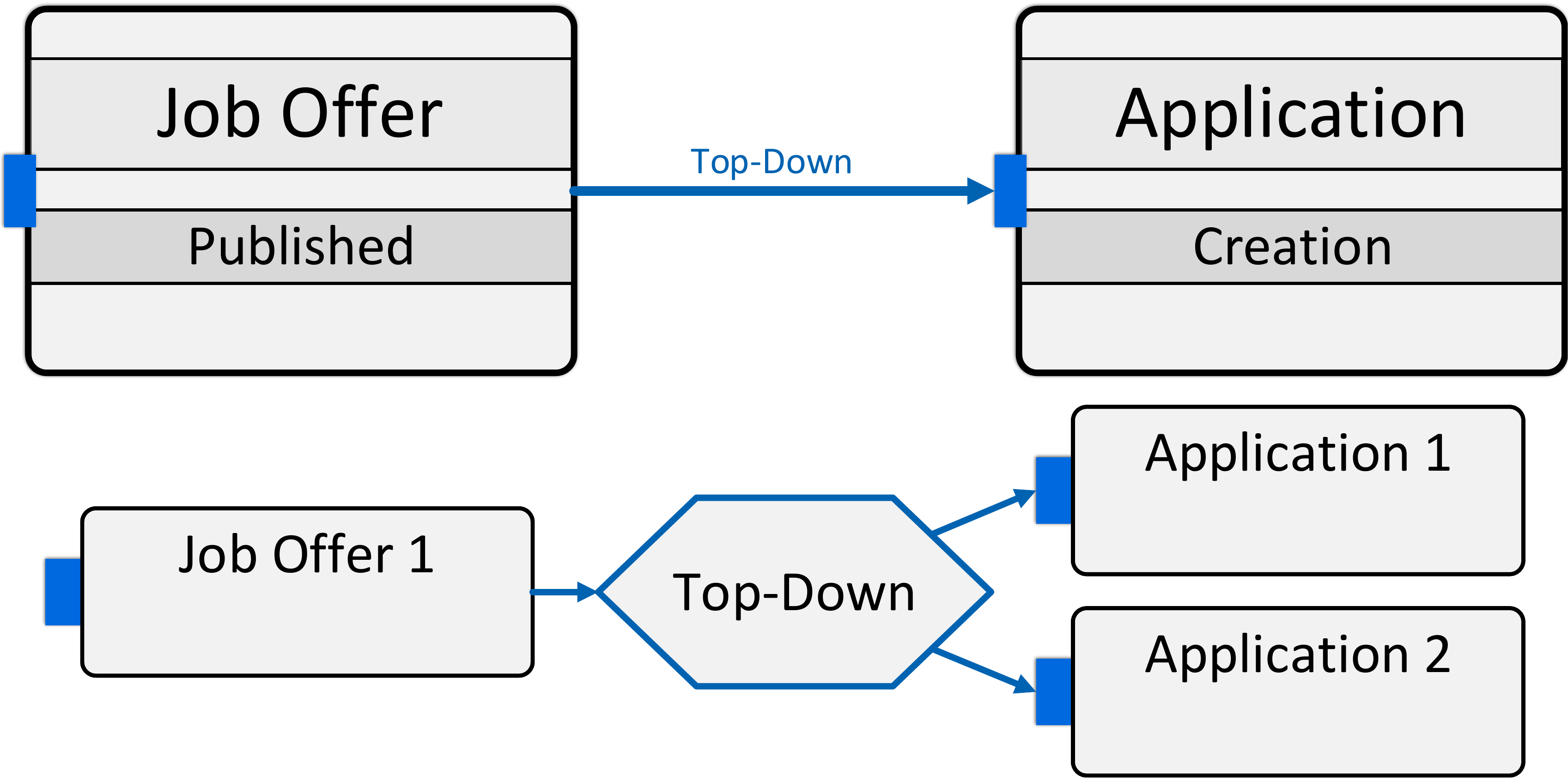}\caption{\label{fig:Example-TopDown-CC}Top-Down Coordination Component Example
I}
\end{figure}

\noindent\begin{minipage}[t]{1\columnwidth}%
\begin{shaded}%
\begin{example}[Top-Down Relationship]
When a $\mathit{Job\:Offer}$ has reached state $\mathit{Published}$,
it is permitted to create $\mathit{Applications}$ for the respective
$\mathit{Job\:Offer}$ (cf. Coordination Constraint \ref{const:coordConstraint1}).
This is illustrated in Figure \ref{fig:Example-TopDown-CC}. More
precisely, $\mathit{Applications}$ are allowed to activate state
$\mathit{Creation}$ if the top-down coordination component is active.
This example is a special case, as $\mathit{Creation}$ is the start
state of an $\mathit{Application}$. In this case, the coordination
process would additionally prevent creating a relation between the
$\mathit{Job\:Offer}$ and an $\mathit{Application}$, if the top-down
coordination component was not fulfilled. In case the state is not
a start state, only the state activation is prevented.\vspace{-1em}
\end{example}
\end{shaded}%
\end{minipage}
\begin{definition}[Top-down Coordination Component Instance]
\label{def:A-top-down-coordination}A top-down coordination component
instance $s_{top\mhyphen down}^{I}$ has the form $(s^{I},\text{\ensuremath{\beta}}_{src}^{I},H_{tar}^{I},\Sigma_{valid}^{T})$
where
\end{definition}
\begin{itemize}
\item $s_{top\mhyphen down}^{I}<:s^{I}$, with $s^{I}$ defined as in Definition
\ref{Def:CoordinationComponentBaseInstance}
\item $\text{\ensuremath{\beta}}_{src}^{I}$ is a coordination step instance,
$\text{\ensuremath{\beta}}_{src}^{I}$ is instantiator (cf. Definition
\ref{Def:CoordinationStepInstance})
\item $H_{tar}^{I}$ is a set of port instances related to $\beta_{src}^{I}$,$\forall\eta^{I}\in H_{tar}^{I}:$$\eta.\beta^{I}\text{.\ensuremath{\omega^{I}}}\twoheadrightarrow\beta_{src}^{I}.\omega^{I}$
(cf. Definition \ref{Def:PortInstance})
\item $\Sigma_{valid}^{T}$ is the set of valid state types from $s^{I}.s^{T}$
\end{itemize}

A top-down coordination component $s_{top\mhyphen down}^{I}$ has
exactly one source coordination step $\text{\ensuremath{\beta}}_{src}^{I}$.
On the target side, $s_{top\mhyphen down}^{I}$ references multiple
port instances $\eta_{i}^{I}\in$$H_{tar}^{I}$ and, by extension,
the respective coordination step instance $\beta^{I}$ of $\eta_{i}^{I}$
and process instance $\beta^{I}.\omega^{I}$. Whether or not a port
instance $\eta_{tar}^{I}$ is referenced in $H_{tar}^{I}$ is determined
by the relation between the respective process instances, i.e., $\forall\eta^{I}\in H_{tar}^{I}:\eta_{tar}^{I}.\beta^{I}.\omega^{I}\twoheadrightarrow\beta_{\mathit{src}}^{I}.\omega^{I}$. 

\noindent\begin{minipage}[t]{1\columnwidth}%
\begin{shaded}%
\begin{example}[Lower-level Instances]
The top-down coordination component for $\mathit{Job\:Offer\:1}$
in Figure \ref{fig:Example-TopDown-CC} references both $\mathit{Application\:1}$
and $\mathit{Application\:2}$. As both have a relation to $\mathit{Job\:Offer\:1}$,
they are contained in the set of lower-level instances $L_{\mathit{Job\:Offer\:1}}^{I}$.\vspace{-1em}
\end{example}
\end{shaded}%
\end{minipage}

\begin{figure}[H]
\centering{}\includegraphics[width=1\columnwidth]{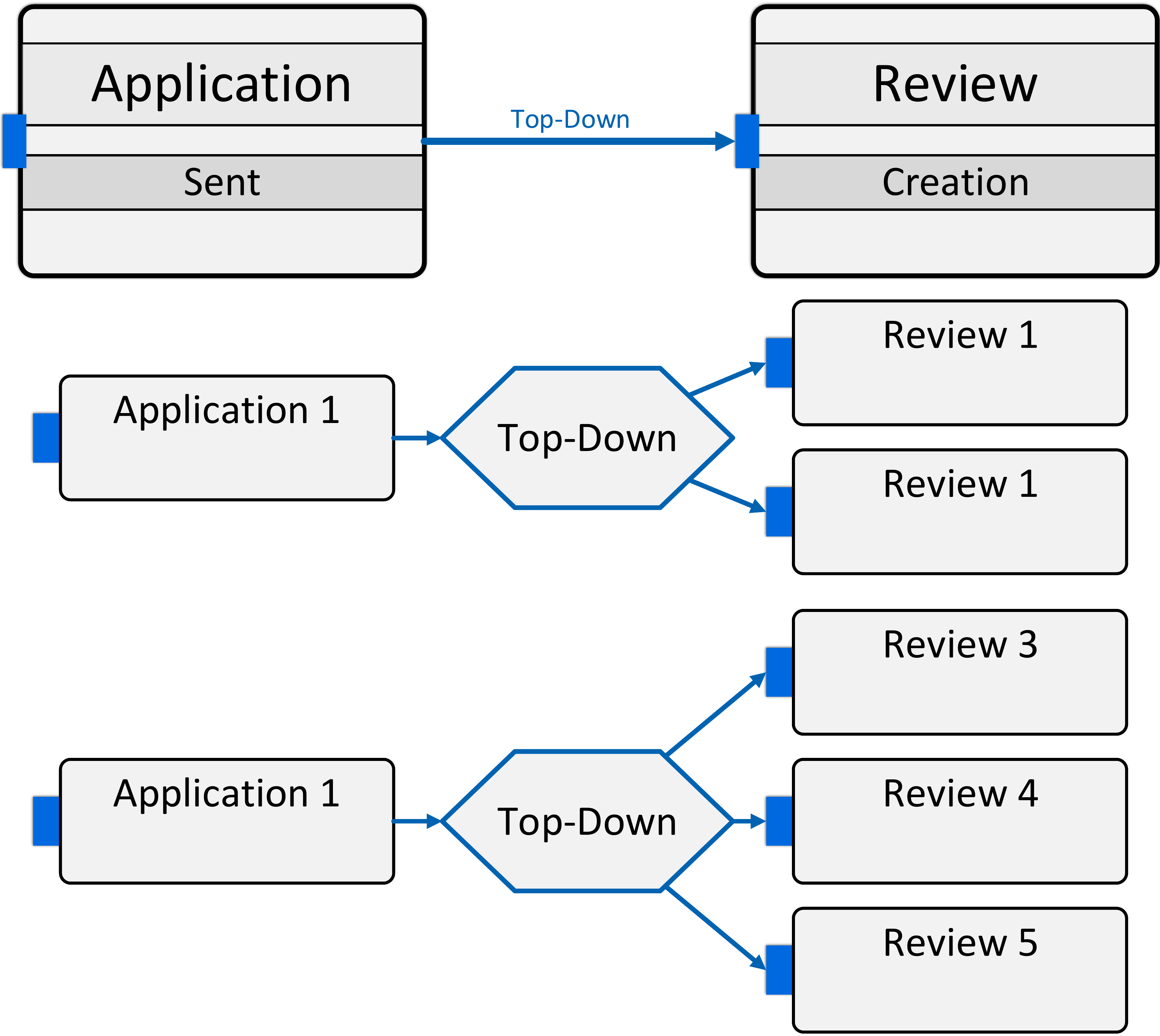}\caption{\label{fig:Example-TopDown-CC-Multi}Top-Down Coordination Component
Example II}
\end{figure}
\noindent\begin{minipage}[t]{1\columnwidth}%
\begin{shaded}%
\begin{example}[Instantiator]
For illustrating the instantiator, the top-down semantic relationship
between $\mathit{Application}$:$\mathit{Sent}$ and $\mathit{Review}$:$\mathit{Creation}$
can be used. There are two $\mathit{Applications}$ and five $\mathit{Reviews}$
in the relational process structure of the running example (cf. Figure
\ref{fig:Relational-Process-Instance}). For each $\mathit{Application}$,
a coordination step instance $\beta^{I}$ is created in the source
container of the respective semantic relationship (cf. Figure \ref{fig:Example-TopDown-CC-Multi}).
As $\beta^{I}$ in the source container is denoted as instantiator,
corresponding top-down coordination components are instantiated as
well. The $\mathit{Reviews}$ are connected to the top-down coordination
components according to their relations to the respective $\mathit{Application}$.\vspace{-0.5em}
\end{example}
\end{shaded}%
\end{minipage}

The instantiation of coordination components on a per-entity basis
enables a fine-grained coordination. This not only applies to top-down
coordination components, but also to the coordination components of
other semantic relationships. However, these components have different
instantiator entities. Each coordination component creates a different
context for its respective instantiator, which allows coordinating
two or more entities independently from each other. The coordination
components hereby take the individual relations of processes into
account, supporting the fulfillment of Challenge 2 \emph{Complex Process
Relations}. Moreover, the way coordination components are employed
directly fulfills Challenge 3 \emph{Local Contexts}. Each coordination
component instance is a representation of such a local context.

\subsubsection{Bottom-Up Semantic Relationship}

Bottom-up semantic relationships are the counterpart of top-down semantic
relationships. Here, the execution status of one process instance
is dependent on the progress of several lower-level processes. A bottom-up
semantic relationship is represented at run-time by a \emph{bottom-up
coordination component} (cf. Definition \ref{def:A-bottom-up-coordination}).
For example, a $\mathit{Job\:Offer}$ may only reach state $\mathit{Position}$
$\mathit{Filled}$ once an $\mathit{Application}$ has reached state
$\mathit{Accepted}$ (cf. Figure \ref{fig:Coordination-Process-for}).
\begin{definition}[Bottom-up Coordination Component Instance]
\label{def:A-bottom-up-coordination}A bottom-up coordination component
instance $s_{bottom\mhyphen up}^{I}$ has the form $(s^{I},B_{src}^{I},\eta_{tar}^{I},\lambda)$
where
\end{definition}
\begin{itemize}
\item $s_{bottom\mhyphen up}^{I}<:s^{I}$, with $s^{I}$ defined as in Definition
\ref{Def:CoordinationComponentBaseInstance}
\item $B_{src}^{I}$ is a set of coordination step instances related to
$\omega_{tar}^{I}$, i.e.,\\
 $\forall\beta^{I}\in B_{src}^{I}:\text{\ensuremath{\beta^{I}.\omega^{I}}}\twoheadrightarrow\eta_{tar}^{I}.\beta^{I}.\omega^{I}$
(cf. Definition \ref{Def:CoordinationStepInstance})
\item $\eta_{tar}^{I}$ is a port instance, $\eta_{tar}^{I}$ is instantiator
(cf. Definition \ref{Def:PortInstance})
\item $\lambda$ is an expression copied from $s^{I}.s^{T}$
\end{itemize}
A bottom-up coordination component $s_{bottom\mhyphen up}^{I}$ references
exactly one port instance $\eta_{tar}^{I}$ on the target side and
multiple coordination step instances $B_{src}^{I}$ on the source
side. A port instance $\eta_{tar}^{I}$ is the instantiator entity.
Note that for both top-down and bottom-up coordination components,
sets $H_{tar}^{I}$ and $B_{src}^{I}$, respectively, may be empty.
The instantiation of the coordination component is independent from
the presence of entities in these sets.

\begin{figure}[h]
\begin{centering}
\includegraphics[width=1\columnwidth]{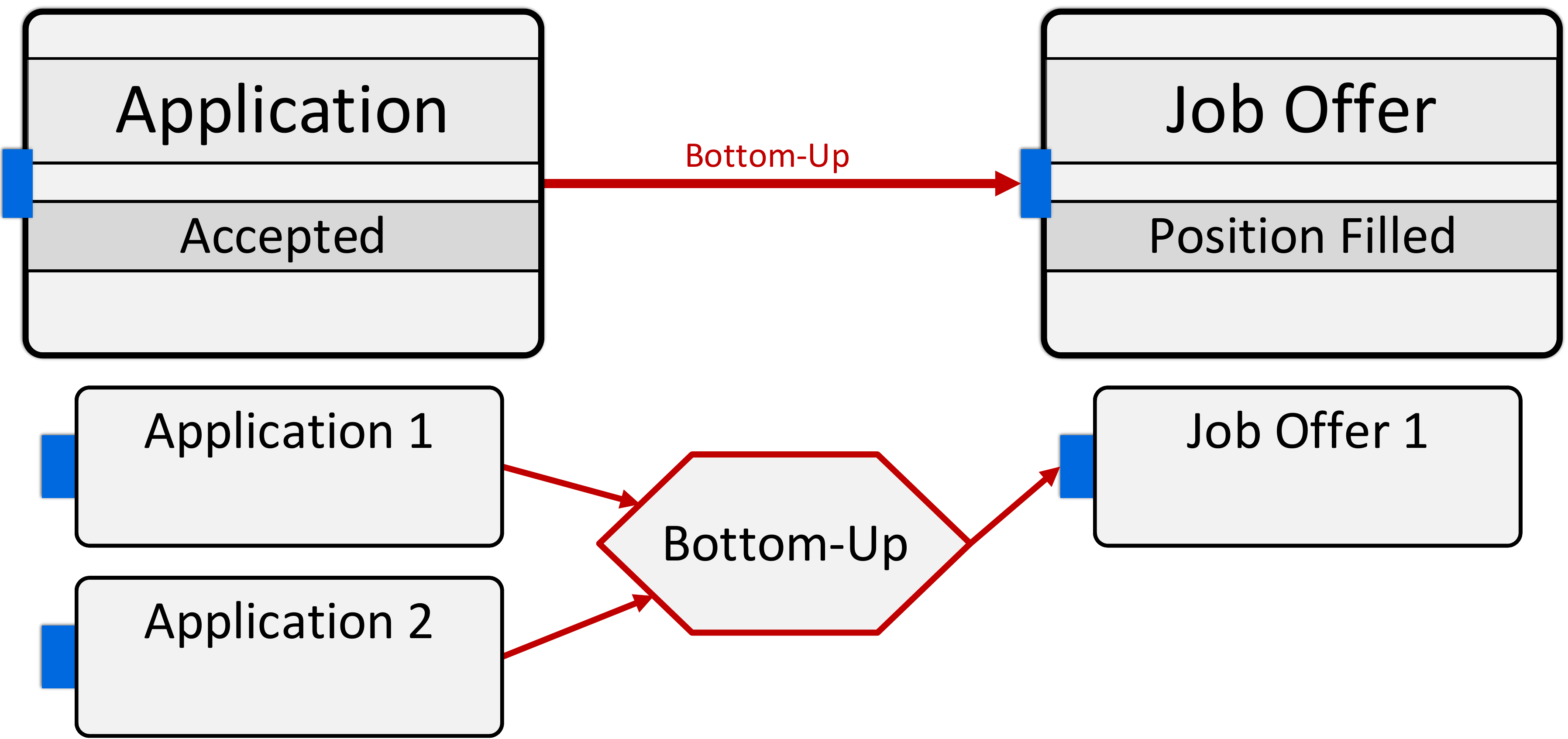}
\par\end{centering}
\caption{\label{fig:Example-BottomUp-CC}Bottom-Up Coordination Component Example}
\end{figure}

Bottom-up coordination components have an expression $\lambda$, which
determines when the coordination component becomes fulfilled (cf.
Example \ref{exa:Bottom-Up Expression}).

\noindent\begin{minipage}[t]{1\columnwidth}%
\begin{shaded}%
\begin{example}[Bottom-Up Expression]
\label{exa:Bottom-Up Expression}The constraint ``there must be
one accepted $\mathit{Application}$ for a $\mathit{Job\:Offer}$
to reach $\mathit{Position\:Filled}$`` requires the definition of
an expression for counting the number of accepted $\mathit{Applications}$.\vspace{-1em}
\end{example}
\end{shaded}%
\end{minipage}

For a full account on expressions refer to \citep{Steinau.2018}.
The expression $\lambda$ is copied from $s^{T}$ due to performance
reasons. Multiple copies of the same expression can be evaluated simultaneously
with different parameters, allowing for parallel execution. These
evaluations would have to be performed sequentially in case of all
bottom-up coordination component instances referencing the single
expression of $s^{T}$. In turn, an avoidable bottleneck would have
been created. Therefore, expressions are copied to each bottom-up
coordination component instance. 

\subsubsection{Transverse Semantic Relationship}

Transverse semantic relationships are employed when multiple process
instances of one type are dependent on multiple process instances
of another type. The two process types, however, do not have a direct
(transitive) relation in the relational process type structure. Instead,
they are connected indirectly by a higher-level process of a third
process type. Both lower level process types must have a (transitive)
relation to this third process type, which is called the \emph{common
ancestor} in this context. At run-time, a \emph{transverse coordination
component} (cf. Definition \ref{sec:Operational-Semantics}) is used
to coordinate instances of each object type.
\begin{definition}[Transverse Coordination Component Instance]
\label{def:A-transverse-coordination}A transverse coordination component
instance $s_{transverse}^{I}$ has the form $(s^{I},\omega_{ca}^{I},B_{src}^{I},H_{tar}^{I},\lambda)$
where
\end{definition}
\begin{itemize}
\item $s_{transverse}^{I}<:s^{I}$, with $s^{I}$ defined as in Definition
\ref{Def:CoordinationComponentBaseInstance}.
\item $\omega_{ca}^{I}$ is the common ancestor, $\omega_{ca}^{I}$ is instantiator
(cf. Definition \ref{Def:ObjectInstance-Normal})
\item $B_{src}^{I}$ is a set of coordination step instances related to
$\omega_{ca}^{I}$, i.e., $\forall\beta^{I}\in B_{src}^{I}:\text{\ensuremath{\beta^{I}.\omega^{I}}}\twoheadrightarrow\omega_{ca}^{I}$
(cf. Definition \ref{Def:CoordinationStepInstance})
\item $H_{tar}^{I}$ is a set of port instances related to $\omega_{ca}^{I}$,
i.e., $\forall\eta^{I}\in H_{tar}^{I}:\text{\ensuremath{\eta^{I}.\beta^{I}.\omega^{I}}}\twoheadrightarrow\omega_{ca}^{I}$,\\
$\forall\text{\ensuremath{\beta}}_{i}^{I}\in B_{src}^{I},\eta_{j}^{I}\in H_{tar}^{I}:$$\text{\ensuremath{\beta_{i}^{I}.\omega^{I}.\omega^{T}}}\neq\eta_{j}^{I}.\beta^{I}.\omega^{I}.\omega^{T}$
(cf. Definition \ref{Def:PortInstance})
\item $\lambda$ is an expression copied from $s^{I}.s^{T}$
\end{itemize}
At both source and target side, transverse coordination components
$s_{transverse}^{I}$ maintain sets : At the source side, a set of
coordination step instances $B_{src}^{I}$ is maintained, and at the
target side a set of port instances $H_{tar}^{I}$. This is different
to top-down or bottom-up coordination components, where either source
$\beta_{\mathit{src}}^{I}$ or target $\eta_{\mathit{tar}}^{I}$ corresponds
to a single entity (cf. Definitions \ref{def:A-top-down-coordination}
and \ref{def:A-bottom-up-coordination}). The referenced processes
in both $B_{src}^{I}$ and $H_{tar}^{I}$ of $s_{transverse}^{I}$
are determined by the common ancestor $\omega_{ca}^{I}$, which is
a single entity. The common ancestor also serves as the instantiator
entity of the transverse coordination component $s_{transverse}^{I}$.

In the running example, a transverse coordination component $s_{transverse}^{I}$
is needed to coordinate $\mathit{Reviews}$ with $\mathit{Interviews}$.
For any $\mathit{Application}$, if a majority of $\mathit{Reviews}$
are in favor of the $\mathit{Application}$, the respective applicant
may be invited for one or more $\mathit{Interviews}$. Thus, multiple
instances of one process type ($\mathit{Interview}$) are dependent
on the instances of another process type ($\mathit{Review}$). Note
that, according to the relational process type structure (cf. Figure
\ref{fig:Exemplary-Relational-Process}), both process types do not
possess a direct relation. Instead, both are lower-level process types
of the $\mathit{Application}$ process type. Consequently, $\mathit{Application}$
instances serve as common ancestor for the transverse coordination
component.

\begin{figure}
\centering{}\includegraphics[width=1\columnwidth]{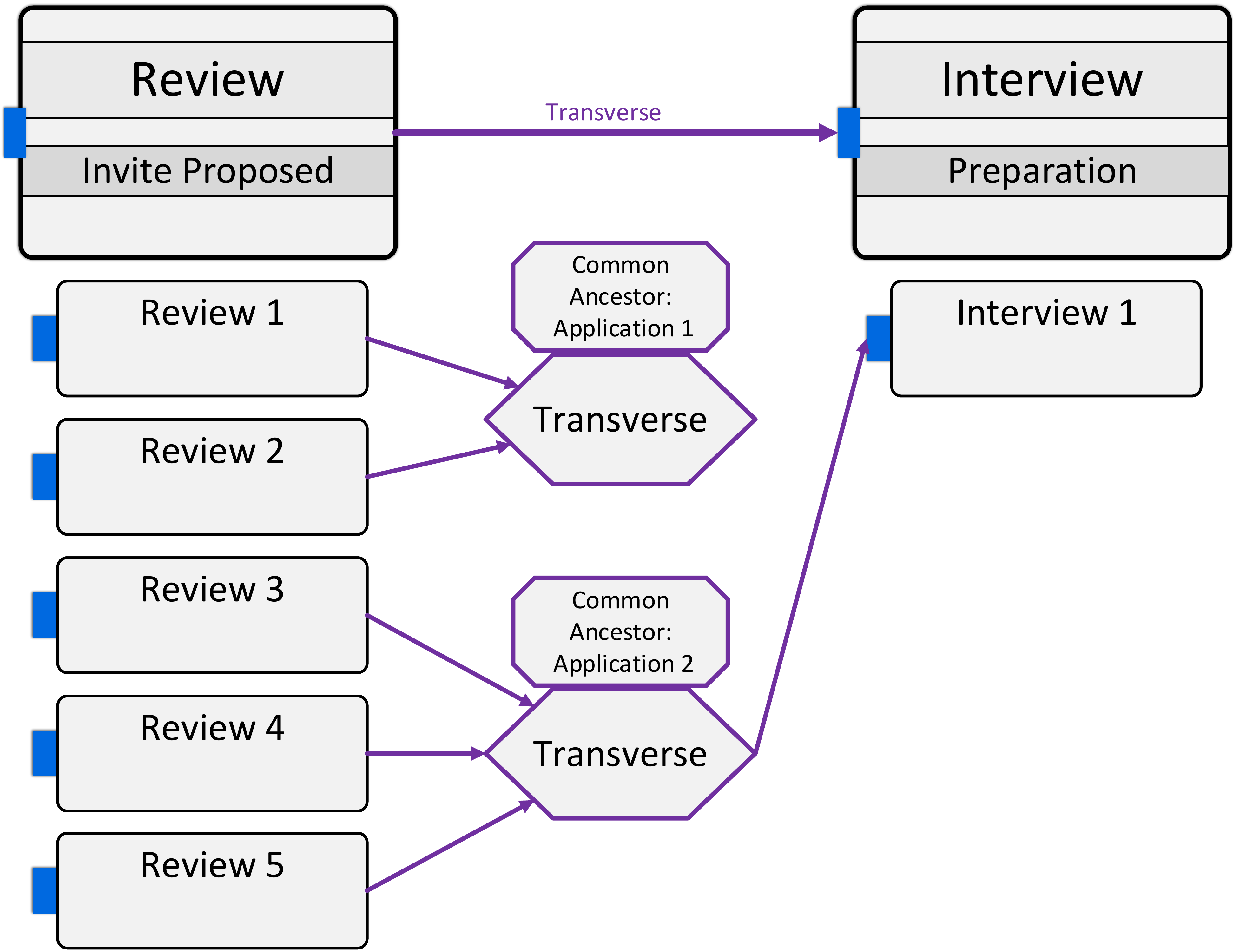}\caption{\label{fig:Example-Tranverse-CC}Transverse Coordination Component
Example}
\end{figure}

\noindent\begin{minipage}[t]{1\columnwidth}%
\begin{shaded}%
\begin{example}[Common Ancestor]
Figure \ref{fig:Example-Tranverse-CC} shows transverse coordination
components for the semantic relationship between $\mathit{Review}$:\LyXZeroWidthSpace$Invite$~$Proposed$
and $\mathit{Interview}$:$\mathit{Preparation}$. As there are two
$\mathit{Applications}$, two transverse coordination components are
instantiated. The connections to the source coordination step and
target port instances are again determined by the relations between
the process instances and the common ancestor.\vspace*{-1em}
\end{example}
\end{shaded}%
\end{minipage}

Similar to bottom-up coordination components, transverse coordination
components can be configured using an expression $\lambda$. For example,
the semantics of the transverse coordination component can be modified
using $\lambda$ to represent ``A majority of reviews must be in
favor of the applicant''.

\subsubsection{Self-Transverse Semantic Relationship}

Self-transverse semantic relationships are a variant of the transverse
semantic relationship. The difference is that both process types on
the source and target side are the same, giving the self-transverse
semantic relationships a different purpose compared to transverse
semantic relationships. Essentially, a self-transverse semantic relationship
corresponds to an $m$-out-of-$n$ choice pattern. It is represented
at run-time by a self-transverse coordination component (cf . Definition
\ref{def:A-self-transverse-coordination}). The self-transverse component
allows $m$ process instances to reach the target state. After $m$
process instances have reached the target state, the other $n-m$
process instances are prevented from reaching the target state.
\begin{definition}[Self-transverse Coordination Component Instance]
\label{def:A-self-transverse-coordination}A self-transverse coordination
component instance $s_{self}^{I}$ has the form $(s^{I},\omega_{ca}^{I},B_{src}^{I},H_{tar}^{I},\lambda)$
where
\end{definition}
\begin{itemize}
\item $s_{self\mhyphen transverse}^{I}<:s^{I}$, with $s^{I}$ defined as
in Definition \ref{Def:CoordinationComponentBaseInstance}
\item $\omega_{ca}^{I}$ is the common ancestor, $\omega_{ca}^{I}$ is instantiator
(cf. Definition \ref{Def:ObjectInstance-Normal})
\item $B_{src}^{I}$ is a set of coordination step instances related to
$\omega_{ca}^{I}$, i.e., $\forall\beta^{I}\in B_{src}^{I}:\text{\ensuremath{\beta^{I}.\omega^{I}}}\twoheadrightarrow\omega_{ca}^{I}$
(cf. Definition \ref{Def:CoordinationStepInstance})
\item $H_{tar}^{I}$ is a set of port instances related to $\omega_{ca}^{I}$,
i.e., $\forall\eta^{I}\in H_{tar}^{I}:\text{\ensuremath{\eta^{I}.\beta^{I}.\omega^{I}}}\twoheadrightarrow\omega_{ca}^{I}$,\\
$\forall\text{\ensuremath{\beta}}_{i}^{I}\in B_{src}^{I},\eta_{j}^{I}\in H_{tar}^{I}:\text{\ensuremath{\beta_{i}^{I}.\omega^{I}.\omega^{T}}}=\eta_{j}^{I}.\beta^{I}.\omega^{I}.\omega^{T}$
(cf. Definition \ref{Def:PortInstance})
\item $\lambda$ is an expression copied from $s^{I}.s^{T}$
\end{itemize}
The formal definition of self-transverse coordination components is
essentially the same as for transverse coordination components. The
deciding difference is that both process types referenced in the source
and target sets $B_{source}^{I}$ and $H_{target}^{I}$ are the same. 

\begin{figure}

\begin{centering}
\includegraphics[width=1\columnwidth]{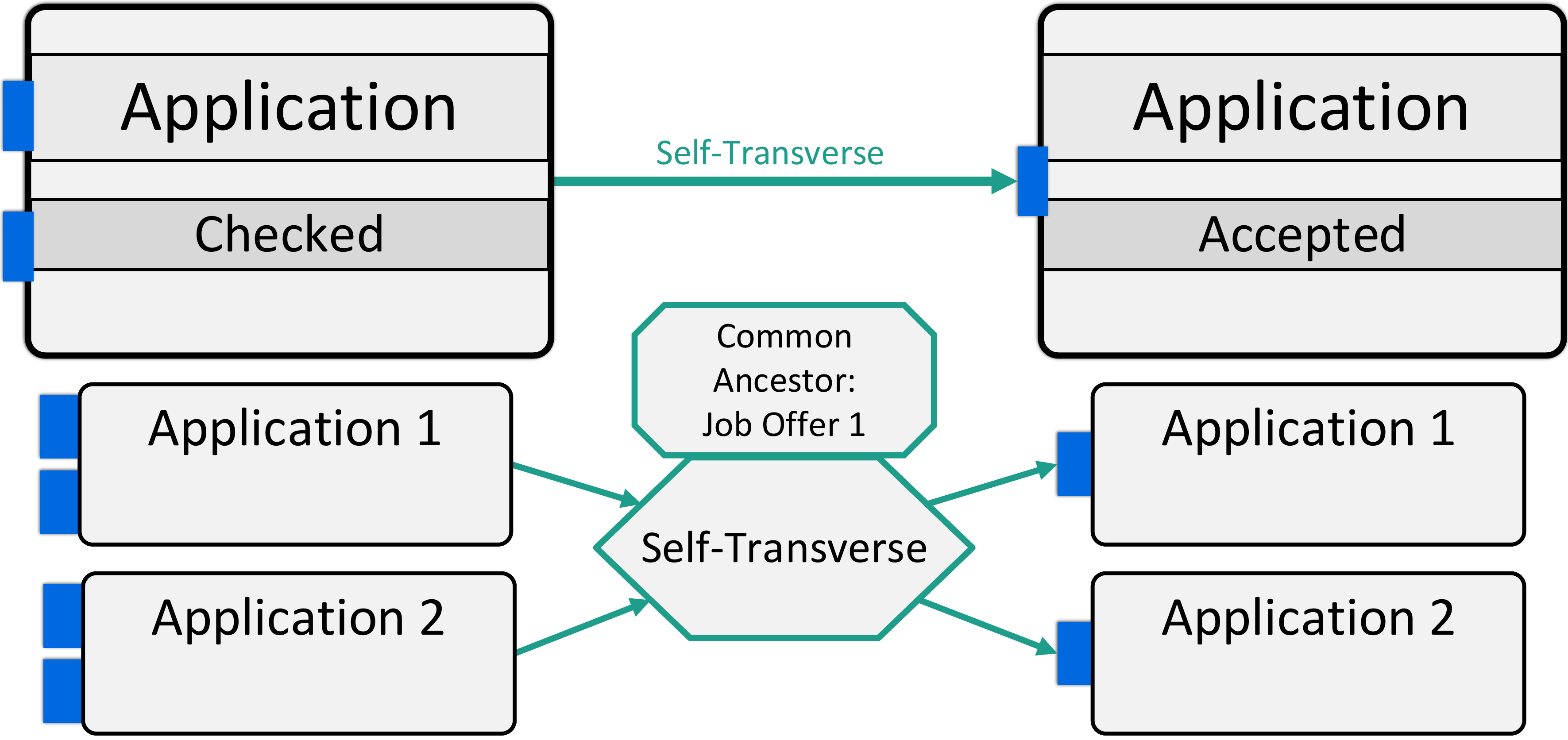}\caption{\label{fig:Example-Self-Transverse-CC}Example of a Self-Transverse
Coordination Component}
\par\end{centering}
\end{figure}

\noindent\begin{minipage}[t]{1\columnwidth}%
\begin{shaded}%
\begin{example}[Self-Transverse Relationship]
The self-transverse coordination component is employed in the coordination
process of the running example between $\mathit{Application}$:$\mathit{Checked}$
and $\mathit{Application}$:$\mathit{Accepted}$ (cf. Figure \ref{fig:Example-Self-Transverse-CC}).
Only one $\mathit{Application}$ may be accepted for a given $\mathit{Job\:Offer}$,
which is achieved by an appropriate expression $\lambda$ and the
self-transverse basic semantics. Once an $\mathit{Application}$ is
accepted, the remaining $\mathit{Applications}$ in state $\mathit{Checked}$
cannot reach state $\mathit{Accepted}$ and must be rejected.\vspace{-0.5em}
\end{example}
\end{shaded}%
\end{minipage}

The self-transverse semantic relationship is the only typo of semantic
relationship that cannot be automatically determined using the relational
process structure. It shares the same characteristics as the self
semantic relationship, namely the same process type as source and
target. Therefore, the modeler of a coordination process has to manually
choose between self semantic relationship and self-transverse semantic
relationship.

\begin{figure*}[t]
\centering{}\includegraphics[width=1\textwidth]{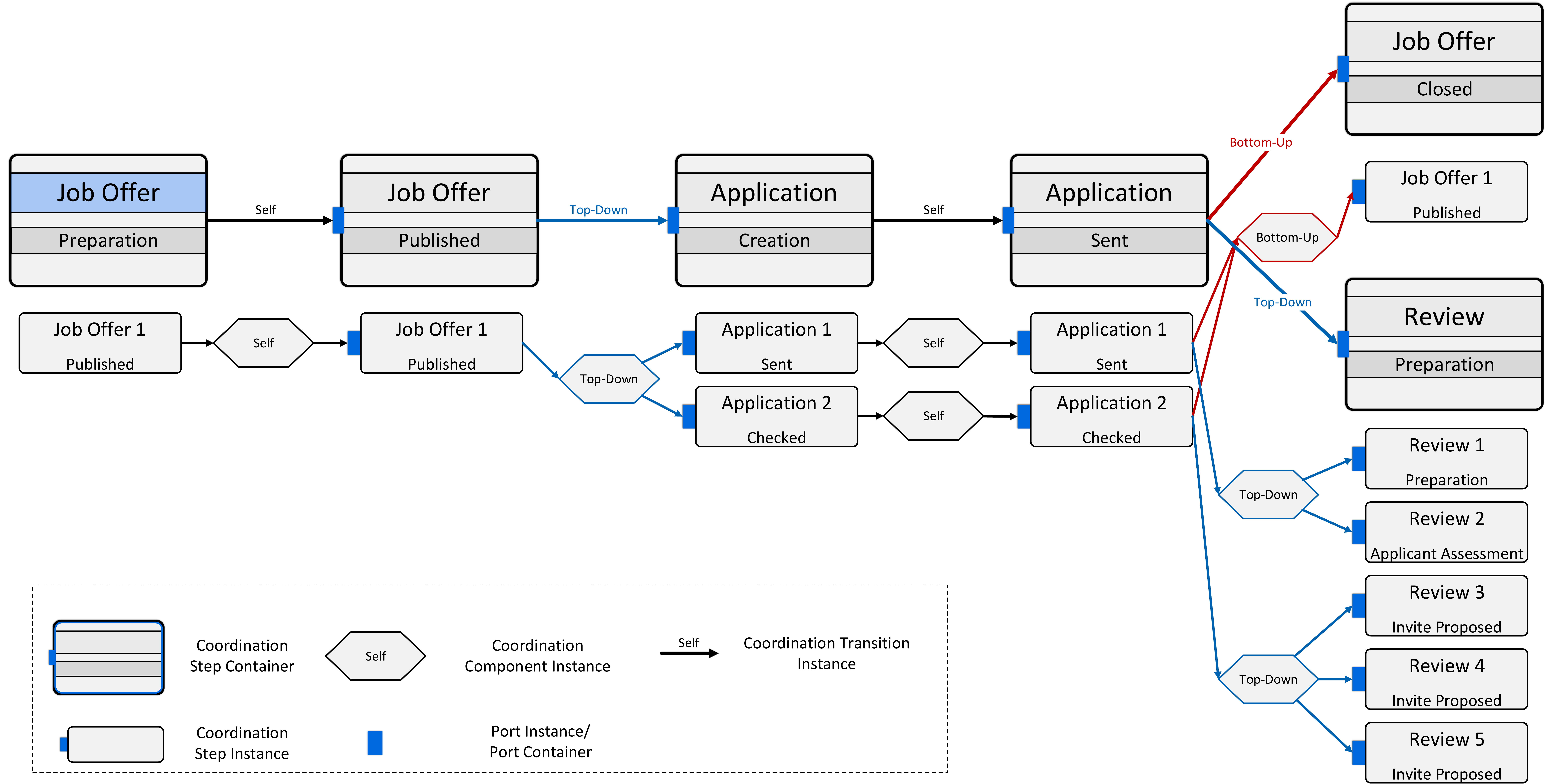}\caption{\label{fig:Partial-container-and-instance-view}Partial container
and instance view of the Job Offer coordination process}
\end{figure*}

\subsection{Conclusions}

This section has shown how a coordination process uses the relational
process structure to correctly build an interconnected graph of coordination
steps, ports, and coordination components. The coordination process
takes the current relational process instance structure and combines
this information with the specified semantic relationships to form
the graph. This graph reacts to changes in the relational process
structure, i.e, the creation or deletion of process instances and
relations, creating or updating coordination components and coordination
steps as needed. The type and number of semantic relationships in
a coordination process instance remain unchanged over the lifetime
of the particular coordination process instance.

\subsubsection{Fulfilling Challenge 2\emph{ Complex Process Relations} and Challenge
3 \emph{Local Contexts}}

In regard to the challenges presented in Section \ref{sec:Challenges-and-Problem},
this representation of process instances in terms of coordination
steps, ports, and coordination components solves Challenge 2 \emph{Complex
Process Relations}. As demonstrated in this section, coordination
processes faithfully replicate the complex relationships between the
numerous process instances in the relational process structure (cf.
Figure \ref{fig:Partial-container-and-instance-view}). Semantic relationships
and their respective coordination components by design take these
complex relationships into account and enforce coordination constraints
based on these relationships. The coordination process further adapts
to changes in the relational process structure by instantiating or
deleting coordination steps and coordination components as required,
or changing the associations between coordination components and coordination
steps according to their changing relations. This enables a high run-time
flexibility when executing individual processes, changing their relations,
as well as creating or deleting process instances. 

Furthermore, replicating coordination components for specific coordination
steps and process instances additionally create local contexts. When
two coordination steps and their associated process instances are
each connected to a coordination component, each coordination step
is within its own local context. This allows enforcing individual
constraints on each of the instances, e.g., two $\mathit{Applications}$
are in different contexts. When one $\mathit{Application}$ is rejected
and one is accepted, different coordination constraints apply to each
$\mathit{Application}$. While the overall set of coordination constraints
is, in principle, applicable to both $\mathit{Applications}$, their
different local contexts alter the relevance of specific subsets of
coordination constraints. Consequently, one $\mathit{Application}$
is treated independently from other $\mathit{Applications}$, fulfilling
Challenge 3 \emph{Local Contexts}.

\section{\label{sec:Operational-Semantics}Markings and Process Rules}

Described in the most general terms, coordination processes must enforce
coordination constraints between different types of processes. As
the processes to be coordinated are represented by a state-based view,
coordination constraints essentially determine whether a state of
a coordinated process is permitted to activate at a given point in
time. This, in turn, is determined by the active states of other,
related processes involved in the respective coordination constraint.

\noindent\begin{minipage}[t]{1\columnwidth}%
\begin{shaded}%
\begin{example}[Application Creation]
An $\mathit{Application}$ may only be created when the corresponding
$\mathit{Job\:Offer}$ is in state $\mathit{Published}$.\vspace*{-0.2cm}
\end{example}
\end{shaded}%
\end{minipage} 

Coordination processes may express coordination constraints by using
combinations of semantic relationships. Semantic relationships, represented
as coordination components, are connected to coordination steps and
ports, which represent the process instances. Section \ref{sec:Representing-Process-Instances}
has shown how a coordination process $c$ uses the relational process
structure $d$ to correctly build an interconnected graph consisting
of coordination steps $\beta$, ports $\eta$, and coordination components
$s$. The coordination process instance $c^{I}$ takes the current
relational process instance structure $d^{I}$ and combines its information
with the specified semantic relationships to form the graph. This
graph reacts to changes in $d^{I}$, i.e, the creation or deletion
of process instances and relations, creating coordination components
and coordination steps as needed. What is still missing is a mechanism
to react to the state changes of the process instances, i.e, to the
process in the relational process instance structure being executed.
Because of the coordination constraints, these state changes may have
impact on the enactment of other process instances.

A coordination process detects when the active state $\sigma_{a}^{I}$
of the process instances changes within its scope, and evaluates all
coordination constraints affected by this change. In the end, this
evaluation results in the coordination process permitting new states
to become activated, or denying other states activation. Hereby, coordination
processes operate according to the \emph{blacklist principle}: An
action is allowed by default unless a coordination constraint specifically
prevents it. This allows for a high degree of flexibility when executing
the coordinated processes, contributing to the fulfillment of Challenge
1 \emph{Asynchronous Concurrency}.

The purpose of the \emph{operational semantics} of a coordination
process is to correctly evaluate and enforce the coordination constraints
specified in a coordination process.

\subsection{\label{subsec:Introducing-Coordination-Step}Coordination Step and
State Markings}

The execution status of coordinated processes, i.e., the active state
$\sigma_{a}^{I}$ of a process instance $\omega^{I}$, changes far
more often than the relational process structure, i.e, the operational
semantics must react to state changes more often than, for example,
to the emergence of a new process instance.

\noindent\begin{minipage}[t]{1\columnwidth}%
\begin{shaded}%
\begin{example}[State Changes]
Consider the relational process structure from Figure \ref{fig:Relational-Process-Instance}.
There are two $\mathit{Applications}$, i.e., $\mathit{Application\:1}$
with active state $\mathit{Sent}$ and $\mathit{Application\:2}$
with active state $\mathit{Checked}$. To each $\mathit{Application}$
belong several $\mathit{Reviews}$ and $\mathit{Interviews}$, each
having their own active states. $\mathit{Application\:1}$ has $\mathit{Review\:1}$
and $\mathit{Review}\:2$ with active states $\mathit{Preparation}$
and $\mathit{Applicant\:Assessment}$, respectively. $\mathit{Application\:2}$
has three $\mathit{Reviews}$ with active state $\mathit{Invite\:Proposed}$
and one $\mathit{Interview}$ with active state $\mathit{Preparation}$.
According to Figure \ref{fig:State-based-views-of}, for each process
instance in the relational process structure to have reached current
status, at least 16 state changes must have occurred, but only 8 new
process instances have emerged. Moreover, as no applicant has been
found yet, more state changes are likely to occur when the processes
are further enacted. This is because few of the process instances
are yet in an end state.\vspace*{-0.2cm}
\end{example}
\end{shaded}%
\end{minipage}

Section \ref{sec:Representing-Process-Instances} describes how process
instances and relations are captured in a coordination process. However,
coordination constraints are not only influenced by processes and
their relations, but also by their execution status, i.e, the active
state of the processes. Therefore, the combination of execution status
and process relations determines whether a coordination constraint
is fulfilled in a coordination process. This combination is represented
in a coordination process by using \emph{markings} for its constituting
entities. Each entity, e.g., coordination component or port, is assigned
a marking that indicates its current status. Formally, Definitions
\ref{Def:CoordinationProcessInstance}-\ref{Def:CoordinationComponentBaseInstance}
are extended (cf. Definition \ref{def:Marking-Extension}).
\begin{definition}[Markings]
\label{def:Marking-Extension}The entities $e$ from Definitions
\ref{Def:CoordinationProcessInstance}-\ref{Def:CoordinationComponentBaseInstance}
are each extended with
\end{definition}
\begin{itemize}
\item $\mu_{e}$ is the marking of the entity $e$, where $\mu_{e}\in$$\,\{\text{\ensuremath{\mathit{Inactive}},}$
$\mathit{Update},$ $\mathit{Active},$ $\mathit{Completed},$ $\mathit{Eliminated\}}$
\end{itemize}
Each marking possesses a specific meaning for each entity type. The
conditions under which the marking is applied varies between entity
types. Table \ref{tab:Coordination-Step-Markings} summarizes the
markings and their meanings for the coordination step instance $\beta^{I}$.
Note that the textual definitions of all markings (cf. Tables \ref{tab:Coordination-Step-Markings}-\ref{tab:Port-Markings})
are abstracted the full formal definitions used in the implementation
of the coordination process approach, but convey the general intentions
behind the marking. The full formal definitions are required to deal
with numerous special cases to guarantee a fully operational coordination
process, which for reasons of comprehensibility have been reduced
to their essentials in this paper.

\begin{table}[h]

\caption{\label{tab:Coordination-Step-Markings}Coordination Step Markings}

\centering{}%
\begin{tabular*}{1\columnwidth}{@{\extracolsep{\fill}}l>{\raggedright}m{0.7\columnwidth}}
\toprule 
Marking $\mu_{\beta}$ & Description\tabularnewline
\midrule
\midrule 
{\footnotesize{}$\mathit{Inactive}$} & {\footnotesize{}The coordination step has not yet been reached, i.e.
none of its ports is active. The corresponding state is not skipped.}\tabularnewline
\midrule 
{\footnotesize{}$\mathit{Update}$} & {\footnotesize{}A change has been triggered, requiring a reevaluation
of the coordination step marking.}\tabularnewline
\midrule 
{\footnotesize{}$\mathit{Active}$} & {\footnotesize{}An attached port has become active. The corresponding
state may be activated, but the state is not yet active.}\tabularnewline
\midrule 
{\footnotesize{}$Completed$} & {\footnotesize{}The corresponding state is activated or confirmed
and an attached port is active or completed.}\tabularnewline
\midrule 
{\footnotesize{}$\mathit{Eliminated}$} & {\footnotesize{}Either the state is skipped or all attached ports
are eliminated.}\tabularnewline
\bottomrule
\end{tabular*}
\end{table}

A coordination step is the only entity in a coordination process that
interacts directly with the state-based view of the coordinated processes.
The marking of a coordination step $\beta^{I}.\mu_{\beta}$ directly
influences state marking $\sigma^{I}.\mu_{\sigma}$ of process $\omega^{I}$,
whereas vice versa marking $\sigma^{I}.\mu_{\sigma}$ directly influences
marking $\beta^{I}.\mu_{\beta}$. Consequently, a state $\sigma_{\mathit{ref}}^{I}$
in a state-based view references all corresponding coordination step
instances $\beta^{I}$ of a coordination process where $\beta.\sigma^{I}=\sigma_{\mathit{ref}}^{I}$.
Having coordination steps as the only entity directly interacting
with state-based views fosters a loose coupling between a coordination
process and the coordinated processes.

State-based views also operate on a markings-based semantics. The
\emph{active state} $\sigma_{a}^{I}$ signifies the state that is
currently executed; there always has to be exactly one state marked
as $\mathit{Activated}$. In case the process has finished executing,
the end state remains marked as $\mathit{Activated}$. Other markings
show whether a state waits to be executed, was executed, or cannot
be executed anymore due to a preceding decision branching. Table \ref{tab:State-Markings}
gives a full overview over state markings $\mu_{\sigma}$ and their
meaning.

\begin{table}[h]
\caption{\label{tab:State-Markings}State Markings}

\centering{}%
\begin{tabular*}{1\columnwidth}{@{\extracolsep{\fill}}l>{\raggedright}m{0.7\columnwidth}}
\toprule 
Marking $\mu_{\sigma}$ & Description\tabularnewline
\midrule
\midrule 
{\footnotesize{}$Waiting$} & {\footnotesize{}The state has not been executed yet. A predecessor
state is activated.}\tabularnewline
\midrule 
{\footnotesize{}$\mathit{Pending}$} & {\footnotesize{}The activation of a state is blocked by an unfulfilled
coordination constraint.}\tabularnewline
\midrule 
{\footnotesize{}$\mathit{Activated}$} & {\footnotesize{}The state is currently active.}\tabularnewline
\midrule 
{\footnotesize{}$Confirmed$} & {\footnotesize{}The state has been successfully executed. A successor
state is activated.}\tabularnewline
\midrule 
{\footnotesize{}$\mathit{Skipped}$} & {\footnotesize{}The state can no longer be executed. A state on an
alternative branch is activated.}\tabularnewline
\bottomrule
\end{tabular*}
\end{table}

If a state $\sigma_{i}^{T}$ of a process is referenced in a coordination
process, i.e., $\exists\beta^{T}\in c^{T}.B^{T}:\beta^{T}.\sigma^{T}=\sigma_{i}^{T}$,
the marking of a corresponding coordination step instance $\beta^{I}$
influences the marking of the state $\sigma_{i}^{I}$. This enables
the enforcement of coordination constraints regarding the coordinated
processes. Coordination processes and the expressed coordination constraints
therefore may restrict the behavior of coordinated processes. For
example, a state may switch directly from marking$\mathit{Waiting}$
to $\mathit{Activated}$ if a state transition is triggered and no
coordination process is involved. If a coordination process is involved,
and if the coordination step instance is not $\mathit{Active}$, the
resulting marking will be $\mathit{Pending}$.

\begin{figure}
\begin{centering}
\includegraphics[width=1\columnwidth]{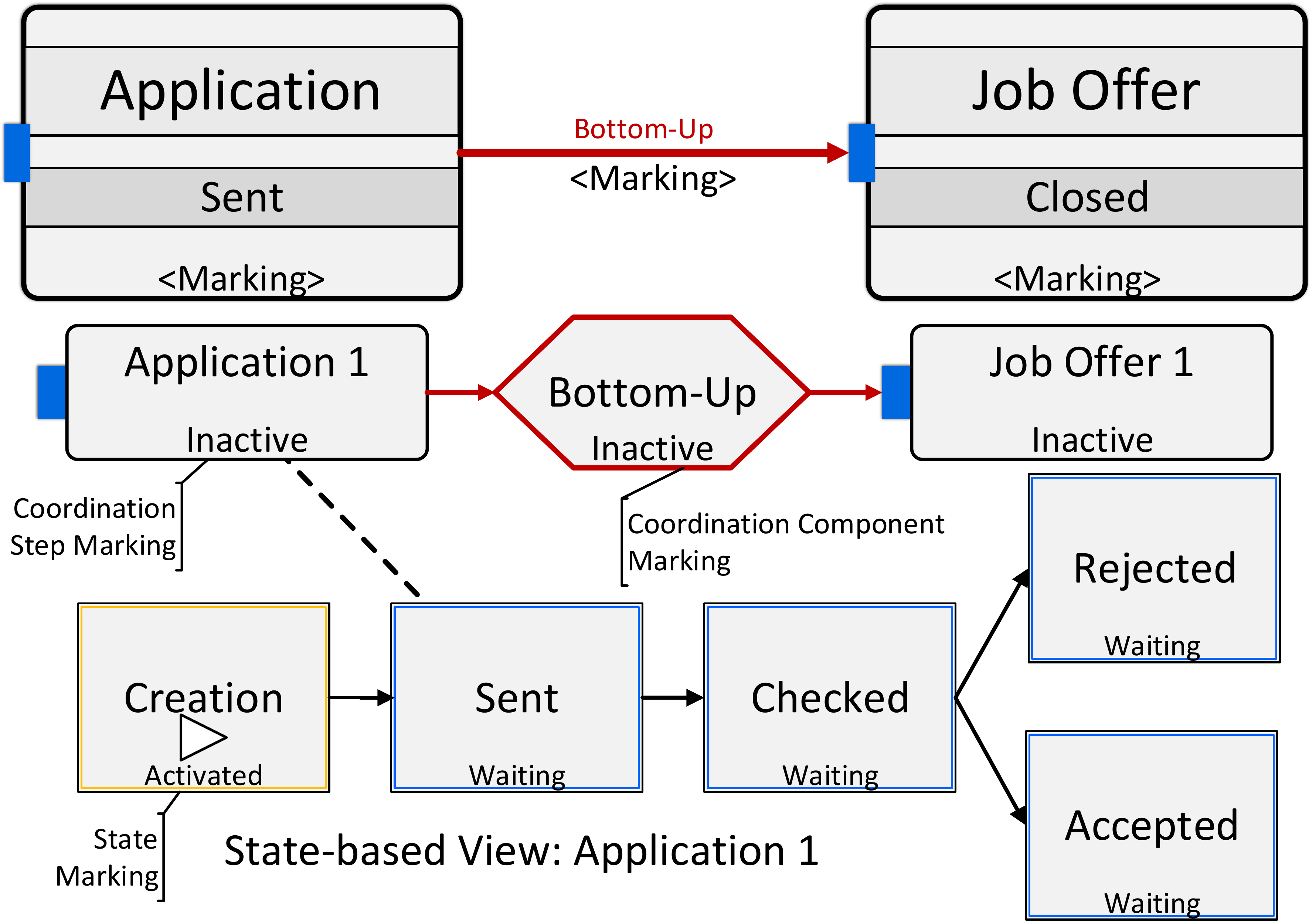}\caption{\label{fig:Markings-Example,-Stage1}Markings Example, Stage 1}
\par\end{centering}
\vspace*{-0.5cm}
\end{figure}

\noindent\begin{minipage}[t]{1\columnwidth}%
\begin{shaded}%
\begin{example}[Inactive Coordination Step]
Figure \ref{fig:Markings-Example,-Stage1} shows an example of coordination
step $\mathit{Application}$:$\mathit{Sent}$ that coordinates state
$\mathit{Sent}$ of process instance $\mathit{Application\:1}$. State
$\mathit{Sent}$ is currently marked as $\mathit{Waiting}$; the start
state of $\mathit{Application\:1}$, i.e., state $\mathit{Creation}$,
is marked as $\mathit{Activated}$. Currently, activation of state
$\mathit{Sent}$ is prevented, as coordination step $\mathit{Application}$:$\mathit{Sent}$
is marked as $\mathit{Inactive}$.\vspace*{-1em}
\end{example}
\end{shaded}%
\end{minipage}

Coordination step containers and coordination transition instances
have markings as well. This is indicated by including placeholder
markings <Marking> in the respective graphical elements (cf. Figure
\ref{fig:Markings-Example,-Stage1}). This intends to show that these
markings exist, but their specific value is irrelevant in the current
context. Port containers have markings as well, which are omitted
for space reasons.

In Figure \ref{fig:Markings-Example,-Stage1}, if the state transition
from $\mathit{Creation}$ to $\mathit{Sent}$ is triggered as $\mathit{Application\:1}$
is executed, the coordination process prevents the immediate activation
of state $\mathit{Sent}$ due to coordination step $\mathit{Application}$:$\mathit{Sent}$
being marked as $\mathit{Inactive}$. In other words, one or more
coordination constraints (not depicted in Figure \ref{fig:Markings-Example,-Stage1})
prevent state $\mathit{Sent}$ from activation. In consequence, state
$\mathit{Sent}$ does not receive marking $\mathit{Activated}$, but
instead is marked as $\mathit{Pending}$. This situation is shown
in Figure \ref{fig:Markings-Example,-Stage2}. The marking $\mathit{Pending}$
of state $\mathit{Sent}$ stops the execution of the $\mathit{Application\:1}$
process until the coordination constraints are fulfilled, i.e., until
the coordination step becomes marked as $\mathit{Active}$.

This, in turn, is determined by the incoming coordination components
of the coordination step instance. Whether or not the incoming coordination
components can be fulfilled is determined by the coordination condition
$\lambda$ of the respective coordination component $s^{I}$. In other
words, the selection which incoming coordination component instance
$s^{I}$ becomes $\mathit{Completed}$ is not made by their source
coordination steps $\beta_{\mathit{source}}^{I}$.

\begin{figure}
\centering{}\includegraphics[width=1\columnwidth]{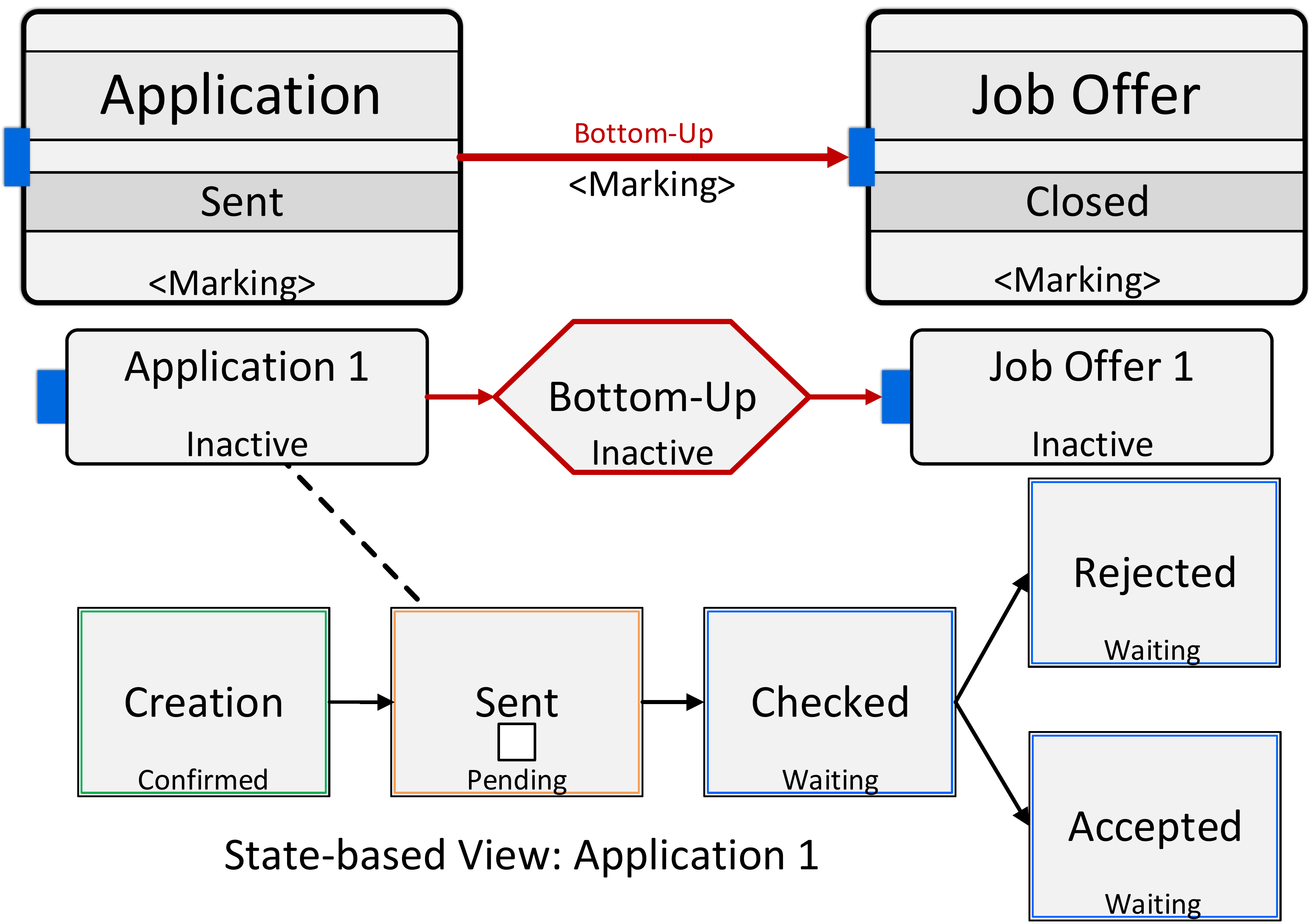}\caption{\label{fig:Markings-Example,-Stage2}Markings Example, Stage 2}
\end{figure}

It is now assumed that, due to a change in the relational process
structure, the incoming coordination constraints of $\mathit{Application}$:$\mathit{Sent}$
become fulfilled. This means the incoming coordination components
of coordination step $\mathit{Application}$:$\mathit{Sent}$ and
its ports, which represent the coordination constraints, allow the
coordination step to be marked as $\mathit{Active}$ (cf. Figure \ref{fig:Markings-Example,-Stage3}).

Immediately after coordination step $\mathit{Application}$:$\mathit{Sent}$
is marked as $\mathit{Active}$, the new status is propagated to the
coordinated process instances and states, i.e., to state $\mathit{Sent}$
of $\mathit{Application\:1}$. In turn, state $\mathit{Sent}$ can
be activated as well, allowing the execution of $\mathit{Application\:1}$
to continue. In this case, the state might be activated before the
corresponding coordination constraints have been fulfilled. Reaching
a state before the corresponding coordination constraints permit its
activation, as shown here, is denoted as \emph{state-first activation},
as opposed to \emph{coordination-first} \emph{activation}. Note that
state-first activation involves an intermediate blocking of process
enactment, signified by marking $\mathit{Pending}$. Coordination-first
activation will be discussed later in this section.

\begin{figure}
\centering{}\includegraphics[width=1\columnwidth]{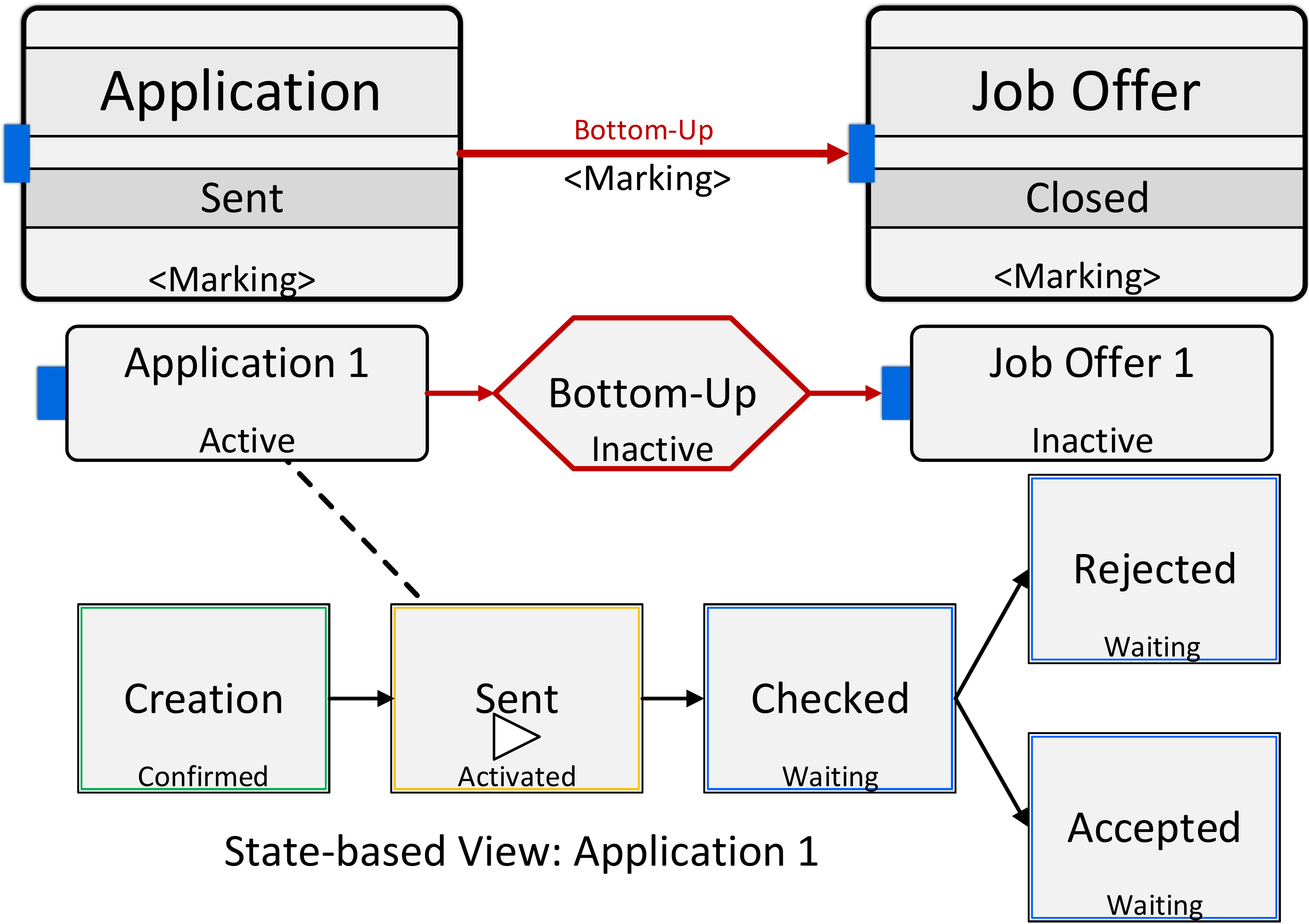}\caption{\label{fig:Markings-Example,-Stage3}Markings Example, Stage 3}
\end{figure}

As state $\mathit{Sent}$ of $\mathit{Application\:1}$ has received
marking $\mathit{Activated}$, another marking change becomes necessary.
According to the definition of marking $\mathit{Completed}$ (cf.
Table \ref{tab:Coordination-Step-Markings}), both conditions are
fulfilled, i.e., ports are active and the referenced state is $\mathit{Activated}$.
Consequently, coordination step $\mathit{Application}$:$\mathit{Sent}$
must be marked as $\mathit{Completed}$. Marking coordination steps
as $\mathit{Completed}$ has a profound impact on its outgoing coordination
components and, by extension, subsequent ports and other coordination
steps. Marking changes of coordination component instances and port
instances are explored in the Section \ref{subsec:Introducing-Coordination-Compone}.

\subsection{\label{subsec:Introducing-Coordination-Compone}Coordination Component
and Port Markings}

Coordination component instances represent semantic relationships
at run-time, where each has different basic semantics (cf Section
\ref{tab:Overview-over-semantic}). When fulfilled, the referenced
state of the target coordination step may become activated, otherwise
the state is prevented from activation or must be skipped when the
target coordination step becomes marked as $\mathit{Eliminated}$.
The indication whether a semantic relationship is fulfilled is determined
through the markings of the corresponding coordination component (cf.
Table \ref{tab:Coordination-Component-Markings}). More specifically,
a semantic relationship is fulfilled when the coordination component
has marking $\mathit{Completed}$.

\begin{table}[h]
\caption{\label{tab:Coordination-Component-Markings}Coordination Component
Markings}

\centering{}%
\begin{tabular*}{1\columnwidth}{@{\extracolsep{\fill}}l>{\raggedright}m{0.7\columnwidth}}
\toprule 
Marking $\mu_{s}$ & Description\tabularnewline
\midrule
\midrule 
{\footnotesize{}$\mathit{Inactive}$} & {\footnotesize{}None of the source coordination steps is completed.}\tabularnewline
\midrule 
{\footnotesize{}$\mathit{Update}$} & {\footnotesize{}A change has been triggered, requiring a reevaluation
of the marking of the coordination component.}\tabularnewline
\midrule 
{\footnotesize{}$\mathit{Active}$} & {\footnotesize{}At least one of the source coordination steps is completed
and the coordination component condition is not fulfilled.}\tabularnewline
\midrule 
{\footnotesize{}$Completed$} & {\footnotesize{}At least one of the source coordination steps is completed
and the coordination component condition is fulfilled.}\tabularnewline
\midrule 
{\footnotesize{}$\mathit{Eliminated}$} & {\footnotesize{}Either all source coordination steps are eliminated
or the coordination component cannot be fulfilled.}\tabularnewline
\bottomrule
\end{tabular*}
\end{table}

Generally, coordination component instances become fulfilled in order
from start step to end step of the respective coordination process.
That means a prerequisite for fulfilling a coordination component
is a path of completed coordination components leading from the start
of the coordination process to the yet unfulfilled coordination component.
Therefore, it may be concluded that this path exists if the source
coordination step instance is marked as $\mathit{Completed}$. This
conclusion is based on the definition of coordination component markings
$\mu_{s}=\mathit{Active}$ and $\mu_{s}=\mathit{Completed}$ (cf.
Table \ref{tab:Coordination-Component-Markings}). It can be proven
inductively that each predecessor coordination component must have
been fulfilled at one point, i.e., been marked as $\mathit{Completed}$,
in order for the source coordination step instance of the current
coordination component instance to be marked as $\mathit{Completed}$.

By default, coordination components are marked as $\mathit{Inactive}$,
indicating that the coordination component may still be fulfilled.
Regarding the meaning, the marking $\mathit{Inactive}$ for a coordination
component instance $s^{I}$ is analogous to the marking $\mathit{Waiting}$
for a state $\sigma^{I}$ (cf. Table \ref{tab:State-Markings}). 

\noindent\begin{minipage}[t]{1\columnwidth}%
\begin{shaded}%
\begin{example}[Completing Coordination Step Instances]
\label{exa:Completing_coordination_step}In previous examples (cf.
Figure \ref{fig:Markings-Example,-Stage1}-\ref{fig:Markings-Example,-Stage3}),
the bottom-up coordination component has been marked as $\mathit{Inactive}$.
With the marking change of the source coordination step to $\mathit{Completed}$,
the conditions for the bottom-up coordination component to be marked
as $\mathit{Active}$ are met (cf. Table \ref{tab:Coordination-Component-Markings}).\vspace*{-1em}
\end{example}
\end{shaded}%
\end{minipage}

Figure \ref{fig:Markings-Example,-Stage4} shows the $\mathit{Completed}$
marking of the $\mathit{Appli}$\-\LyXZeroWidthSpace$\mathit{cation\:1}$
coordination step as well as the $\mathit{Active}$ marking of the
bottom-up coordination component.

Coordination components that are connected to the same source coordination
step $\beta_{\mathit{src}}^{I}$ are all marked as $\mathit{Active}$
upon marking the coordination step instance $\beta_{\mathit{src}}^{I}$
as $\mathit{Completed}$. This is due to the fact that any coordination
component $s^{I}\in\beta_{\mathit{src}}^{I}.S_{out}^{I}$ requires
the completion of all previous coordination components as well, i.e.,
the coordination components that are on the path from the start coordination
step of the coordination process to the source coordination step $\beta_{\mathit{src}}^{I}$.
In consequence, each outgoing coordination component instance $s^{I}$
must be allowed to become fulfilled. 

This basically constitutes AND-split semantics, as known from, for
example, BPMN 2.0. AND-split semantics for outgoing coordination components
enables the fulfillment of all the subsequent coordination components.
Whether or not the coordination components can be fulfilled is determined
solely by the coordination condition $\lambda$ of the respective
coordination component $s^{I}$. In other words, the selection which
outgoing coordination component instances $s^{I}$ become $\mathit{Active}$
is not done by the source coordination step $\beta_{\mathit{src}}^{I}$.
Consequently, there is no counterpart to XOR-split or OR-split semantics
in a coordination process, as these are not required or achieved by
different means.

\begin{figure}
\centering{}\includegraphics[width=1\columnwidth]{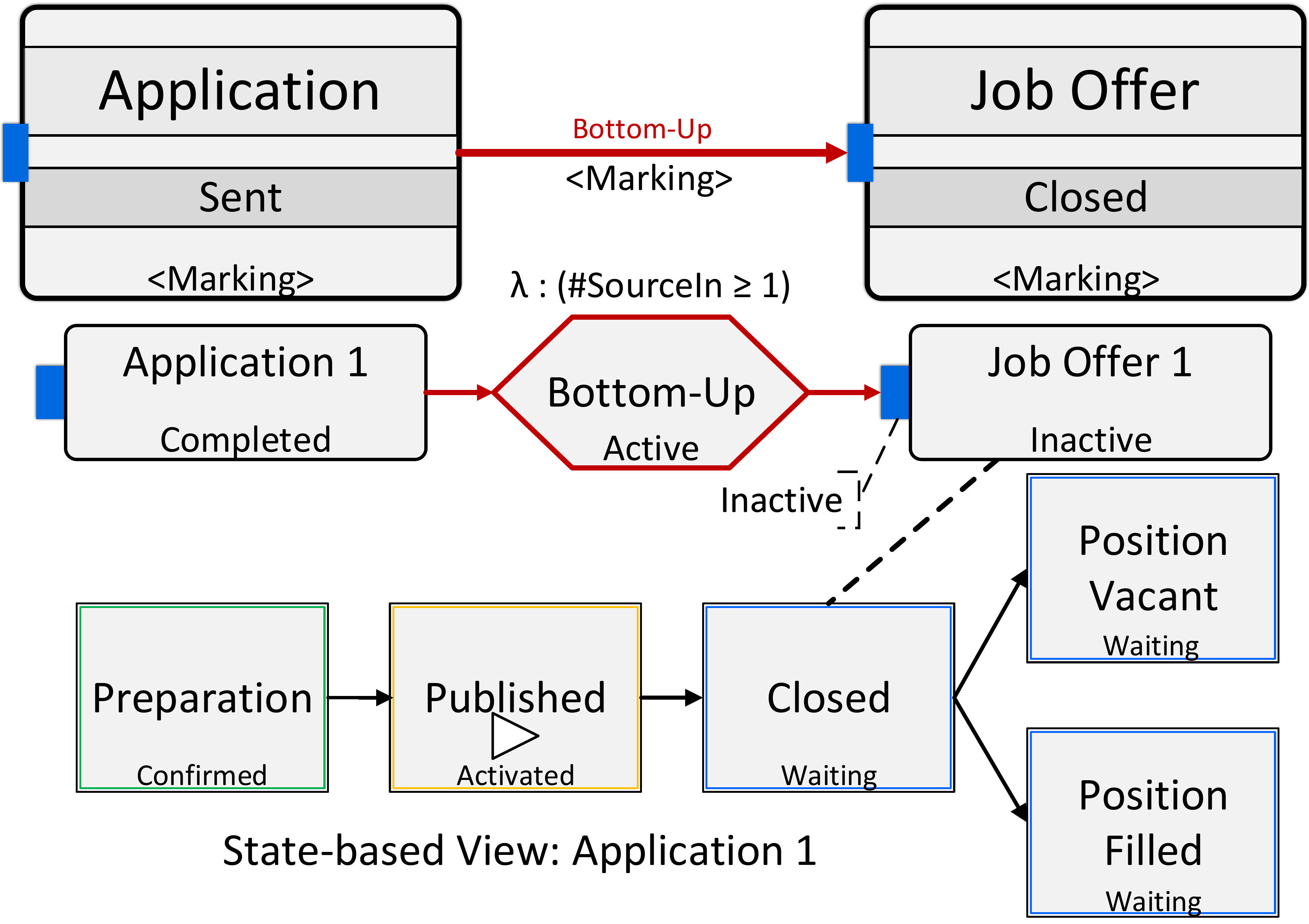}\caption{\label{fig:Markings-Example,-Stage4}Markings Example, Stage 4}
\end{figure}

\noindent\begin{minipage}[t]{1\columnwidth}%
\begin{shaded}%
\begin{example}[Unfulfilled Coordination Component Instance]
The process $\mathit{Job\:Offer\:1}$ is however not yet allowed
to activate state $\mathit{Closed}$ (cf. Figure \ref{fig:Markings-Example,-Stage4}).
The associated coordination step and its attached port are both still
marked as $\mathit{Inactive}$. In order to change that, the bottom-up
coordination component is required to be marked as $\mathit{Completed}$,
which, in turn, depends on whether or not the coordination component
condition (cf. Table \ref{tab:Coordination-Component-Markings}) is
fulfilled.\vspace{-0.5em}
\end{example}
\end{shaded}%
\end{minipage}

For a bottom-up coordination component, the coordination condition
is an expression with a Boolean return value (cf. Definition \ref{def:A-bottom-up-coordination}).
The concrete expression $\lambda$ that has been defined for the coordination
component from Figure \ref{fig:Markings-Example,-Stage4} is $\mathit{\lambda}=\mathit{\#SourceIn\,+\mathit{\#SourceAfter}}\geq1$.
The function $\mathit{\#SourceIn}$ counts the number of coordination
steps $\beta^{I}$ in $s_{\mathit{bottom-up}}^{I}.B_{source}^{I}$,
for which the referenced state $\beta^{I}.\sigma^{I}$ is $\mathit{Activated}$.
Function $\mathit{\#SourceAfter}$ works analogously for state marking
$\mathit{Confirmed}$. The coordination step count must be at least
1 for $s_{\mathit{bottom-up}}^{I}$ to be fulfilled, i.e., to mark
$s_{\mathit{bottom-up}}^{I}$ as $\mathit{Completed}$. In other words,
at least one $\mathit{Application}$ must be in state $\mathit{Sent}$
or must have progressed to a subsequent state. 

\noindent\begin{minipage}[t]{1\columnwidth}%
\begin{shaded}%
\begin{example}[Coordination Component Fulfillment]
The evaluation of expression $\lambda$ is triggered upon marking
$s_{\mathit{bottom-up}}^{I}$ as $\mathit{Active}$. As $s_{\mathit{bottom-up}}^{I}.B_{source}^{I}$
contains exactly one coordination step $\beta^{I}$ ($\mathit{Application\:1}$)
being marked as $\mathit{Completed}$, $\lambda$ is fulfilled and
$s_{\mathit{bottom-up}}^{I}$ is immediately re-marked as $\mathit{Completed}$.\vspace{-1em}
\end{example}
\end{shaded}%
\end{minipage}

The evaluation of $\lambda$ is not only triggered by $s_{\mathit{bottom-up}}^{I}$
becoming marked as $\mathit{Active}$. Numerous changes within the
coordination process can affect $s_{\mathit{bottom-up}}^{I}$, including,
but not limited to, other marking changes of coordination steps in
$s_{\mathit{bottom-up}}^{I}.B_{\mathit{source}}^{I}$ or the emergence
or deletion of processes and their associated coordination steps.
Generally, each of these changes may affect the fulfillment of any
coordination component $s^{I}$ and the associated coordination condition.
Furthermore, changes may also revert a previous fulfillment of a coordination
component. How a coordination process generally deals with these numerous
influences is described in Section \ref{subsec:The-Update-Marking}.

Marking $s_{\mathit{bottom-up}}^{I}$ as $\mathit{Completed}$ triggers
a change of the target port marking. Ports play a vital role in coordination
processes, as they allow combining different semantic relationships
to express complex coordination constraints. Connecting coordination
components to the same port constitutes AND-join-semantics, i.e.,
all coordination components $\eta^{I}.S_{in}^{I}$ must have marking
$\mathit{Completed}$ for a port $\eta^{I}$ to become $\mathit{Active}$.
If at least one of the ports $H^{I}$ of a coordination step $\beta^{I}$
is active, $\beta^{I}$ becomes active as well, constituting OR-join-semantics.
An overview of port markings is given in Table \ref{tab:Port-Markings}.

\begin{table}[h]
\caption{\label{tab:Port-Markings}Port Markings and their description}

\centering{}%
\begin{tabular*}{1\columnwidth}{@{\extracolsep{\fill}}l>{\raggedright}m{0.7\columnwidth}}
\toprule 
Marking $\mu_{\eta}$ & Description\tabularnewline
\midrule
\midrule 
{\footnotesize{}$\mathit{Inactive}$} & {\footnotesize{}None of the incoming coordination components are marked
as completed}\tabularnewline
\midrule 
{\footnotesize{}$\mathit{Update}$} & {\footnotesize{}A change has been triggered, requiring a reevaluation
of the marking of the port}\tabularnewline
\midrule 
{\footnotesize{}$\mathit{Active}$} & {\footnotesize{}All incoming coordination components are marked as
completed}\tabularnewline
\midrule 
{\footnotesize{}$Completed$} & {\footnotesize{}The referenced coordination step is marked as completed}\tabularnewline
\midrule 
{\footnotesize{}$\mathit{Eliminated}$} & {\footnotesize{}Either the referenced coordination step is eliminated,
or all incoming coordination steps are eliminated}\tabularnewline
\bottomrule
\end{tabular*}
\end{table}

\noindent\begin{minipage}[t]{1\columnwidth}%
\begin{shaded}%
\begin{example}[Port and Coordination Step Activation]
\label{exa:port_and_coordination_step_activation}There is only one
incoming coordination component for coordination step $\mathit{Job\:Offer\:1}$
in Example \ref{fig:Markings-Example,-Stage5}, which is marked as
$\mathit{Completed}$. According to the port marking semantics (cf.
Table \ref{tab:Port-Markings}),\vspace{-1em}
\end{example}
\end{shaded}%
\end{minipage} %
\noindent\begin{minipage}[t]{1\columnwidth}%
\begin{shaded}%
the port of coordination step $\mathit{Job\:Offer\:1}$ is marked
as $\mathit{Active}$ (cf .Figure \ref{fig:Markings-Example,-Stage5}).
According to Table \ref{tab:Coordination-Step-Markings}, the coordination
step for $\mathit{Job\:Offer\:1}$ is then marked as $\mathit{Active}$
as well. Consequently, $\mathit{Job\:Offer\:1}$ may now activate
state $\mathit{Closed}$ as desired; it is no longer blocked by the
coordination process. \end{shaded}%
\end{minipage}

This is a sample of \emph{coordination-first} \emph{activation}. Here,
the coordination process permits state activation even before the
respective process intends to activate the state. Accordingly, there
is no intermediate blocking as in the case of state-first activation.
Upon reaching the state, the state may immediately become marked as
$\mathit{Activated}$.

\begin{figure}
\centering{}\includegraphics[width=1\columnwidth]{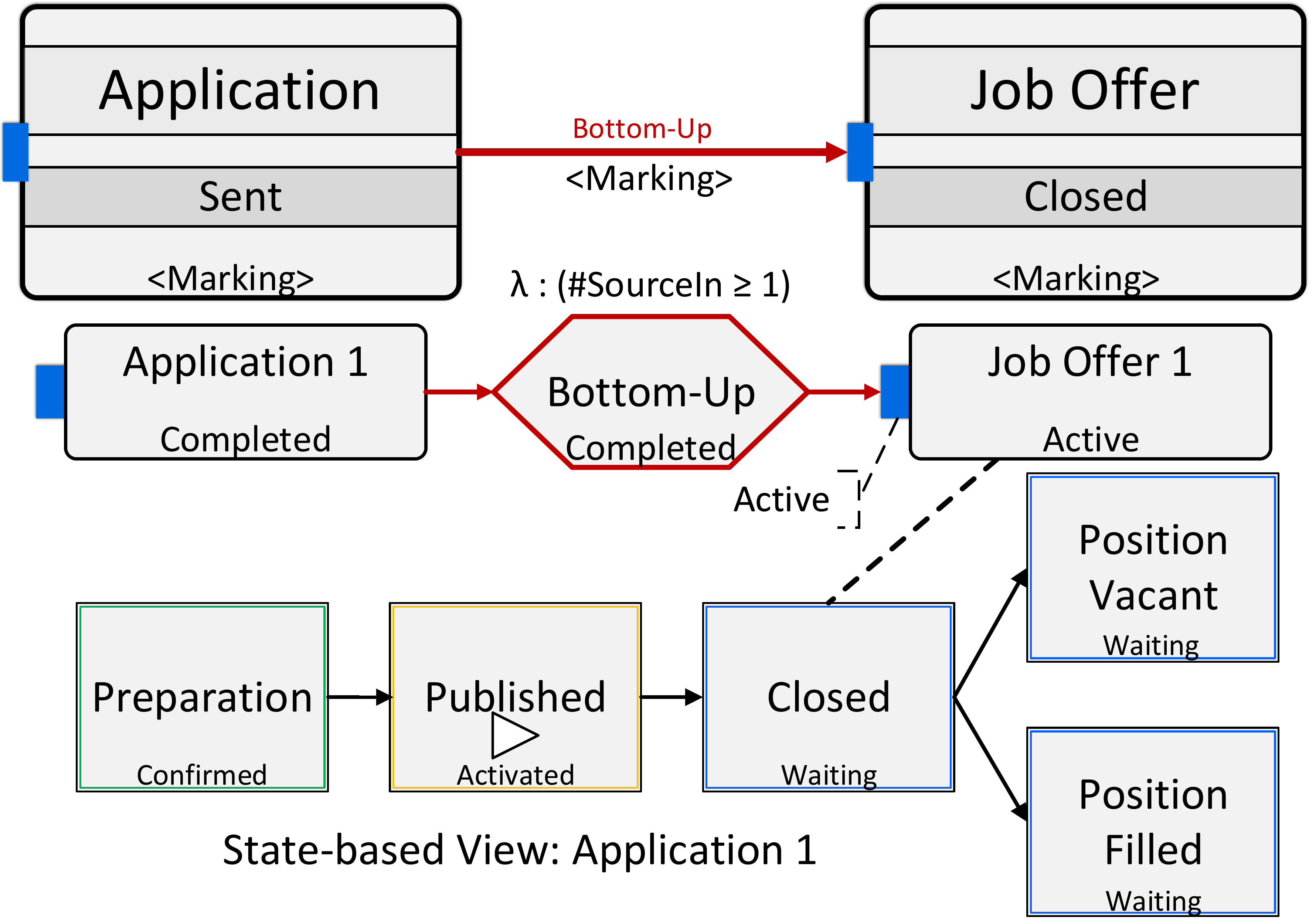}\caption{\label{fig:Markings-Example,-Stage5}Markings Example, Stage 5}
\end{figure}

What has been described in Section \ref{subsec:Introducing-Coordination-Compone}
can be considered the linear execution of a coordination process.
The coordination process prescribes the coordination step instances
to be activated successively, i.e., from start to end of the coordination
process. In practice however, the involved process instances usually
do not follow the prescribed path of a coordination process neatly.
As previously mentioned, processes are executed concurrently and asynchronously,
which, for example, results in a plethora of ways a corresponding
semantic relationship may come into fulfillment. Moreover, processes
may be created and deleted at any point in time, which potentially
has profound impact on a coordination process as well. In a nutshell,
coordination processes \emph{require the capabilities to correctly
and flexibly react to a wide variety of situations}. The necessary
measures taken to make coordination processes flexible are described
in the next sections.

\subsection{\label{subsec:The-Update-Marking} Process Rules}

The change in markings of different entities of a coordination process
is governed by the application of \emph{process rules}. Process rules
constitute a specific variant of Event-Condition-Action (ECA) rules
tailored to coordination processes. In essence, a process rule, i.e.,
its evaluation, is triggered upon the change of a marking of an entity
(event). The process rule then checks markings of the triggering entity
(e.g., a coordination step instance) or other entities (e.g., ports)
(condition), and assigns new markings to the triggering entity or
other entities (action). If conditions are true and actions are performed,
the process rule is said to \emph{apply. }Formally, a process rule
as in Definition \ref{def:process_rule}.
\begin{definition}[Process Rule]
\label{def:process_rule} A process rule $r_{e^{T}}$ has the form
$(n,e^{T},Q,F$) where
\begin{itemize}
\item $n$ is the name of the process rule
\item $e^{T}$ refers to a type, denoted as the context type
\item $Q$ is a set of preconditions $q$
\item F is a set of effects $f$
\end{itemize}
\end{definition}
\begin{figure*}
\includegraphics[width=1\textwidth]{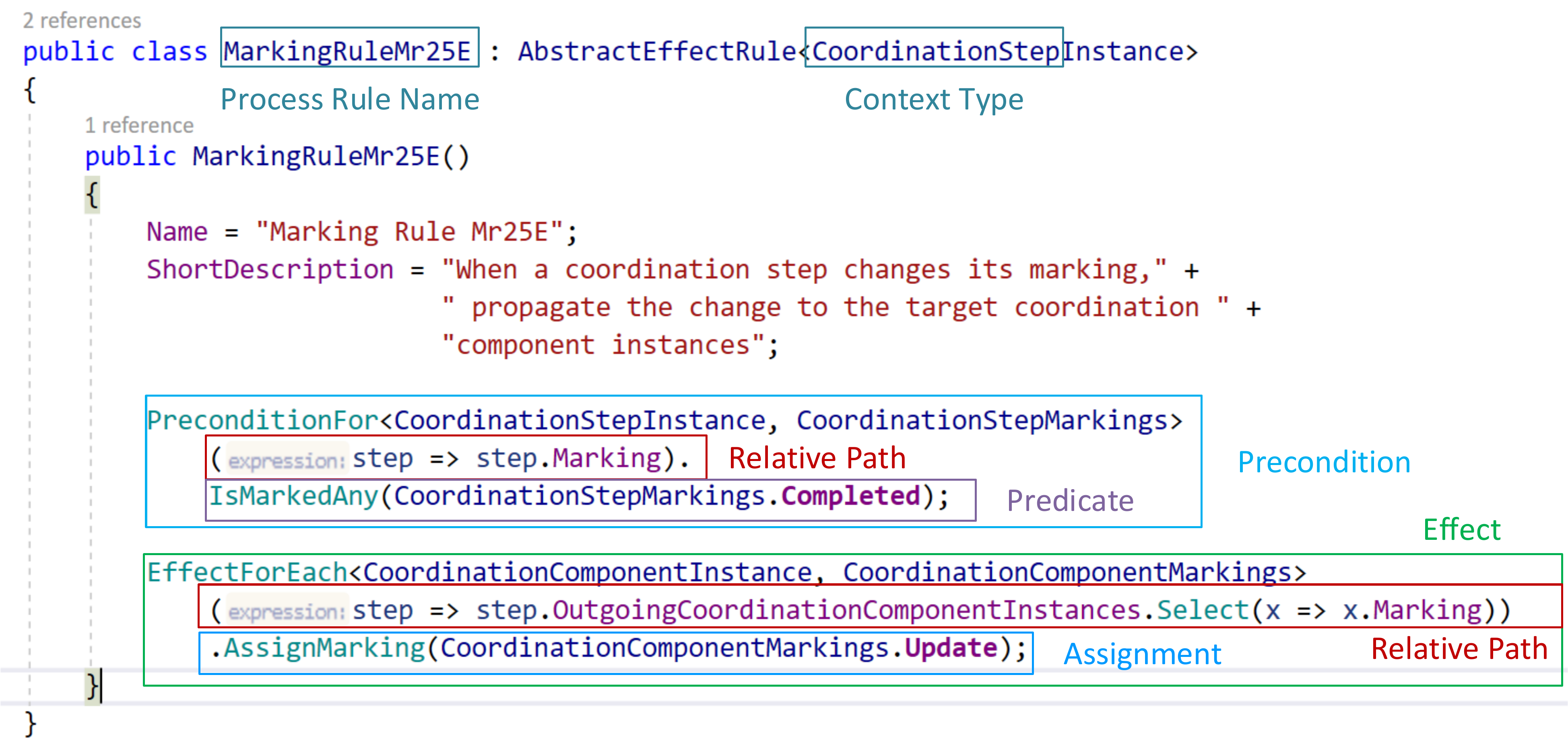}\caption{\label{fig:Process-Rule-as}Process Rule as used in the PHILharmonicFlows
implementation}
\end{figure*}

Figure \ref{fig:Process-Rule-as} shows a process rule as implemented
in PHILharmonicFlows. Process rules are defined in context of a specific
entity type $e^{T}$, called the context type. Only events that are
raised by instances of this specific type may trigger this process
rule. This represents an implicit precondition for triggering this
process rule, always in addition to any further (explicit) preconditions
$Q$. An application of a process rule is always evaluated for a specific
instance of type $e^{T}$, the \emph{context instance} $e^{I}$. An
example of this behavior is given in Example \ref{exa:process_rule_type_constraint}.
\noindent\begin{minipage}[t]{1\columnwidth}%
\begin{shaded}%
\begin{example}[Process Rule Context Type]
\label{exa:process_rule_type_constraint}A process rule $A$ has
\emph{coordination step type} defined as its context type. Consequently,
only an event raised by an instance of this type, i.e., a \emph{coordination
step instance,} may trigger this process rule. For evaluating whether
process rule A can be applied, the coordination step instance is the
context instance. Consequently, instances of other types, e.g., port
instances and coordination transition instances, need not be checked
for whether process rule $A$ may be triggered.\vspace{-1em}
\end{example}
\end{shaded}%
\end{minipage}

This implicit precondition induced by the context type has two advantages:
First, it greatly reduces the search space for possible rule application.
Only process rules with one specific context type $e^{T}$ must be
checked for further preconditions $Q$, as opposed to all process
rules. So applicable process rules may be found faster. Second, the
specification of preconditions and effects can be simplified. In addition
to the context type, process rules consists of a set of \emph{preconditions}
and a set of \emph{effects. }Preconditions correspond to the \emph{condition}
of ECA-Rules and effects correspond to the \emph{action}. Condition
and action have been renamed to precondition and effect to avoid ambiguity
with other concepts named condition and action in context of coordination
processes. Preconditions are defined in Definition \ref{def:precondition}.
\begin{definition}[Precondition]
\label{def:precondition}A precondition $q$ has the form $(\mathit{path},\mathit{predicate)}$
where t
\begin{itemize}
\item $\mathit{path}$ refers to a set of entity instances, relative to
context instance$e^{I}$
\item $\mathit{predicate}$ tests a condition on the entity referred through
$\mathit{path}$
\end{itemize}
\end{definition}
The expression $\mathit{predicate}$ returns a boolean value upon
evaluation of the precondition and also represents the return value
of the precondition as a whole. A process rule is applied when all
preconditions $Q$ of a process rule $r_{e^{T}}$ evaluate to true,
i.e., $\bigwedge r_{e^{T}}.Q=\mathit{true}$. If all preconditions
are true, the process rule is said to \emph{match}. Simple matching
implies that no effects have yet been activated.

The evaluation of the $\mathit{predicate}$ is performed on each entity
obtained through $path$. The function $\mathit{path}$ maps from
the context instance $e^{I}$ to any set of entities belong to the
coordination process graph $c^{I}$. $\mathit{Paths}$ are defined
in Definition \ref{def:process_rule_path}.
\begin{definition}[Path]
\label{def:process_rule_path} A $\mathit{path}$ is a sequence of
entities $e$ composed with the member access operator ($.$) or the
collection access operators $([]$), starting from a base entity $e$.
The collection access operator $A[p]$ applies each partial path $p$
on the each element $a$ of set $A$. A partial path has no base entity
defined and can therefore not be fully evaluated unless a base entity
is provided. Paths are evaluated from left to right, with the base
entity at the left.
\end{definition}
$\mathit{Paths}$ allow navigating the coordination process graph
and are essential for defining preconditions and effects. Consider
again Example \ref{exa:Completing_coordination_step}, where a coordination
step instance $\beta^{I}$, marked as $\mathit{Completed}$, leads
to the outgoing coordination step instances $S_{out}^{I}$ becoming
marked as $\mathit{Active}$. The elicitation of a process rule that
accomplishes this is presented in Example \ref{exa:process_rule_elicitation_1}.

\noindent\begin{minipage}[t]{1\columnwidth}%
\begin{shaded}%
\begin{example}[Process Rule Elicitation I]
\label{exa:process_rule_elicitation_1}Upon marking a specific coordination
step instance $\beta^{I}$ as $\mathit{Completed}$, it is possible
for the associated coordination component instances $S_{out}^{I}$
to become marked as $\mathit{Active}$. The context entity type $e^{T}$
is a coordination step type $\beta^{T}$. The event is a change in
the marking of the context instance $\beta^{I}.\mu_{\beta}$. The
first precondition consists of checking whether the coordination step's
new marking is $\mathit{Completed}$.
\end{example}
\begin{itemize}
\item $\beta^{I}.\mu_{\beta}$$\:==\:$$\mathit{Completed}$
\end{itemize}
The $\mathit{path}$ is $\beta^{I}.\mu_{\beta}$, which navigates
from the coordination step instance to its marking. $==$$\mathit{Completed}$
represents the predicate and compares the value of $\mu_{\beta}$
and marking literal $\mathit{Completed}$ for equality. For demonstration
purposes and additional safety, it can be checked whether the outgoing
coordination component instances $S_{out}^{I}$ are marked as either
$\mathit{Active}$ or $\mathit{Completed}$. If the outgoing coordination
component instances $S_{out}^{I}$ are already marked as $\mathit{Active}$,
applying the (not yet specified) effects $F$ of the process rule
is redundant. Furthermore, if, for any reason, the coordination component
instances were marked as $\mathit{Completed}$, it would constitute
a violation of the intended semantics of the marking $\mathit{Completed}$.
This violation of the $\mathit{Completed}$ semantics occurs for both
coordination step instances and coordination component instances.
This would be caught by developers due to the non-application of this
process rule and subsequently rectified. The corresponding second
precondition is as follows:
\begin{itemize}
\item $\beta^{I}.S_{\mathit{out}}^{I}[.\mu_{s}]$$\:\neq\text{\ensuremath{\mathit{Active}}}\,\wedge\neq\mathit{Completed}$
\end{itemize}
The path $\beta^{I}.S_{\mathit{out}}^{I}[.\mu_{s}]$ contains the
collection access operator with the partial path $.\mu_{s}$. The
operator obtains the marking $\mu_{s}$ from each coordination component
$s$ in $\beta^{I}.S_{\mathit{out}}^{I}$, resulting in a collection
of markings. Note that this is not a set, as duplicate markings shall
be retained. The predicate $\neq\text{\ensuremath{\mathit{Active}}}\,\wedge\neq\mathit{Completed}$
is then evaluated on each marking in the collection, obtaining a collection
of Boolean \emph{true} or \emph{false} values. Finally, the collection
of Booleans is aggregated into a single Boolean value using AND, representing
the result of the precondition.\end{shaded}%
\end{minipage}

$\mathit{Paths}$ are defined relative to a base entity for easier
specification by a developer and performant evaluation at run-time.
Moreover, the simple specification of $\mathit{paths}$ are the primary
driver behind the many mutual references of the definitions of the
constituting elements of a coordination process. Simply getting from
a coordination step to the next coordination step through the coordination
components and their target ports is both performant in the evaluation
of the process rule and simple to specify by a developer. This is
one contributing factor to Challenge 5 \emph{Manageable Complexity}.

For efficiency reasons, the preconditions are evaluated sequentially
using \emph{short-circuit evaluation}. That means, if for any reason
a precondition evaluates to $\mathit{false}$, process rule evaluation
is aborted and the process rule is not applied. This is possible as
all preconditions must evaluate to $\mathit{true}$ for the process
rule to apply.

If all preconditions of process rule are true, i.e., the process rule
matches, all effects defined for the process rule are applied. Effects
can be specified for any entity reachable by a $\mathit{path}$ from
the context instance $e^{I}$. Effects are defined in Definition \ref{def:effect}.
\begin{definition}[Effect]
\label{def:effect}An effect $f$ has the form $(\mathit{path}$,\\
$assignment$) where
\begin{itemize}
\item $\mathit{path}$ refers to a set of instances, relative to context
instance$e^{I}$
\item $\mathit{assignment}$ assigns a value to the entity referred through
$\mathit{path}$
\end{itemize}
The $assignment$ of an effect usually consists of assigning a specific
marking to an entity. There are, however, other assignments that are
used rarely. These rare assignments are omitted here for the sake
of brevity. Analogous to preconditions, the entity or entities to
which the assignment is made is determined by a $\mathit{path}$.
Example \ref{exa:process_rule_elicitation_2} continues the process
rule elicitation by defining effects (cf. Example \ref{exa:process_rule_elicitation_1}).
\end{definition}
\noindent\begin{minipage}[t]{1\columnwidth}%
\begin{shaded}%
\begin{example}[Process Rule Elicitation II]
\label{exa:process_rule_elicitation_2}Continuing the process rule
specification of Example \ref{exa:process_rule_elicitation_1}, the
process rule shall mark outgoing coordination components as $\mathit{Active}$
when a coordination step instance becomes marked as $\mathit{Completed}$.
So far, the context instance and preconditions have been defined:
\end{example}
\begin{itemize}
\item Context type $e^{T}$ is a coordination step type $\beta^{T}$.
\item Preconditions $Q$ are 
\begin{itemize}
\item $\beta^{I}.\mu_{\beta}$$\:==\:$$\mathit{Completed}$
\item $\beta^{I}.S_{\mathit{out}}^{I}[.\mu_{s}]$$\:\neq\text{\ensuremath{\mathit{Active}}}\,\wedge\neq\mathit{Completed}$
\end{itemize}
\end{itemize}
When both preconditions are true for a given context instance $\beta^{I}$,
the following effect is defined:
\begin{itemize}
\item Effects $F$ are
\begin{itemize}
\item $\beta^{I}.S_{\mathit{out}}^{I}[.\mu_{s}]$$\:\coloneqq\text{\ensuremath{\mathit{Active}}}$
\end{itemize}
\end{itemize}
The effect assigns marking $\mathit{Active}$ to all outgoing coordination
component instances $s^{I}$ in $S_{\mathit{out}}^{I}$. The path
$\beta^{I}.S_{\mathit{out}}^{I}[.\mu_{s}]$ obtains the marking for
each outgoing coordination component instance. The expression $\:\coloneqq\text{\ensuremath{\mathit{Active}}}$
represents the assignment, with $\coloneqq$ as the assignment operator
and $\mathit{Active}$ as a marking literal.\end{shaded}%
\end{minipage}

\subsection{Events and Snapshots}

The evaluation of process rules is initiated by events. Events are
either \emph{external} or \emph{internal} in respect to a coordination
process instance $c^{I}$. An external event is a state change of
a coordinated process instance or the creation of a new relation between
a process instance and the coordinating process instance of the coordination
process. Internal events mostly cover events raised by the change
of a marking. Formally, an event $\epsilon$ is described in Definition
\ref{def:event}.
\begin{definition}[Event]
\label{def:event} An event $\epsilon$ has the form $(t_{\epsilon},g_{\epsilon},e)$
where
\begin{itemize}
\item $t_{\epsilon}$ is the type of the event
\item $g_{\epsilon}$ is the origin of the event, $g_{\epsilon}\in\mathit{\{ext,int\}}$
\item $e$ is the entity that raised $\epsilon$
\end{itemize}
The set of all events is $E$
\end{definition}
The type $t_{\epsilon}$ includes types for signifying marking changes,
state changes, or relation creations. State changes and relation creations
are the main external events that influence a coordination process
instance. For the sake of brevity, the full list of event types is
omitted here. The origin $g_{\epsilon}$ denotes whether the event
is external ($\mathit{ext}$) or internal ($\mathit{int}$). Some
event types may either be external or internal. Lastly, the entity
$e$ that raised the event is retained. This ties into the context
type and context instance of the process rules, as it can be easily
determined whether a specific process rule needs to be evaluated for
this event.

In order to capture the interplay of coordination process instances,
events, and process rules, the concept of \emph{snapshot} is needed.
A snapshot of a coordination process instance $c^{I}$ captures all
entities, markings, properties and references between entities at
a specific point in the execution of a coordination process instance.
Formally, a snapshot $\chi$ is described in Definition \ref{defsnapshot}.
\begin{definition}[Snapshot]
\label{defsnapshot} A snapshot $\chi$ has the form $(c^{I},t)$
where
\begin{itemize}
\item $c^{I}$ is a fixed copy of a coordination process instance 
\item $t$ is a logical timestamp
\end{itemize}
The tuple $(\chi,E)$ denotes a snapshot and a set of new events $E$
\end{definition}
With process rules, events, and snapshots defined, the interplay between
these concepts can be described. The general principle is as follows:
The occurrence of an event $\epsilon$ associated with a snapshot
$\chi$ triggers the evaluation of process rules. If a rule matches,
it is applied and a new snapshot $\chi'$ and possibly new events
$\epsilon_{1}..\epsilon_{n}$ are created. 

\begin{figure}

\includegraphics[width=1\columnwidth]{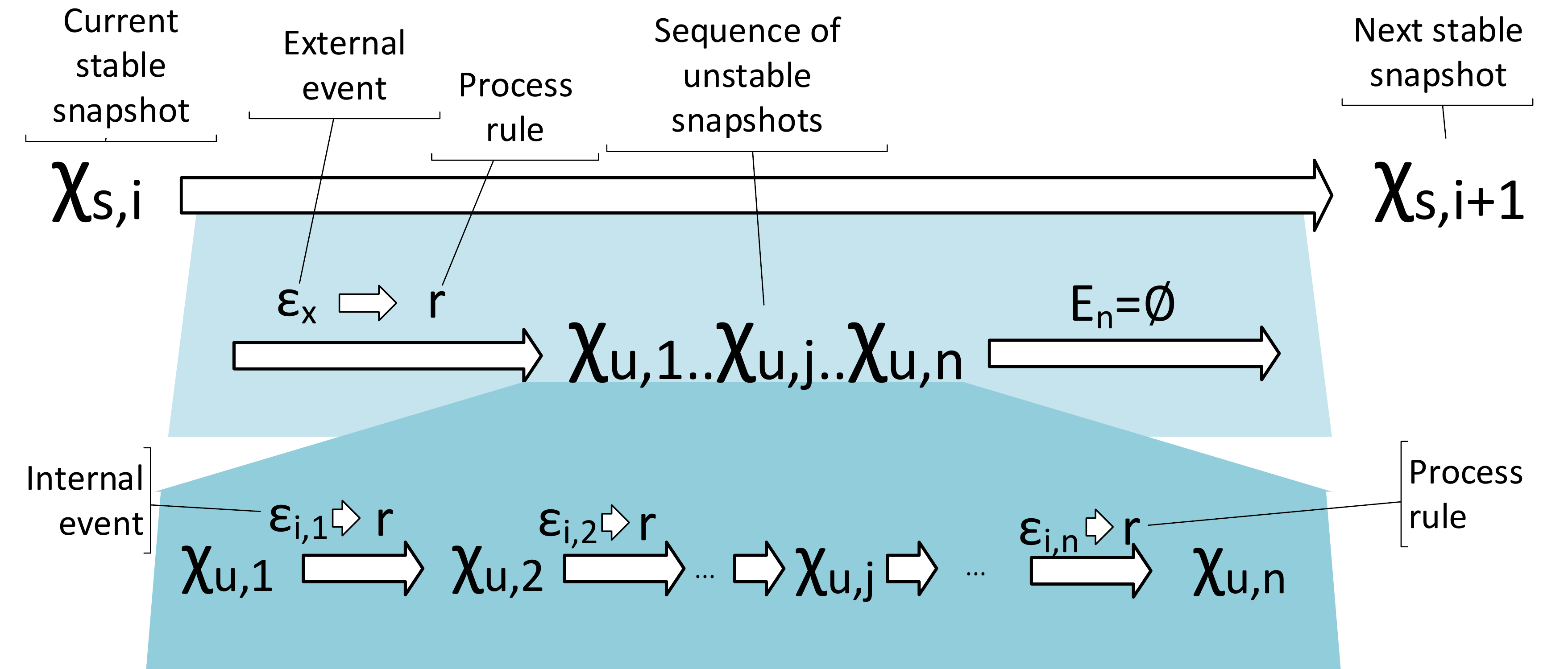}\caption{\label{fig:Process-Rule-Application}Process Rule Application and
Snapshots}

\end{figure}

Due to the presence of events in $E,$snapshots $\chi$can be distinguished
into \emph{stable snapshots }(denoted $\chi_{s}$) and \emph{unstable
snapshots} (denoted as $\chi_{u}$). In an unstable snapshot $\chi_{u}$,
the set of events $E$ is non-empty and populated with internal events.
Unstable snapshot require further process rule applications. In a
stable snapshot $\chi_{s}$, the set of event $E$ is empty and no
more process rules may be applied. The following sequence describes
how snapshots, events, and process rules interact to execute a coordination
process instance $c^{I}$.
\begin{enumerate}
\item Execution starts with a stable snapshot $\chi_{s,i}$ of $c^{I}$
and an empty set of events $E_{i}=\emptyset$. Unless a new external
event occurs, $(x_{s,i},E_{i})$ remains unchanged and no execution
takes place
\item A new external event $\epsilon_{x}$ occurs.
\begin{enumerate}
\item A process rule reacts to $\epsilon_{x}$ , takes the type $t_{\epsilon}$
and the snapshot $\chi_{s,i}$ and assigns new markings $\mu$ to
corresponding entities. 
\item Assigning new markings creates a new snapshot $\chi_{j}$ of the coordination
process instance $c^{I}$.
\item The marking changes raise new internal events, which are added to
$E_{i}$, creating $E_{j}$. Due to $E_{i+1}$ being non-empty, $\chi_{j}$
is unstable, i.e., $\chi_{u,j}$
\end{enumerate}
\item The current snapshot ($\chi_{u,j},E_{j}$) may be unstable. More process
rules may be applied
\begin{enumerate}
\item If $E_{j}=\emptyset$ continue at (4). If $E_{j}\neq\emptyset$ ,
continue at (3b) 
\item An event is removed from $E_{j}$. The event triggers a process rule
evaluation. 
\item If a process rule matches, this process rule is applied. This assigns
new markings $\mu$ to entities of $c^{I}$ , thus creating a new
snapshot $\chi_{u,j+1}$. The marking changes also create new internal
events $\epsilon_{1},\epsilon_{2},..$ with $E_{j+1}=E_{j}\cup\{\epsilon_{1},\epsilon_{2},..\}$
\item Continue at (3) with ($\chi_{u,j},E_{j}$)= ($\chi_{u,j+1},E_{j+1}$)
\end{enumerate}
\item A new stable snapshot $(\chi_{s,i+1},E)$= ($\chi_{u,j},E_{j}$) is
created as $E_{j}=\emptyset$. 
\end{enumerate}
The processing moves a coordination process instance $c^{I}$ from
one stable snapshot to the next stable snapshot. In between these
stable snapshots, usually many unstable snapshots occur. When events
no longer lead to the successful application of process rules and
consequently no more internal events are raised, the set of events
$E$ becomes empty and a new stable snapshot $\chi_{s,i+1}$ emerges.
Figure \ref{fig:Process-Rule-Application} exemplifies this graphically. 

In other words, the sequential assignment of markings leads to a \emph{process
rule cascade}, i.e., the repeated triggering of process rules by other
process rules. This cascade passes through a coordination process
upon an initial change and assigns new markings in its wake. Sections
\ref{subsec:Introducing-Coordination-Step} and \ref{subsec:Introducing-Coordination-Compone}
describe one process rule cascade (cf. Figures \ref{fig:Markings-Example,-Stage1}-\ref{fig:Markings-Example,-Stage5}).
Process rules and process rule cascades drive the enactment of a coordination
process.

The cascade stops once a stable state, i.e., snapshot, of the coordination
process has been achieved, i.e., no more marking changes are caused
by the cascade. Marking change events are considered internal events
of the coordination process. Another application of process rules
therefore must be triggered from outside the coordination process
with a new (external) event. The prevalent external events that affect
a coordination process are the state change of a coordinated process
or the addition or deletion of a process in the relational process
structure that is within the scope of the coordination process. Consequently,
changing the state of a coordinated process instance triggers further
process rule applications.

\subsection{The ``Update'' Marking}

From the exemplary coordination process execution described in Sections
\ref{subsec:Introducing-Coordination-Step} and \ref{subsec:Introducing-Coordination-Compone},
it may be concluded that markings of, for example, a coordination
component, directly influence the marking of its target port. In other
words, it is possible to express the dependencies of markings as rules
in the form of ``If entity X has marking A, entity Y must be marked
as B'', where A and B are placeholders for markings and X and Y are
entities of the coordination process. For example, when the port of
coordination step $\mathit{Job\:Offer\:1}$ is marked as $\mathit{Active}$,
a process rule is triggered which marks the coordination step $\mathit{Job\:Offer\:1}$
as $\mathit{Active}$ as well. This is denoted as determining markings
\emph{from the outside}.

The assumption that applying markings from the outside might work
is grounded in the structure of the exemplary coordination process
execution itself. \emph{One} coordination step $\mathit{Application\:1}$
is connected to exactly \emph{one} bottom-up coordination component,
which, in turn, is connected to \emph{one} port and \emph{one} coordination
step corresponding to $\mathit{Job\:Offer\:1}$. In short, the coordination
process graph is simple and linear. In general however, the structure
of the coordination process graph is not always linear. The complex
relations of the coordinated processes are mapped to entities of the
coordination process, which mirror the same complex relationships.
For example, a coordination step may have multiple attached ports
(one-to-many relationship), and the ports may have multiple coordination
components. The coordination components may be, in addition, of different
type. On the other side, coordination steps may have multiple outgoing
coordination component. The resulting coordination process graph is
not linear, but branched out (cf. Figure \ref{fig:Partial-container-and-instance-view}).

Generally, it is impossible for a port to correctly prescribe the
marking of its attached coordination step, i.e, apply a marking from
the outside. In order for the port to prescribe the correct marking
for the coordination step, the port requires at least knowledge of
all its sibling ports and their markings that are attached to the
particular coordination step. A sibling port is a port that is attached
to the same coordination step. Only by knowing at least the markings
of the sibling ports, a port is able to determine the correct marking
for the coordination step.

\begin{figure}
\centering{}\includegraphics[width=1\columnwidth]{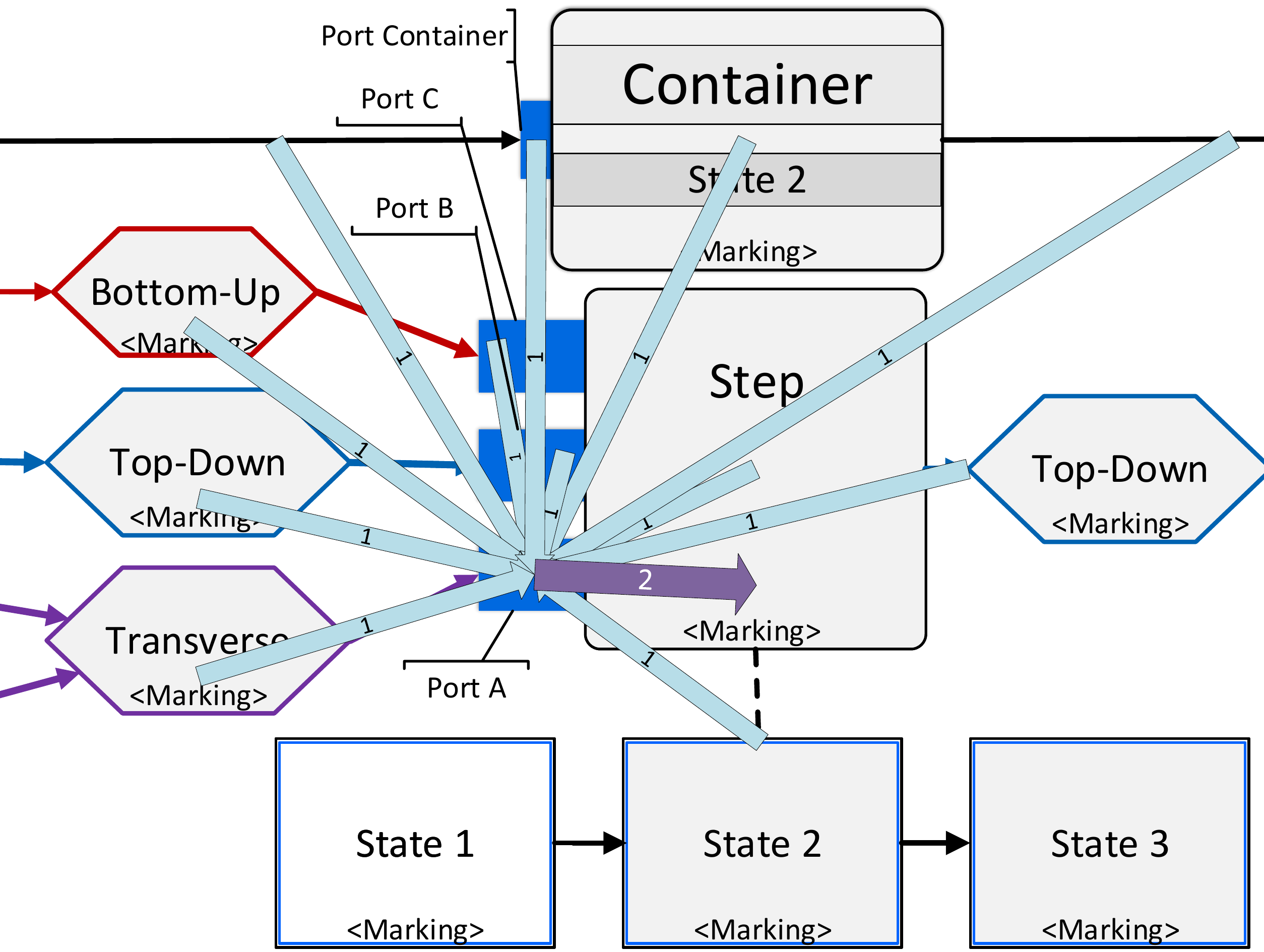}\caption{\label{fig:Marking-Application-from-outside}Marking Application \emph{from
the outside}}
\end{figure}

However, the marking of a coordination step is not only determined
by its ports. For dealing with more complex and advanced scenarios
(some are described in Section \ref{subsec:Flexibility}), the port
may require information from potentially all the surrounding entities
of the attached coordination step. In other words, the marking of
a coordination step may be influenced by the markings of the following
entities:
\begin{enumerate}
\item the attached ports
\item the\emph{ incoming} coordination components belonging to the attached
ports
\item the coordination step container
\item the port instance containers belonging to the coordination step container
\item the incoming coordination transitions belonging to the port instance
containers
\item the coordination process
\item the state of the corresponding process instance
\item the \emph{outgoing} coordination components
\item the outgoing coordination transitions belonging to the coordination
step container
\end{enumerate}
For the other entities of a coordination process (cf. Definitions
\ref{Def:CoordinationStepInstanceContainer}-\ref{def:A-self-transverse-coordination}),
a similar list of influencing entities exists. This influence is not
simply restricted to a coordination step instance. In Figure \ref{fig:Marking-Application-from-outside},
it is exemplarily shown for port $A$ which of the neighboring entities
may be relevant for determining the current marking of the step instance.
First, the marking of port $A$ is changed and the port must set the
new marking of the attached coordination step. Consequently, port
$A$ gathers the necessary information from the neighboring entities
of the coordination step, i.e., port $A$ looks at their markings
(cf. arrows labeled ``1'' in Figure \ref{fig:Marking-Application-from-outside}).
From the gathered information, the resulting marking of the coordination
step is determined and assigned directly to the coordination step
by port $A$ (cf. arrow labeled ``2'' in Figure \ref{fig:Marking-Application-from-outside}).
Once the marking of the coordination step changes, the same procedure
is triggered on the coordination step, assigning markings to, for
example, outgoing coordination components.

Moreover, a marking change of an entity, e.g., port $A$ in Figure
\ref{fig:Marking-Application-from-outside}, may affect the markings
of multiple other entities. In case the port determines the new marking
of each of the affected entities, the assignment of markings is highly
inefficient, as the port may become a bottleneck. Because of the design
of markings and process rules, the marking of a particular entity
is fully determined by the markings of its neighboring entities. This
applies to any entity of a coordination process, creating a mutual
interplay between entities due to changes of their markings.

As the marking of an entity is fully described by the markings of
its neighboring entities, there is a better possibility to assign
markings: This is designated as marking assignment \emph{from the
inside}, as opposed to the previous marking assignment from the outside.
Here, the assignment of a new marking for an entity is split into
two rule applications: 
\begin{enumerate}
\item When an entity's marking changes, a \emph{notification rule} is tasked
with notifying neighboring entities of the change. Returning to the
example from Figure \ref{fig:Marking-Application-from-outside}, when
the marking of port $A$ changes, a notification rule informs the
attached coordination step that its marking must be reevaluated due
to the marking change of port $A$. This notification consists of
marking the coordination step as $\mathit{Update}$ (cf. Table \ref{tab:Coordination-Step-Markings}). 
\item The marking $\mathit{Update}$ is temporary and is only used to trigger
an \emph{update rule} that overwrites the $\mathit{Update}$ marking
with the new correct marking for the coordination step. The update
rules take over the task to gather the information from the neighboring
entities for assigning the correct marking to the coordination step.
\end{enumerate}
Essentially, the coordination step assigns its own marking via an
update rule, hence it is called marking assignment \emph{from the
inside}. Once the correct marking for the coordination step has been
determined, notification rules are triggered by the marking change
to inform other entities of the change. Consequently, a rule cascade
may be triggered until no more marking changes occur.

\begin{figure}
\centering{}\includegraphics[width=1\columnwidth]{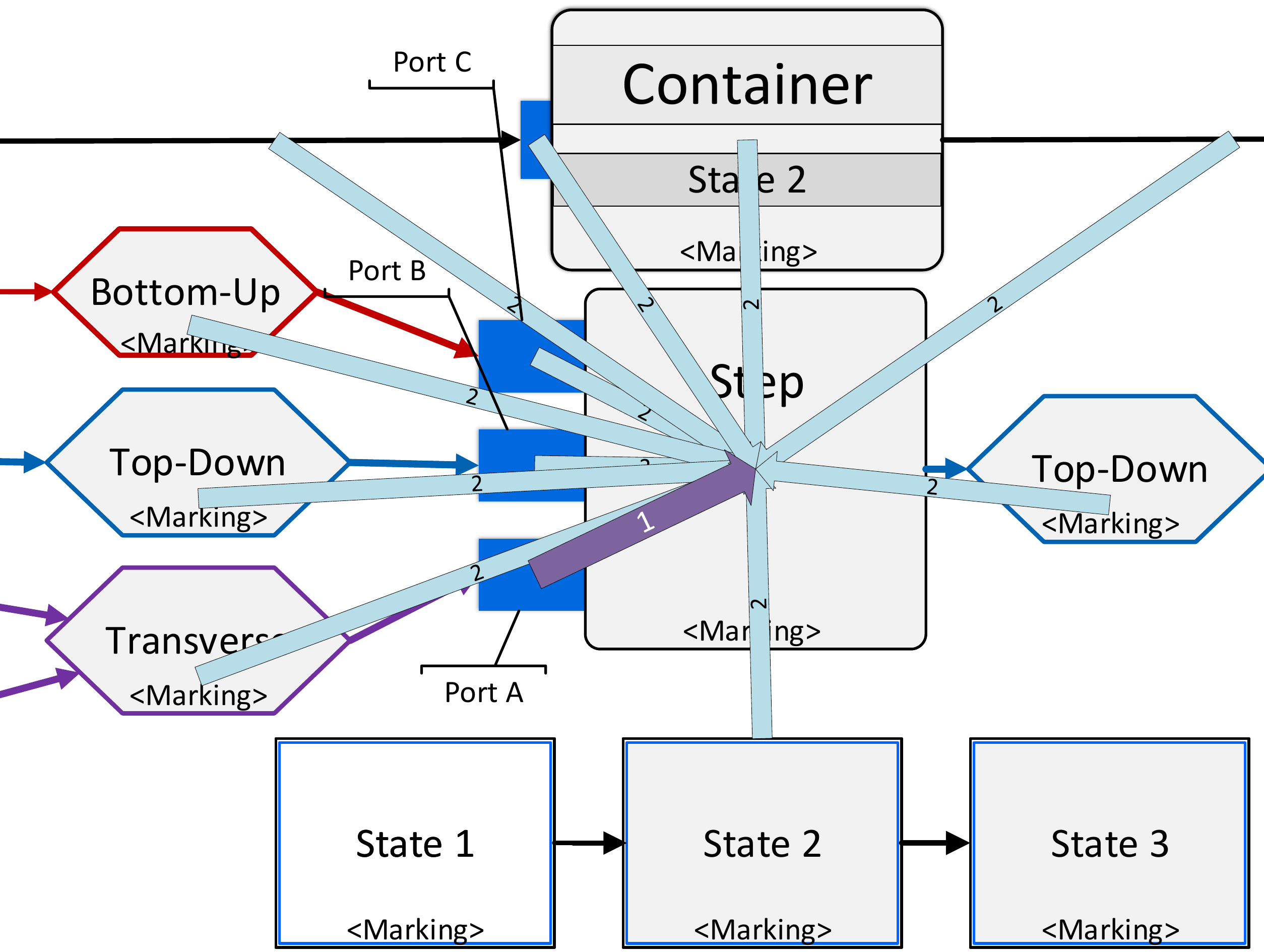}\caption{\label{fig:Marking-Application-from-inside}Marking Application \emph{from
the inside}}
\end{figure}

Considering the marking change of port $A$ in Figure \ref{fig:Marking-Application-from-outside},
Figure \ref{fig:Marking-Application-from-inside} shows the same situation
with marking assignment from the inside. As soon as the marking change
of port $A$ occurs, a notification rule marks the corresponding coordination
step as $\mathit{Update}$ (see the arrow labeled ``1'' in Figure
\ref{fig:Marking-Application-from-inside} ). The coordination step
itself now employs update rules to determine its new marking (see
the arrows labeled ``2'' in Figure \ref{fig:Marking-Application-from-inside}
).

The division into notification rules and update rules brings notable
benefits to coordination processes. In no particular order, the benefits
include:
\begin{itemize}
\item Process rules become much simpler, as each exactly fulfills one task.
Designing the rules that govern process execution becomes easier,
and unwanted side effects can be minimized.
\item The execution of a coordination process becomes more robust. Overlap
between rules or mistakes in process rules that cause interruptions
in rule cascades can be easily detected and rectified, resulting in
fewer errors.
\item Process rules overall become better maintainable. It is immediately
traceable which process rule assigned a specific marking, enabling
faster debugging in the implementation.
\item Feature extensions for coordination processes, which affect the process
rules, are easier to implement. Additional process rules can easily
be integrated into the existing process rule set; additional markings
can be introduced without necessarily breaking existing process rules.
\item The notification and update rule model allows realizing a flexible
execution of coordination processes more easily. When an update is
triggered by setting the marking of an entity to $\mathit{Update}$,
the resulting marking is determined by the markings of the neighboring
entities. With update rules, a specific rule can be specified for
any contingency, i.e., any combination of markings in the neighboring
entities.
\end{itemize}
The capability of entities in a coordination process to react to any
contingency is one of the greatest strengths of the concept. The coordination
of multiple processes running concurrently requires a high degree
of flexibility on part of the coordination process. The basic flexibility
is delivered by markings, process rules and the overall operational
semantics. Section \ref{subsec:Flexibility} gives a rough overview
over advanced flexibility requirements in regard to coordination processes.

\subsection{\label{subsec:Flexibility}Flexibility}

The coordination of a myriad of concurrently running process instances
is not straightforward and linear. Over the course of their execution,
several situations may arise that affect their coordination in profound
and non-standard ways. Careful analysis of these situations permitted
to incorporate concepts and adaptations into coordination processes.
These adaptations allow coordination processes to deal with these
situations in a flexible manner. The following unsorted list briefly
describes advanced features of coordination processes for which flexibility
is paramount. Note that, for the sake of brevity, this list is not
intended as a full account of all advanced flexibility features comprised
in a coordination process.

\subsubsection*{Coordination Process Graph Alterations}

In classical processes (e.g., a BPMN process), the process graph of
a process instance remains largely unaltered during its execution.
By contrast, a coordination process graph is designed to change due
to the addition of new coordination steps, ports, or coordination
components. When adding a new entity to the process graph, the existing
entities may have to change their markings accordingly. This is taken
into account by the process rules governing the execution of a coordination
process.

\subsubsection*{Arrangement Linking}

In this paper, it has been assumed that only a single process instance
with no pre-existing relations may be added to a relational process
structure at a point in time. However, it may occur that a process
instance that becomes linked to a process structure already has pre-existing
relations. This is denoted as an \emph{arrangement}, i.e., a specific
configuration of process instances and relations (cf. Section \ref{sec:Enacting-Coordination-Processes}).

\noindent\begin{minipage}[t]{1\columnwidth}%
\begin{shaded}%
\begin{example}[Arrangement Linking]
An $\mathit{Application}$ has been reviewed, i.e., it is related
to several \emph{completed} $\mathit{Review}$ processes, but the
$\mathit{Reviews}$ show that another $\mathit{Job\,Offer}$ than
the current one is a better fit for the applicant. Therefore, the
arrangement of the $\mathit{Application}$ and its $\mathit{Reviews}$
is moved to another $\mathit{Job\,Offer}$ by creating a new relation
between the $\mathit{Application}$ and the other $\mathit{Job\,Offer}$.
The $\mathit{Review}s$ are now related as well to the $\mathit{Job\,Offer}$
due to transitive relations.\vspace*{-0.2cm}
\end{example}
\end{shaded}%
\end{minipage} The challenge for the coordination process here is that, first, multiple
process instances are linked at the same time, and second, that some
of these process instances have been already executed to some degree.
The operational semantics of a coordination process takes both challenges
into account and can handle them correctly. This can work even when
the newly linked arrangement violates some of the constraints of the
coordination process. In this case, the coordination process assumes
that the execution status of the linked processes is legit and adapts
accordingly. This might result in seemingly paradoxical situations,
e.g., a state might be marked as $\mathit{Activated}$, but all the
ports and coordination steps associated with the state are marked
as $\mathit{Eliminated}$. However, coordination processes still handle
these situations correctly, e.g., preventing deadlocks. Arrangement
linking is a unique feature of coordination processes and relational
processes and not found in other approaches.

\subsubsection*{Dead Path Elimination}

When a coordination constraint, i.e., a coordination component, can
no longer be fulfilled, the coordination component is marked as $\mathit{Eliminated}$.
Consequently, successor coordination components and coordination steps,
in general, cannot be marked as $\mathit{Active}$ anymore and must
be marked as $\mathit{Eliminated}$ as well. Therefore, starting from
the eliminated coordination component, the coordination process performs
a \emph{dead path elimination.} A process rule cascade is triggered,
marking entities as $\mathit{Eliminated}$ where it can be ensured
that they are not able to be activated in the future. For entities
for which it is still possible that they can be marked as $\mathit{Active}$,
e.g., a coordination step having several ports, the dead path elimination
stops.

\subsubsection*{Reverse Dead Path Elimination}

Due to the dynamics of process execution and the emergence and deletion
of process instances, the elimination of entities in a coordination
process is not entirely final. When circumstances that led to a dead
path elimination change, it becomes possible that coordination constraints
can be fulfilled again. Obviously, this makes the dead path elimination
obsolete. Consequently, the dead path elimination must be reversed
by marking the affected entities as $\mathit{Inactive}$ instead of
$\mathit{Eliminated}$. Again, a reverse dead path elimination is
realized as a process rule cascade in the coordination process.

\subsubsection*{Backwards Transitions}

Generally, the coordinated processes move forward, i.e., states become
successively activated from start to end of the process. Therefore,
a coordination process has a natural progression from start to end
as well, that is in line with the progression of the coordinated processes.
However, it is possible that a coordinated process regresses in its
execution state, i.e., a predecessor state to the currently active
state is activated at a later point in time. 

While state-based views do not allow looping, they allow for \emph{backwards
transitions}, i.e., transitions can go opposite the normal direction
for transitions from start to end of a process. Backwards transitions
allow activating a previous state. In practice, this is very relevant,
as oftentimes mistakes may be made that must be rectified afterwards.
Otherwise, the mistake may persist, resulting in an overall faulty
process execution. By going back to a previous state with a backwards
transition, a user is able to correct a mistake made during the execution
of that state.

As active states are directly relevant for the activation of constraints,
activating a previous state in a coordinated process creates some
challenges for the overall consistency. 

\noindent\begin{minipage}[t]{1\columnwidth}%
\begin{shaded}%
\begin{example}[Backward transitions]
\label{exa:Backwards-Transitions}Suppose an $\mathit{Application}$
has been accepted, i.e., respective $\mathit{Reviews}$ and $\mathit{Interviews}$
fulfill a coordination constraint that allows for acceptance. If,
for some reason, one of the $\mathit{Reviews}$ goes back to a previous
state, the coordination\vspace*{-1em}
\end{example}
\end{shaded}%
\end{minipage}

\noindent\begin{minipage}[t]{1\columnwidth}%
\begin{shaded}%
constraint is no longer fulfilled. Thus, it can be argued that the
acceptance of the $\mathit{Application}$ is no longer valid.\end{shaded}%
\end{minipage}

The simplest solution is to forbid backwards transitions. However,
this is unrealistic in practice. Therefore coordination processes
do not implement this rigid enactment strategy.

Another possible solution to this issue would be to force the $\mathit{Application}$
to go back to a previous state as well. This would enable the overall
consistency of the coordination constraints in the coordination process,
as the $\mathit{Application}$ is no longer accepted. However, forcing
coordinated processes to go back to previous states may have enormous
detrimental implications. Depending on which coordinated process goes
back to which previous state, potentially a large number of coordination
constraints may become invalidated. In turn, this may force a large
number of other coordinated processes to regress as well to keep a
consistent state. In consequence, the progress of a significant number
of process instances might be lost. Additional information loss might
incur, depending on how the regress to a previous state is handled
beneath the state-based view of the process instances. In principle,
the lost progress can be regained afterwards by re-doing the necessary
work.

Due to the possibly large ramifications, coordination processes handle
backwards transitions differently. As the coordination constraint
that allowed the $\mathit{Application}$ to be accepted was true at
one point in time, a regress of a $\mathit{Review}$ or $\mathit{Interview}$
will not force the $\mathit{Application}$ to regress as well (cf.
Figure \ref{fig:Coordination-Process-for}, coordination constraints
\raisebox{.5pt}{\textcircled{\raisebox{-.9pt} {4}}} and \raisebox{.5pt}{\textcircled{\raisebox{-.9pt} {6}}}).
Instead, the $\mathit{Application}$ keeps its current status, but
the coordination constraint becomes invalidated. In consequence, eventual
future attempts to accept the same $\mathit{Application}$ again,
after the $\mathit{Application}$ regressed to a previous state, would
fail due to the unfulfilled constraint. Note that an $\mathit{Application}$
may still be manually transitioned into a previous state if desired.
Overall, this solution is perceived to be favorable as it allows for
backwards transitions.

In general, it may occur during the execution of a coordination process
that a state of a coordinated lifecycle process is marked as $\mathit{Activated}$,
yet the coordination steps for the state are $\mathit{Eliminated}$
or $\mathit{Waiting}$. According to the coordination constraints,
the state should not be marked as $\mathit{Activated}$, which is
an inconsistency between coordination process and the coordinated
process. Example \ref{exa:Backwards-Transitions} describes only one
situation where such an inconsistency might arise.

In general regarding any inconsistencies, coordination processes take
the status of the coordinated process as truth in regard to resolving
inconsistencies, i.e., the marking $\mathit{Activated}$ for a state
of the process is valid. This is based on the fact that in order for
the state to be currently marked as $\mathit{Activated}$, the activation
must have been allowed by a coordination process at one point during
the past execution of the particular process. Consequently, subsequent
changes in regard to the fulfillment of coordination constraints after
the activation of the state have no impact on the state being activated.
However, if the coordinated process is executed further, activations
of successor states are of course subject to their respective coordination
constraints and may only activate if the coordination constraints
permit the activation.

\noindent\begin{minipage}[t]{1\columnwidth}%
\begin{shaded}%
\begin{example}[Coordination Inconsistencies]
Suppose an $\mathit{Application}$ is in state $\mathit{Checked},$for
which coordination constraints requires at least three $\mathit{Reviews}$
to provide a verdict, i.e, three $\mathit{Reviews}$ or more must
be in either state $\mathit{Reject\;Proposed}$ or $\mathit{Invite\:Proposed}$.
Currently, the $\mathit{Application}$ has three $\mathit{Reviews}$
in the required states, the coordination constraints currently allow
the activation of state $\mathit{Checked}$, which also has been activated.
Now consider the deletion of one $\mathit{Review}$, with the effect
that the coordination constraint becomes unfulfilled. The $\mathit{Application}$
remains in state $\mathit{Checked}$, as the state has been legitimately
activated. However, if the $\mathit{Application}$ attempts to transition
to a successor state of $\mathit{Checked}$ (either state $\mathit{Accepted}$
or $\mathit{Rejected})$, the coordination constraints prevent activation
and instead the state is marked as $\mathit{Pending}$ (cf. Coordination
Constraint \ref{const:coordConstraint5}) of Example \ref{exa:Rundown}).
\vspace*{-1em}
\end{example}
\end{shaded}%
\end{minipage}

The successor state to $\mathit{Checked}$ cannot be activated as
there are not enough $\mathit{Reviews}$ overall to obtain a conclusive
verdict. Note that the involvement of sufficient $\mathit{Reviews}$
in the coordination constraints of the successor states is coincidental.
In other business processes, the coordination constraints for successor
states may be unrelated to the coordination constraints of the current
state.

Coordination constraints becoming fulfilled and unfulfilled is one
aspect of the dynamics that coordination processes accomplish. The
changes over time regarding the coordination constraints create issues
regarding the consistency between coordination constraints and coordinated
processes. A suitable resolution for these inconsistencies consequently
involves the fact that coordination constraints may change over time.For
future versions of the coordination process concept, it is conceivable
to let a modeler choose the method to resolve inconsistencies, e.g.,
a modeler may choose that accepted $\mathit{Applications}$automatically
return to a previous state if the coordination constraints for $\mathit{Accepted}$
are no longer met.

\subsubsection*{}

\subsection{Conclusion}

In summary, the execution of a coordination process is based on the
markings of its various constituting entities. Each marking has a
specific significance for the overall business process execution,
e.g., particular actions are allowed or disallowed depending on the
current markings. The driver behind the execution of a coordination
process is a set of process rules that governs how markings change.
To provide the required flexibility to deal with numerous different
situations, e.g., emergence or deletion of a process instance, process
rules are divided into notification rules and update rules.

The way markings and process rules are organized, together with the
dynamic process instance representation, conceptually fulfills Challenge
1 \emph{Asynchronous Concurrency} in the sense that process execution
is impacted minimally by a coordination process. In principle, in
absence of a coordination process, the execution of any process instance
of a relational process structure is fully unrestricted. This means
any process may activate states according to its state-based view
without interference. If a coordination process exists, the activation
of a state may be prevented, but only if several conditions apply.
\begin{enumerate}
\item A coordination step must reference the process and a specific state
before any coordination constraints may be enforced. Without such
a reference, the execution of a process cannot be restricted. 
\item At the time the coordinated process wants to activate that state,
the coordination constraints imposed by the coordination process must
be unfulfilled. 
\end{enumerate}
Only if these conditions apply, the activation of a state may be prevented.
In all other cases, a coordination process does not affect the execution
of the coordinated process. In other words, coordination processes
operate according to the blacklist principle: Every action is allowed
unless specifically prevented by a coordination process. This allows
for a high degree of freedom in executing the process instances while
still enforcing all the necessary coordination constraints.

However, for the complete fulfillment of Challenge 1 \emph{Asynchronous
Concurrency}, it must be shown that processes may indeed run concurrently
and asynchronously, otherwise the minimal impact of coordination processes
would be rendered moot. Section \ref{sec:Technical-Implementation}
shows that processes may run fully asynchronously and concurrently
in the PHILharmonicFlows process management system.

\subsubsection{Fulfilling Challenge 5 \emph{Manageable Complexity}}

Regarding Challenge 5 \emph{Managing Complexity}, the concepts that
have been described in Sections \ref{sec:Enacting-Coordination-Processes}-\ref{sec:Operational-Semantics}
are hardly noticed by an end user executing the coordinated processes.
Effectively, the only effect of these concepts for the end user is
whether or not a coordination process allows activating a state, and
if not, why it is prevented from activation. Consequently, except
for this, a coordination process and its execution may be fully hidden
from an end user at run-time. Thereby, the complexity perceived by
an end user at run-time is very low, though the complexity of the
run-time coordination process is high.

Regarding the complexity for a process modeler at design-time, a detailed
understanding of coordination processes, as presented in this paper,
is overall beneficial, though not required. A process modeler would
need to know the definition and usage of semantic relationships and
how multiple process instances are dealt with in a coordination process.
Furthermore, the definition and usage of each process modeling element
and its properties. This is analogous to BPMN 2.0 \citep{ObjectManagementGroup.2011},
where a modeler requires knowledge about the function of each modeling
element, for example an inclusive gateway, yet not how this functionality
is exactly provided by the BPMN process engine. There are significantly
less process modeling elements for coordination processes and object-aware
process management that for BPMN. Therefore, the argument can be made
that coordination processes possess less or at the very least equal
complexity in regard to modeling as BPMN 2.0 while simultaneously
fulfilling all challenges. This fulfills Challenge 5 \emph{Manageable
Complexity} from the viewpoint of the process modeler.

For developers, the PHILharmonicFlows process engine (cf. Section
\ref{sec:Technical-Implementation}) is the proof that implementation
of all concepts and challenges is feasible. The PHILharmonicFlows
process engine fully supports all challenges as stated in this article
and all requirements imposed by object-aware process management. Therefore.
the complexity of implementing coordination processes and its related
concepts is manageable, fully supporting Challenge 5 \emph{Manageable
Complexity} from the viewpoint of the developer.

In general, for both run-time and design-time , the complexity perceived
by end users, process modelers, and developers is significantly lower
than the actual complexity of the concept. As such, Challenge 5 \emph{Managing
Complexity} can be regarded as accomplished.

\subsubsection*{}

\section{\label{sec:Technical-Implementation}Technical Implementation and
Evaluation}

Coordination processes originate in the object-aware process management
paradigm \citep{Kunzle.2011}. The concepts of object-aware process
management, i.e., objects, lifecycle processes, relations, and coordination
processes, have been implemented in the PHILharmonicFlows prototype.
It has been shown that coordination processes impact process execution
as minimally as possible, as demanded by Challenge 1 \emph{Asynchronous
Concurrency}. In order to completely fulfill Challenge 1, it remains
to show that PHILharmonicFlows is indeed capable of executing processes
fully asynchronously and concurrently, which is achieved with actors
and microservices (cf. Section \ref{subsec:Actors-and-Microservices}).
Otherwise, coordination processes would not fulfill Challenge 1 \emph{Asynchronous
Concurrency} to full extent. Finally, Challenge 4 \emph{Immediate
Consistency} is discussed in Section \ref{subsec:Coordination-Process-Performance}.

The challenges and operational semantics stand in the context of the
re-implementation of the prototype of PHILharmonicFlows. The initial
prototypical implementation of PHILharmonicFlows as an object-aware
process management system was developed from 2008 to 2012. It involved
functional design- and run-time of the basic concepts of object-ware
process management, though many advanced features could not be realized
due to the technology available at the time. These imposed severe
limitations on the functional and technical capabilities of the prototype.

In 2015, a fully new implementation internally named ``Proteus''
was started, leveraging the emerging concepts of \emph{microservices}
\citep{Andrews.2018}. Microservices allow for a scalable and performant
execution of object-aware processes, resolving many of the previous
limitations. However, this also rendered almost all of the existing
codebase of the previous prototype obsolete. The re-implementation
of PHILharmonicFlows also paved the way for the development of advanced
features such as ad-hoc changes for object aware-process management
\citep{Andrews.2019} and a hyper-scalable run-time environment \citep{Andrews.2018}.
This new and improved microservice-based implementation continues
using the branding PHILharmonicFlows. The new paradigm of microservices
also required to re-think and adapt many concepts of object-aware
process management, accounting for the new requirements imposed by
this fundamental change of using microservices. Concepts such as object
lifecycles , the relational process structure, and coordination processes
were extended and also adapted, with a major focus on the run-time
\citep{Steinau.2017,Steinau.2018,Steinau.2018b}. These extensions
and adaptation both represent evolutionary and revolutionary changes.
In particular, the operational semantics of  coordination processes
required a substantial overhaul and, therefore, pose a novel contribution.
This overhaul of the operational semantics simultaneously drove and
was driven by the re-implementation of the PHILharmonicFlows system,
constituting a high practicality.

\subsection{\label{subsec:Actors-and-Microservices}Actors and Microservices}

With PHILharmonicFlows, much effort has been put into development
to create a scalable process management system that supports many
concurrently running processes \citep{Andrews.2018}. The paradigm
of object-aware process management is uniquely suited for this, as
its conceptual elements, e.g., objects and their lifecycle processes,
can be represented with actor theory as individual \emph{actors} \citep{Agha.1988,Hewitt.2010}.
PHILharmonicFlows uses a variant of actors with reliable messaging.
In essence, an actor is an independent entity that consists of a message
queue and a store for arbitrary data. An actor may receive messages
from other actors or from external sources, and process them using
data contained in the message and data from its store. 

\begin{figure}
\centering{}\includegraphics[width=1\columnwidth]{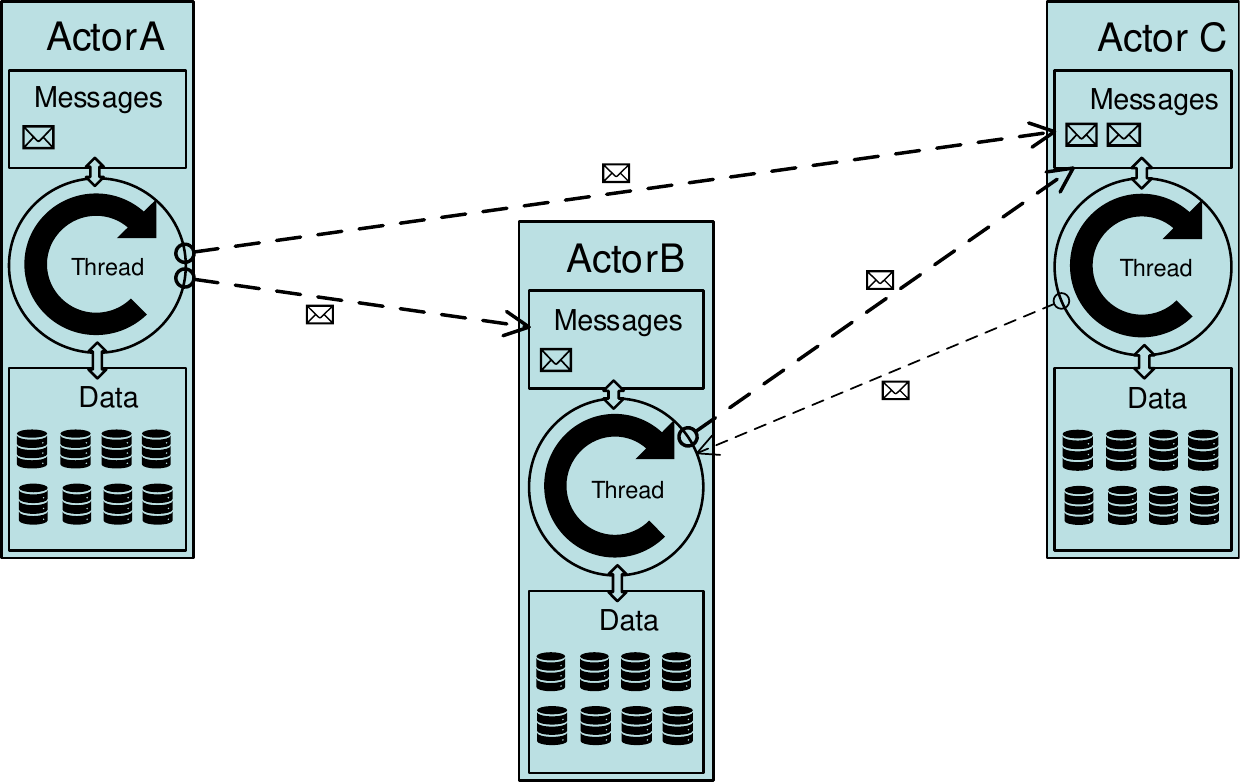}\caption{\label{fig:Schematic-Actors-and}Schematic Actors and their communication}
\end{figure}

An actor may only work on exactly one task at a time, i.e., it runs
conceptually on one single computational thread. An actor servicing
a message may only work on that one message, all other messages are
put in the queue until the current message has been serviced. Messages
are not fire-and-forget, but have callbacks. An \emph{actor system}
is realized by having multiple actors of different types that express
different functionality. In this system, actors then may run concurrently
and in parallel. Due to using a single computational thread and message
queuing, most of the concurrency problems regarding persistence and
computation, e.g., race conditions and dirty reads/writes, are not
present in an actor system. Moreover, actors may communicate asynchronously.

PHILharmonicFlows is realized as such an actor system. Each object
instance, together with its lifecycle process instance and its attributes,
is implemented as one actor. A lifecycle process of an object conforms
to the definition of process type (cf. Definition \ref{Def:ObjectType-Normal}).
Coordination processes are actors as well, but of different actor
type. Figure \ref{fig:Schematic-Actors-and} shows a schematic view
of actors and their communication. In particular, an actor may involve
other actors when servicing a request, as required data or functionality
may be located in other actors. Note that in Figure \ref{fig:Schematic-Actors-and},
Actor A is servicing an external request, depicted by the message
in its message queue and the outgoing communication from its thread.

\begin{figure}
\centering{}\includegraphics[width=1\columnwidth]{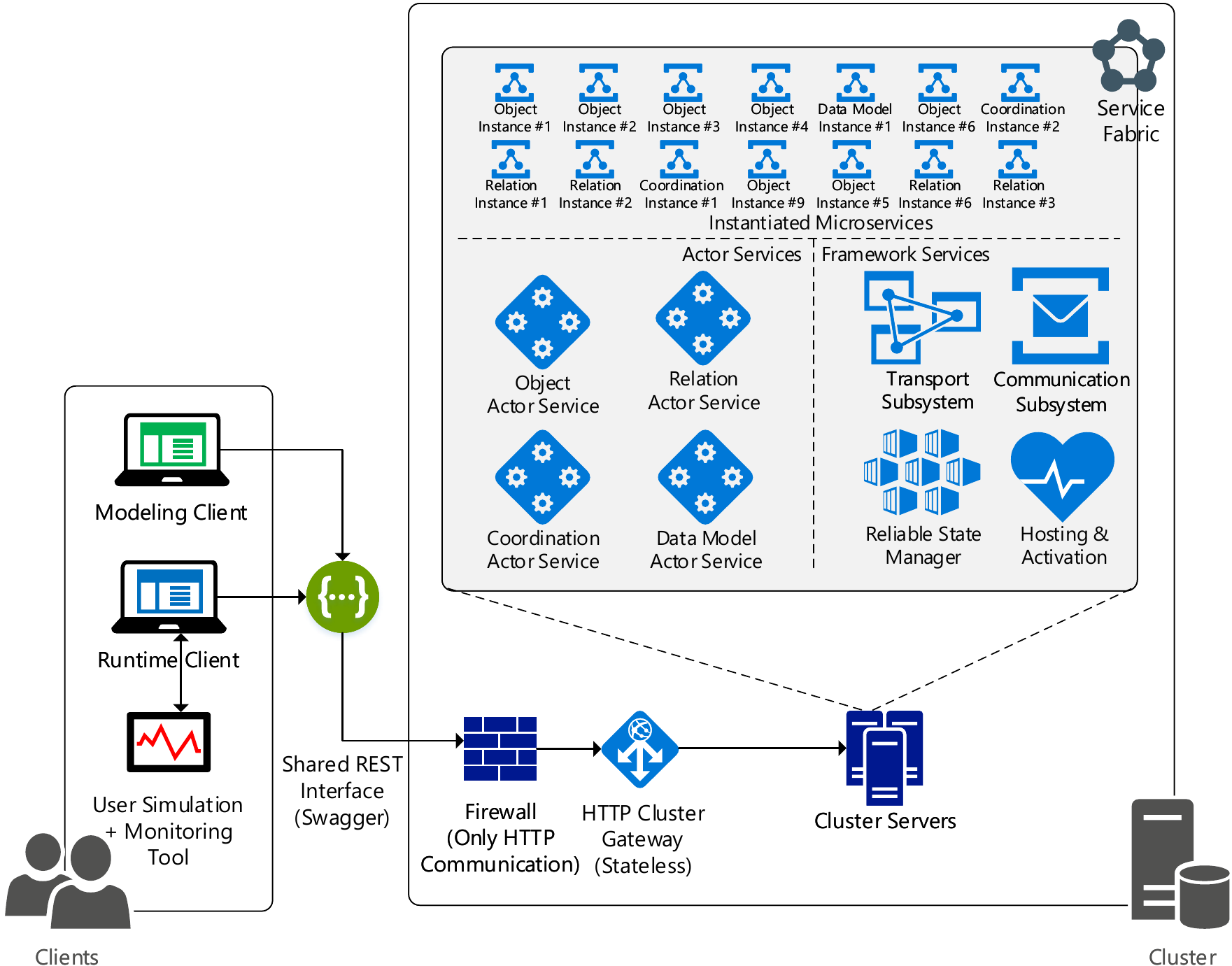}\caption{\label{fig:PHILharmonicFlows-Architecture}PHILharmonicFlows Architecture}
\end{figure}

Each actor in the PHILharmonicFlows system is realized as a \emph{microservice}
using Microsoft's \emph{Azure Service Fabric Framework}\footnote{https://docs.microsoft.com/en-us/azure/service-fabric/}.
Azure Service Fabric combines microservices with the actor paradigm,
and is therefore an ideal technical framework for building PHILharmonicFlows.
The overall architecture of the PHILharmonicFlows system can be seen
in Figure \ref{fig:PHILharmonicFlows-Architecture}. As shown in \citep{Andrews.2018},
PHILharmonicFlows scales very well horizontally, i.e., across distributed
machines or a cloud.

Microservices may run concurrently or in parallel by definition. As
each process is implemented as a microservice, logically the concurrent
execution of process instances is therefore guaranteed.

Still, the implementation must also enable the asynchronous interactions
between processes. Any object lifecycle process is, in principle,
independent from any other object lifecycle process, i.e., there is
no synchronization between any kind of processes. The only exception
is the coordination process. Conceptually, semantic relationships
enable the asynchronous execution of the processes. Semantic relationships
are represented by actor data and several communication patterns between
actors to represent their functionality at a fundamental level. Their
implementation uses the message stores and message exchange capability
of the actors. Therefore, semantic relationships are abstractions
over multiple, conditional series of messages between actors. As actors
are inherently capable of asynchronous communication, the implementation
of semantic relationships enables asynchronous process interactions
as well. Therefore, true asynchronous execution of interdependent
lifecycle processes is enabled by PHILharmonicFlows. 

In summary, PHILharmonicFlows is capable of executing a multitude
of processes concurrently and in parallel due to the use of the Azure
Service Fabric Framework to implement the concepts. Asynchronous communication
is enabled by the underlying actors. Therefore, Challenge 1 \emph{Asynchronous
Concurrency} is fully supported by coordination processes as implemented
in PHILharmonicFlows.

\subsection{\label{subsec:Coordination-Process-Performance}Coordination Process
Performance}

The challenge that has not been discussed so far is Challenge 4 \emph{Immediate
Consistency}. It is essentially concerned with the performance and
partly with the correctness of a coordination approach. For coordination
processes, it has been shown in Sections \ref{sec:Representing-Process-Instances}
and \ref{sec:Operational-Semantics} how the coordination process
is kept up to date on a qualitative level. For the purpose of this
paper it is assumed that coordination processes work correctly, i.e.,
they are updated correctly as required and they send correct updates
to the coordinated processes (i.e., the lifecycle processes). Therefore,
only the actual performance of coordination processes is important
in this context.

To prove that coordination processes fulfill Challenge 4 \emph{Immediate
Consistency,} the following experiment was set up. It must be demonstrated
that the execution time of the processes in a process structure that
is coordinated by a coordination process is not significantly larger
than the same process executions without the involvement of a coordination
process. In other words, coordination processes do not create a significant
overhead.

\begin{figure*}
\centering{}\includegraphics[width=1\textwidth]{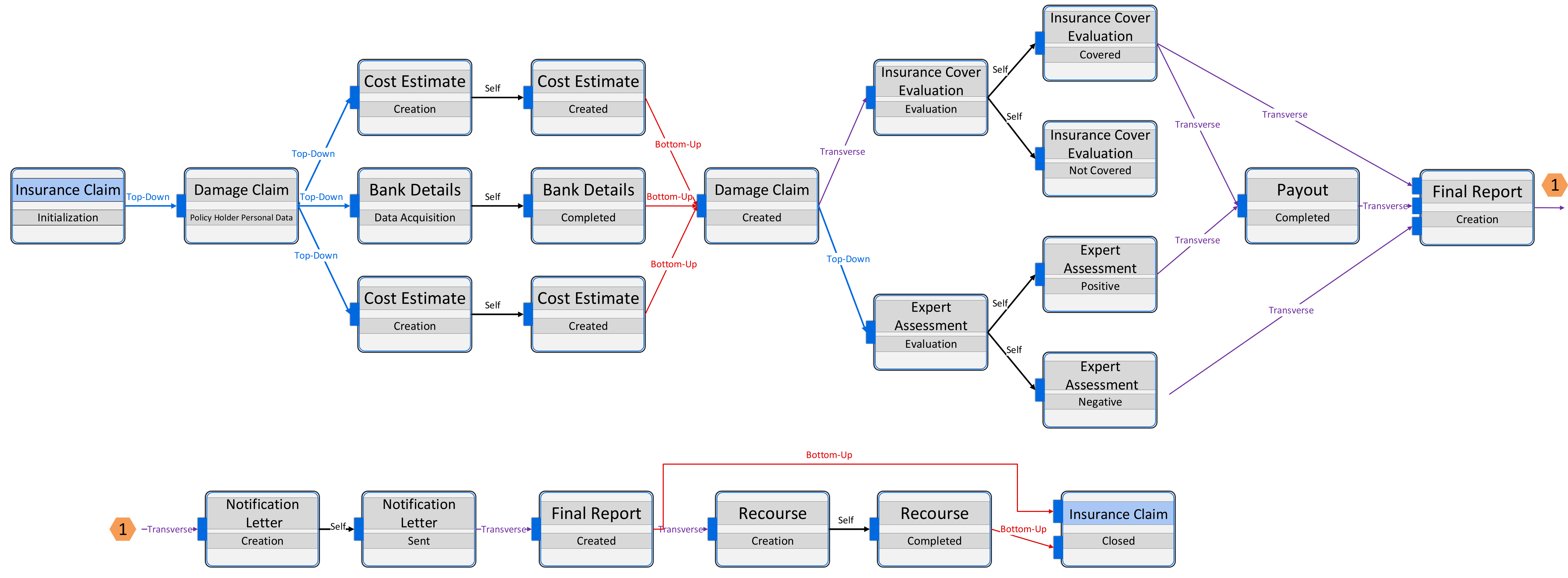}\caption{\label{fig:Insurance-Claim-Coordination-process}Coordination process
of the insurance claim model}
\end{figure*}

For the quantitative assessment in form of performance measurements,
two different PHILharmonicFlows business process models were defined
- the recruitment business process from the running example (cf. Example
\ref{exa:Running-Example}) and an insurance claim business process
(cf. Figure \ref{fig:Insurance-Claim-Coordination-process}). Both
possess exactly one coordination process, but the insurance claim
model is slightly larger. Additionally, a variant of each business
process model exists where the coordination process has been entirely
removed from the model, i.e., only objects and their lifecycle processes
remain. This means that no coordination constraints can be enforced
at run-time.

\begin{table}
\centering{}\caption{\label{tab:Execution-sequence-operations}Execution sequence operations}
\begin{tabular*}{1\columnwidth}{@{\extracolsep{\fill}}>{\centering}p{0.48\columnwidth}>{\raggedright}p{0.45\columnwidth}}
\toprule 
{\scriptsize{}Function} & {\scriptsize{}Description}\tabularnewline
\midrule
\midrule 
{\scriptsize{}$\mathit{InstantiateProcess}(\omega^{T})$} & {\scriptsize{}Creates a new instance of process type $\omega^{T}$}\tabularnewline
\midrule 
{\scriptsize{}$\mathit{LinkInstances(\omega_{1}^{I},\omega_{2}^{I}})$} & {\scriptsize{}Creates a new relation instance between process instances
$\omega_{1}^{I}$ and $\omega_{2}^{I}$.}\tabularnewline
\midrule 
{\scriptsize{}$\mathit{ChangeAttributeValue(\omega^{I}\text{,\ensuremath{\phi^{T}}},v)}$} & {\scriptsize{}Writes value $v$ of an attribute instance $\phi^{I}$that
has type $\phi^{T}$ of process $\omega^{I}$}\tabularnewline
\midrule 
{\scriptsize{}$\mathit{CommitTransition(\omega^{I},\tau^{T})}$} & {\scriptsize{}Commits transition $\tau^{I}$ that has type $\tau^{T}$
of process $\omega^{I}$. Implies a state change of $\omega^{I}$}\tabularnewline
\bottomrule
\end{tabular*}
\end{table}

Furthermore, for each business process model, an execution sequence
was defined that resembles a fairly standard and sufficiently complex
execution of the business processes. An execution sequence defines
a series of \emph{actions}, describing at which points process instances
are created or deleted, or when they change state. In detail, an execution
sequence action is created using one of the following functions (cf.
Table \ref{tab:Execution-sequence-operations}) and supplying it with
concrete parameter values.

Function $\mathit{InstantiateProcess}(\omega^{T})$ creates a new
process instance, given a corresponding process type $\omega^{T}$.
Function $\mathit{LinkInstances(\omega_{1}^{I},\omega_{2}^{I}})$
takes two process instances $\omega_{1}^{I}$ and $\omega_{2}^{I}$
as arguments and creates a relation $\pi^{I}$ between them, provided
a corresponding relation type was specified at design-time. Moreover,
function $ChangeAttributeValue(\omega^{I}\text{,\ensuremath{\phi^{T}}},v)$
writes value $v$ to the attribute instance $\phi^{I}$ of process
instance $\omega^{I}$. Attribute instances are part of an object
instance in PHILharmonicFlows and can be uniquely identified by its
type $\phi^{T}$, which occurs only once in each object instance $\omega^{I}$.
In case $\phi^{I}$ already has a value, the value is overwritten
with $v$. Finally, function $\mathit{CommitTransition(\omega^{I},\tau^{T})}$
causes a state change, i.e., after committing the transition, the
active state $\sigma_{a}^{I}$ is the target of the transition. 

It is possible to execute processes using only these four main functions
(cf. Table \ref{tab:Execution-sequence-operations}). This is due
to the underlying data-driven lifecycle processes in PHILharmonicFlows,
which the business process models are based on. Lifecycle processes
are enacted by acquiring data values $v$ for attribute instances
$\phi^{I}$. The details of lifecycle process execution have been
shown in \citep{Steinau.2019b}. Based on the four functions, an execution
sequence is designed that realizes a full business process execution
involving multiple process instances which are being coordinated by
a coordination process.

The execution sequence describes how instances of both business process
model variants are executed, i.e., any instance of the business process
model with and without the coordination process performs the same
actions in the same order. The execution sequence is designed to not
violate any coordination constraints, in order to achieve identical
results even when there is no coordination process involved. Otherwise,
in one case, an action may be blocked by a coordination process, whereas
in case of a missing coordination process, the action would be allowed,
creating different results and therefore bias in the performance measurements.
A full description of both business process models with high-resolution
graphs, together with the detailed execution sequences and their descriptions,
has been made available\footnote{The data can be found at https://bit.ly/2DvFZvk}.

As PHILharmonicFlows supports parallel and concurrent (lifecycle)
process execution as enabled by the actor microservices, performance
measurements follow the \emph{guidelines for measuring the performance
of parallel computing systems}, as defined in \citep{Andrews.2018}.
Here, the experiment reuses the exact methodology from \citep{Andrews.2018}
and is therefore not replicated in detail here for the sake of brevity.
The general idea is to dynamically determine the number of runs $n$
needed to achieve a given confidence interval $\mathit{CI}$ for the
value that is measured.

\begin{table*}[t]
\caption{\label{tab:Performance-Measurements-of}Performance Measurements of
Two Process Models}

\centering{}{\footnotesize{}}%
\begin{tabular*}{1\textwidth}{@{\extracolsep{\fill}}cccccccccc}
\toprule 
{\scriptsize{}\#} & {\scriptsize{}Model} & {\scriptsize{}Instances} & {\scriptsize{}Actions} & {\scriptsize{}$CI\:t_{\mathit{exec}}^{nc}$} & {\scriptsize{}$n^{nc}$} & {\scriptsize{}Confidence} & {\scriptsize{}$CI\:t_{\mathit{exec}}^{c}$} & {\scriptsize{}$n^{c}$} & {\scriptsize{}Confidence}\tabularnewline
\midrule
\midrule 
{\scriptsize{}1} & {\scriptsize{}Recruitment} & {\scriptsize{}96} & {\scriptsize{}289} & {\scriptsize{}{[}00:01:769, 00:01:882{]}} & {\scriptsize{}6} & {\scriptsize{}96.88} & {\scriptsize{}{[}00:03:619, 00:03:720{]}} & {\scriptsize{}6} & {\scriptsize{}96.88}\tabularnewline
\midrule 
{\scriptsize{}2} & {\scriptsize{}Recruitment} & {\scriptsize{}32} & {\scriptsize{}82} & {\scriptsize{}{[}00:00:502, 00:00:530{]}} & {\scriptsize{}6} & {\scriptsize{}96.88} & {\scriptsize{}{[}00:00:911, 00:00:978{]}} & {\scriptsize{}6} & {\scriptsize{}96.88}\tabularnewline
\midrule 
{\scriptsize{}3} & {\scriptsize{}Insurance} & {\scriptsize{}18} & {\scriptsize{}168} & {\scriptsize{}{[}00:01:069 - 00:01:140{]}} & {\scriptsize{}6} & {\scriptsize{}96.88} & {\scriptsize{}{[}00:01:420, 00:01:478{]}} & {\scriptsize{}6} & {\scriptsize{}96.88}\tabularnewline
\bottomrule
\end{tabular*}{\footnotesize\par}
\end{table*}

The measured value is the execution time $t_{\mathit{exec}}^{c}$
of the execution sequence with coordination by by a coordination process.
For comparison, $t_{\mathit{exec}}^{nc}$ denotes the time for enacting
the execution sequence without a coordination process. $t_{\mathit{exec}}^{c}$
and $t_{\mathit{exec}}^{nc}$ are the summation of the execution time
of each individual action in the execution sequence.

As stated in Section \ref{sec:Challenges-and-Problem}, immediate
consistency is achieved if users do not experience delay when issuing
two consecutive actions. The book ``Usability Engineering'' \citep{Nielsen.1993}
contains three tiers for response times of software systems with respect
to the user experience, stated as follows:
\begin{itemize}
\item \textbf{0.1 second} is about the limit for having the user feel that
the system is reacting instantaneously, meaning that no special feedback
is necessary except to display the result. 
\item \textbf{1.0 second} is about the limit for the user's flow of thought
to stay uninterrupted, even though the user will notice the delay.
Normally, no special feedback is necessary during delays of more than
0.1 but less than 1.0 second, but the user does lose the feeling of
operating directly on the data. 
\item \textbf{10 seconds} is about the limit for keeping the user's attention
focused on the dialogue. For longer delays, users will want to perform
other tasks while waiting for the computer to finish, so they should
be given feedback indicating when the computer expects to be done.
Feedback during the delay is especially important if the response
time is likely to be highly variable, since users will then not know
what to expect.
\end{itemize}
Ideally, for optimally fulfilling Challenge 4 \emph{Immediate Consistency},
each action in the execution sequences should require less than 0.1
second for execution. Formally, several goals are defined for Challenge
4 \emph{Immediate Consistency. }The primary goal is stated as follows: 
\begin{primarygoal}[Primary]
\label{primary-goal}Each action in the execution sequence should
remain below 100ms of execution time
\end{primarygoal}
If this goal is unattainable, a secondary goal is defined:
\begin{primarygoal}[Secondary]
Each action in the execution sequence should remain below 1000 ms
of execution time
\end{primarygoal}
The overall execution times of both business processes ``Recruitment''
and ``Insurance'' are less important regarding Challenge 4 \emph{Immediate
Consistency}. Nevertheless, they are reported in Table \ref{tab:Performance-Measurements-of}
as they give an overall impression of the performance of the PHILharmonicFlows
process engine. All execution times are given as standard intervals
of the format $\mathit{[lower,upper]}$, where time has the format
$\mathit{[mm:ss:fff]}$. Three scenarios \#1-3 were run, two times
using the recruitment business process (i.e, the running example),
and one time the insurance business process, with the coordination
process shown in Figure \ref{fig:Insurance-Claim-Coordination-process}.
Scenario \#1 is the recruitment business process, where 5 $\mathit{Applications}$
are submitted and reviewed, each having 3-5 $\mathit{Reviews}$. Scenario
\#2 is a cut-down version of Scenario \#1, where only one $\mathit{Application}$
is submitted and subsequently accepted to fill the position. Scenario
\#3 is an insurance business process, comprising one instance of each
of the eighteen process types.

As can be seen in Table \ref{tab:Performance-Measurements-of}, running
the same execution sequence with a coordination process takes roughly
twice the time compared to running it without any coordination process.
However, given the number of process instances (96) and actions (289)
in Scenario \#1, total execution time manages to remain below the
4 second mark. Scenarios \#2 and \#3 have less actions and, consequently,
achieve better total execution times. Note that the execution sequences
produce very consistent results, as it takes only the minimum number
of runs (6) to obtain the necessary confidence level of $\geq95$\%.

The execution sequences have been designed to resemble what can be
considered fairly standard process executions. Still, the sequences
try to prolong process execution. Whenever branches may be chosen
during decisions, the execution sequence chooses the longer path.
Furthermore, the execution sequences take no advantage of executing
processes concurrently, as in principle enabled by PHILharmonicFlows
engine. Each sequence simulates a single user, executing each action
of the execution sequence strictly sequentially. Note that this does
not prevent the PHILharmonicFlows engine from using any parallelism,
as it has been designed for high parallelism regardless of how an
execution sequence is structured. Still, the execution sequences constitute
a worst case as far as the concurrent execution of processes is concerned.

\begin{figure*}[t]
\centering{}\includegraphics[width=1\textwidth]{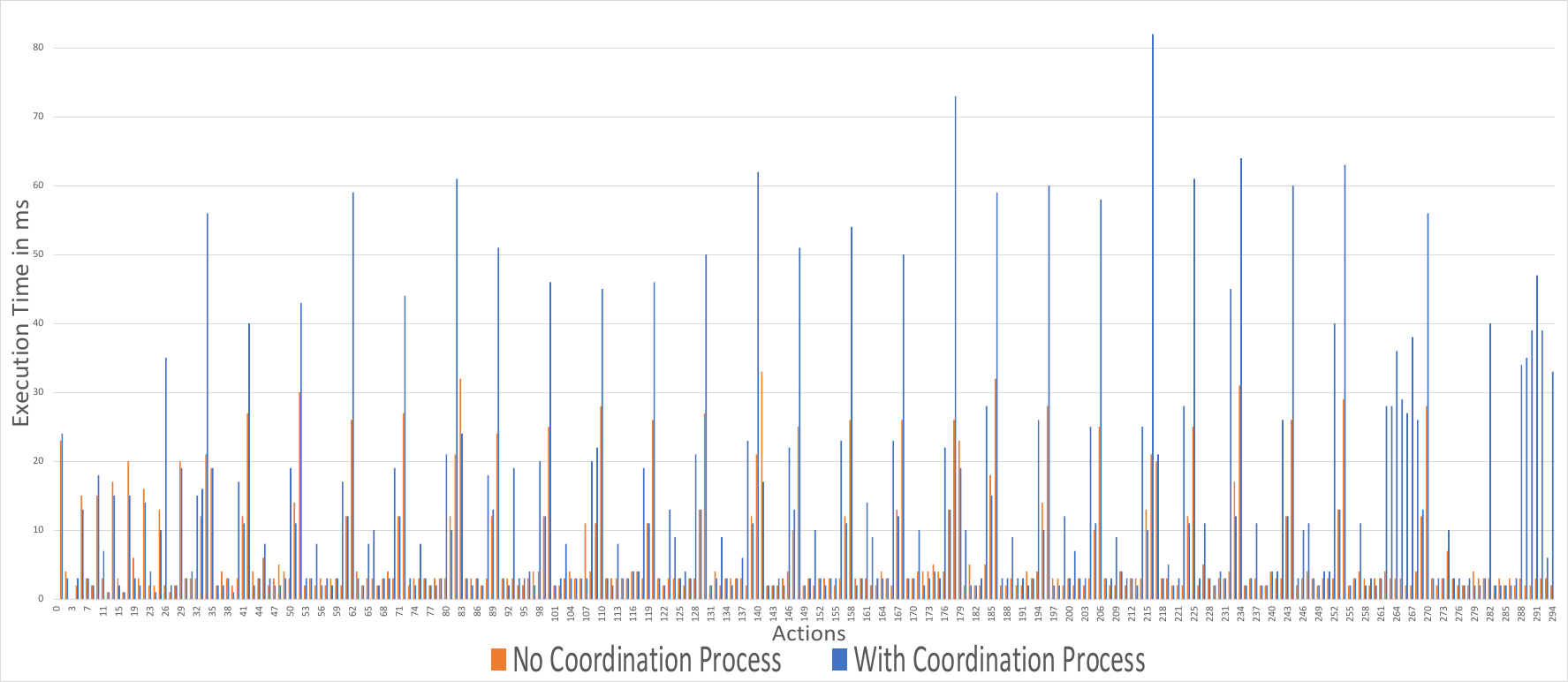}\caption{\label{fig:Comparison-of-Execution}Comparison of Execution Times
with and without a coordination process per action}
\end{figure*}

\begin{table*}
\centering{}\caption{\label{tab:Statistical-Data-for}Statistical Data for the Last Run
of Scenario \#1 Recruitment}
{\scriptsize{}}%
\begin{tabular*}{1\textwidth}{@{\extracolsep{\fill}}ccccccccccc}
\toprule 
 & \multicolumn{2}{c}{{\scriptsize{}Total Time}} & \multicolumn{2}{c}{{\scriptsize{}Average}} & \multicolumn{2}{c}{{\scriptsize{}Median}} & \multicolumn{2}{c}{{\scriptsize{}Minimum}} & \multicolumn{2}{c}{{\scriptsize{}Maximum}}\tabularnewline
\midrule 
 & {\scriptsize{}No CP} & {\scriptsize{}CP} & {\scriptsize{}No CP} & {\scriptsize{}CP} & {\scriptsize{}No CP} & {\scriptsize{}CP} & {\scriptsize{}No CP} & {\scriptsize{}CP} & {\scriptsize{}No CP} & {\scriptsize{}CP}\tabularnewline
\midrule 
{\scriptsize{}$\mathit{InstantiateProcess}(\omega^{T})$} & {\scriptsize{}00:378} & {\scriptsize{}00:393} & {\scriptsize{}00:012} & {\scriptsize{}00:012} & {\scriptsize{}00:011} & {\scriptsize{}00:011} & {\scriptsize{}00:010} & {\scriptsize{}00:010} & {\scriptsize{}00:021} & {\scriptsize{}00:025}\tabularnewline
\midrule 
{\scriptsize{}$\mathit{ChangeAttributeValue(\omega^{I}\text{,\ensuremath{\phi^{T}}},v)}$} & {\scriptsize{}00:344} & {\scriptsize{}00:411} & {\scriptsize{}00:003} & {\scriptsize{}00:003} & {\scriptsize{}00:002} & {\scriptsize{}00:003} & {\scriptsize{}00:001} & {\scriptsize{}00:001} & {\scriptsize{}00:009} & {\scriptsize{}00:010}\tabularnewline
\midrule 
{\scriptsize{}$\mathit{CommitTransition(\omega^{I},\tau^{T})}$} & {\scriptsize{}00:223} & {\scriptsize{}01:291} & {\scriptsize{}00:003} & {\scriptsize{}00:017} & {\scriptsize{}00:003} & {\scriptsize{}00:011} & {\scriptsize{}00:002} & {\scriptsize{}00:002} & {\scriptsize{}00:007} & {\scriptsize{}00:047}\tabularnewline
\midrule 
{\scriptsize{}$\mathit{LinkInstances(\omega_{1}^{I},\omega_{2}^{I}})$} & {\scriptsize{}00:738} & {\scriptsize{}01:529} & {\scriptsize{}00:024} & {\scriptsize{}00:049} & {\scriptsize{}00:023} & {\scriptsize{}00:048} & {\scriptsize{}00:021} & {\scriptsize{}00:023} & {\scriptsize{}00:030} & {\scriptsize{}00:083}\tabularnewline
\bottomrule
\end{tabular*}
\end{table*}

Moreover, the business process models used for the measurements require
a high amount of coordination. Especially, the recruitment example
shows very tight coordination. The business process model comprises
4 processes with 5 states each, and of these 20 states in total, 4
are not subject to coordination by a coordination process (cf. Figures
\ref{fig:State-based-views-of} and \ref{fig:Coordination-Process-for}).
As such, the business process model nearly maximizes the amount of
coordination, leading to almost another worst case for the total execution
time. The insurance scenario is less tightly coordinated. In light
of these detrimental conditions, a maximum execution time of less
than 4 seconds for scenario \#1 is satisfactory.

All performance measurements have been run on a Lenovo T470p notebook.
The notebook features an Intel(R) Core(TM) i7-7700HQ 4 Core/8 Thread
CPU running at base clock 2.80GHz in stock configuration. The CPU
was neither overclocked, nor undervolted nor locked to a specific
frequency. The notebook has 16 GB RAM DDR3-2400 and an SSD. Software-wise,
it runs Windows 10 Pro x64 v1903 and Visual Studio Enterprise in the
most up-to-date version (v16.0.9, as of October 15th, 2019), and the
debug-compiled, up-to-date PHILharmonicFlows software. The performance
measurements were performed with the notebook plugged in, using best
performance mode of Windows 10. 

For Challenge 4 \emph{Immediate Consistency}, the performance measurements
captured the execution time of each action during the last run of
Scenario \#1. Figure \ref{fig:Comparison-of-Execution} shows a chart
of Scenario \#1, comparing the execution times with a coordination
process to the execution times without using a coordination process.

Figure \ref{fig:Comparison-of-Execution} shows that coordination
processes only delay some actions. The actions derived from function
$\mathit{CommitTransition(\omega^{I},\tau^{T})}$ experience significantly
high\-er execution times if coordinated by coordination processes.
Moreover, actions that use $\mathit{LinkInstances(\omega_{1}^{I},\omega_{2}^{I}})$
require additional execution time because of the coordination processes,
though not as much time as with function $\mathit{CommitTransition(\omega^{I},\tau^{T})}$.
Actions derived from function $\mathit{ChangeAttributeValue(\omega^{I}\text{,\ensuremath{\phi^{T}}},v)}$
and function $\mathit{InstantiateProcess}(\omega^{T})$ do not experience
higher execution times.

\begin{figure*}
\includegraphics[width=1\textwidth]{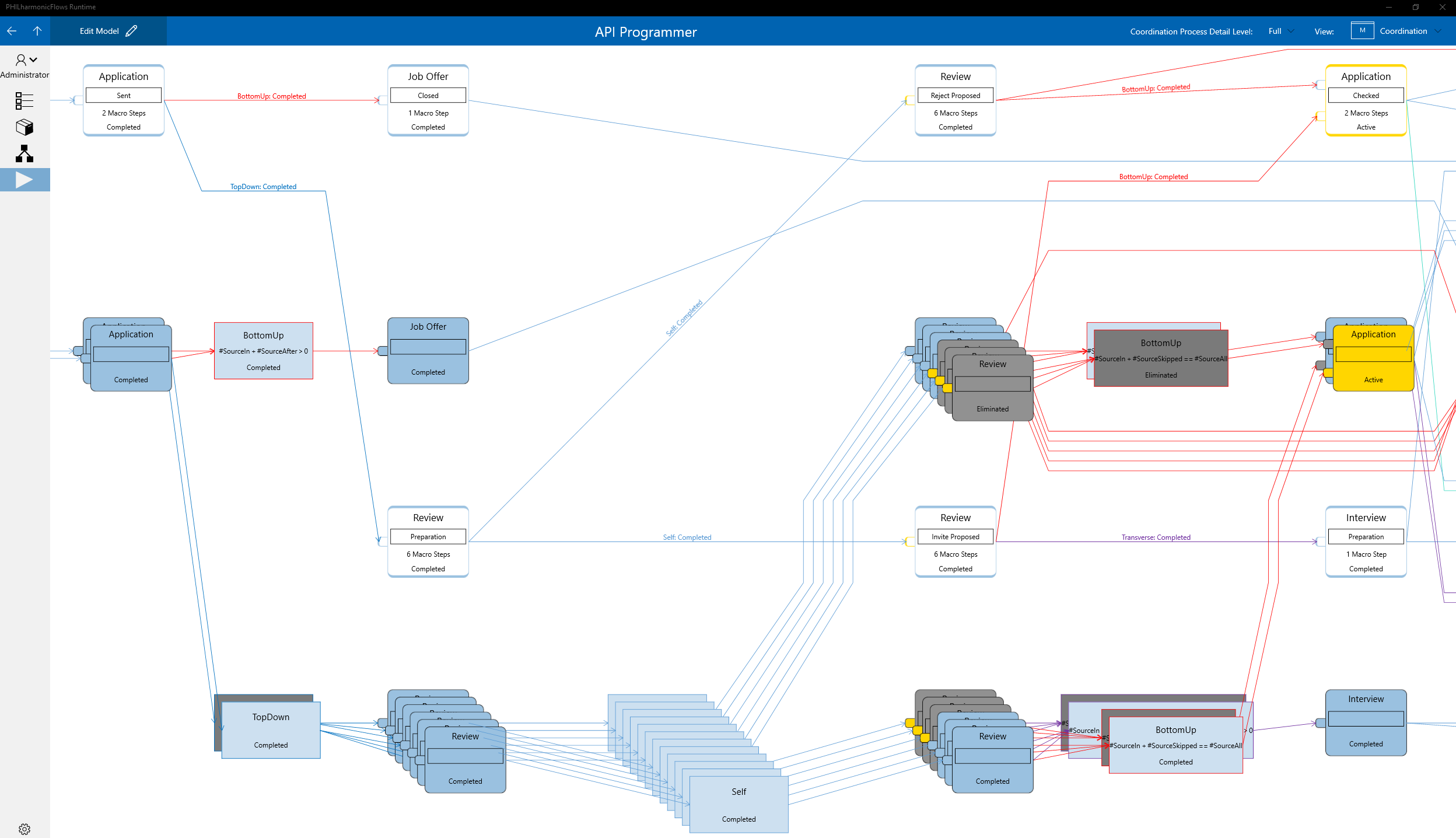}\caption{\label{fig:Screenshot-from-a}Screenshot from a Coordination Process
Execution}
\end{figure*}

Regarding the execution times of each individual action, it can be
noticed that coordination processes do have a significant impact,
sometimes doubling or tripling the required execution time of a particular
action (cf. Figure \ref{fig:Comparison-of-Execution}). However, the
maximum execution time of any action in the execution sequence is
83 ms involving a coordination process. In consequence, the primary
goal has been completely achieved (cf. Goal \ref{primary-goal}),
i.e., each action does not go over 100ms of execution time even with
a coordination process involved.

Table \ref{tab:Statistical-Data-for} summarizes metric scores related
to the performance measurements. As these are based on a single run,
there is no variance and the interval notation of Table \ref{tab:Performance-Measurements-of}
is not needed. The data is obtained from the last (i.e., sixth) run
of Scenario \#1 with and without a coordination process, as the total
execution time is guaranteed to be within bounds and therefore representative. 

Again, the time for a run of $\mathit{LinkInstances(\omega_{1}^{I},\omega_{2}^{I}})$
with a coordination process is noted as 83 ms at maximum, confirming
the chart in Figure \ref{fig:Comparison-of-Execution}. Table \ref{tab:Statistical-Data-for}
also reveals that the coordination process predominantly affects execution
times of functions $\mathit{LinkInstances(\omega_{1}^{I},\omega_{2}^{I}})$
and $\mathit{CommitTransition(\omega^{I},\tau^{T})}$. Moreover, it
can be observed from the averages and the median that the total execution
time is comprised of many actions with very low execution times and
a few other actions with very high execution times.

Consequently, a favorable user experience is ensured.

In summary, it has been shown that each action in several execution
sequences remains below 100ms of execution time, fulfilling the primary
Goal \ref{primary-goal}. This is valid for several different scenarios
involving different models and instance counts. In conclusion, Challenge
4 \emph{Immediate Consistency} may be regarded as fulfilled, as there
are scenarios which meet the requirements. With the fulfillment of
the remaining Challenge 4, all five Challenges are fulfilled simultaneously
by object-aware process management and PHILharmonicFlows.

\textcolor{red}{}

\subsection{Execution of an Object-aware Business Process}

In order to showcase that coordination processes and their implementation
are working, a screencast\footnote{https://vimeo.com/389004670} has
been created. It shows the execution of the recruitment business process
introduced in Example \ref{Running-Example:-Recruitment}. The recruitment
process starts with the creation of a $\mathit{Job\:Offer}$ process
and its publication. Several $\mathit{Applications}$ are created
and subsequently reviewed. Throughout the screencast, it is checked
whether the coordination process enforces the coordination constraints
(cf. Example \ref{Running-Example:-Recruitment}). For one $\mathit{Application}$,
an $\mathit{Interview}$ is created, and in the following the $\mathit{Application}$
is accepted. Other $\mathit{Applications}$ then must be rejected.
The screencast concludes with the $\mathit{Job\:Offer}$ reaching
the $\mathit{Position\:Filled}$ end state. Figure \ref{fig:Screenshot-from-a}
shows a screenshot of the Runtime Tool depicting the same process
as in the screencast.

The intention of the screencast is to demonstrate the complexity of
the coordination of interdependent business processes and to emphasize
that the presented challenges occur in practice. Furthermore, it is
intended to show that the concepts presented in this paper manage
to properly coordinate interdependent processes, creating a purposeful
overall business process. The screencast shows that the presented
concepts are already implemented in working software.

\subsection{Case Study: E-learning Platform}

PHILharmonicFlows is applied to a sophisticated real world scenario
involving an e-learning platform. PHILharmonicFlows is used to realize
an e-learning platform called Phoodle, which supplants the regular
Moodle e-learning platform in a lecture. Phoodle was used throughout
the whole semester to support the lecture. More specifically, Phoodle
implemented the following basic functionality. The highlighted words
are relevant objects with lifecycles.
\begin{enumerate}
\item Creation of \emph{lectures} and associated elements, such as \emph{exercises}.
\item Supplementary material to the lectures, such as slides or videos,
are available as a \emph{download} from the platform.
\item \emph{Students} should be allowed to register for and \emph{attend}
lectures and participate in exercises
\item Supervisors create exercises and publish these on a weekly basis
\item Students create solution \emph{submission}s for these exercises.
\item Tutors grade the submissions from the students and give feedback.
\end{enumerate}
\begin{figure*}
\centering{}\includegraphics[width=1\textwidth]{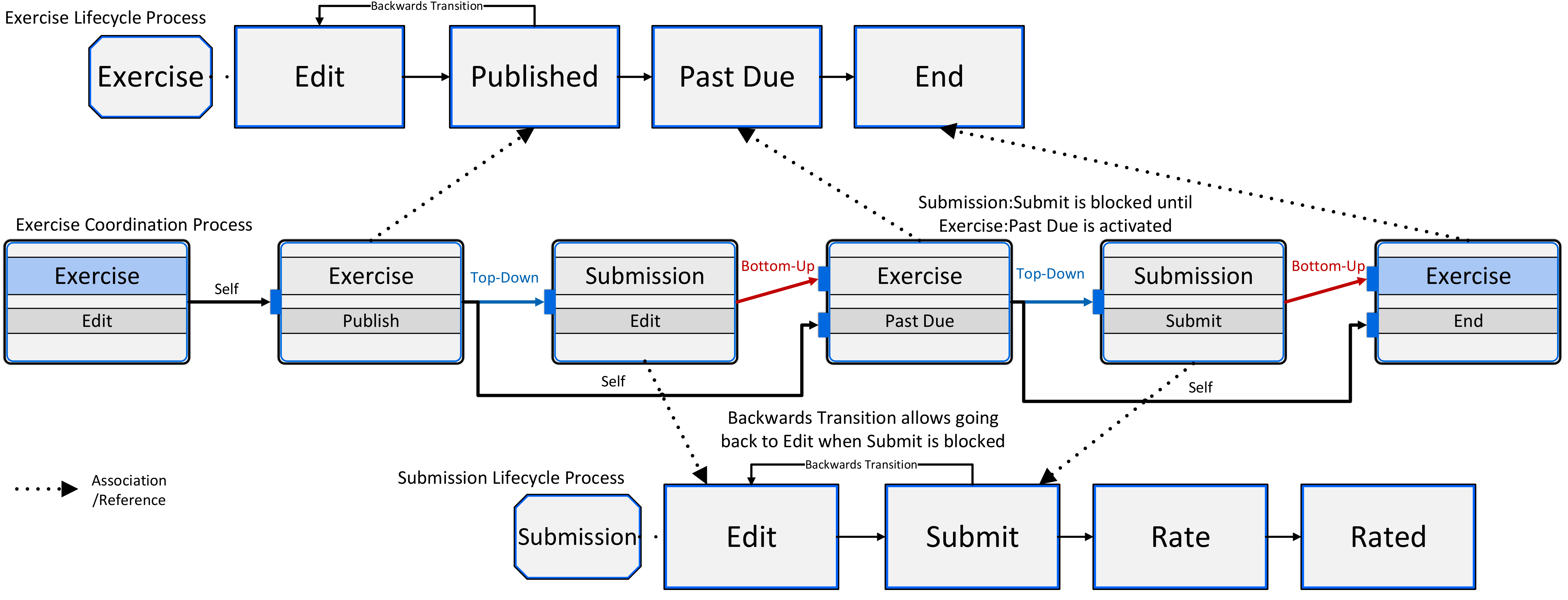}\caption{\label{fig:Coordination-Process-for-Phoodle}Coordination Process
for Phoodle}
\end{figure*}

The primary intention with Phoodle was to research how well the complexity
of an object-aware business process can be concealed from end users.
Furthermore, it was of interest which issues arise when using completely
generic business process management software to implement real-world
software. The Phoodle e-learning platform is realized entirely as
a PHILharmonicFlows business process. More specifically, one instance
of the Phoodle business process was used. The web-based front-end
is completely generic as well, except for a configuration option for
connecting to the Phoodle business process instance on the PHILharmonicFlows
server.

\begin{figure}
\begin{centering}
\includegraphics[width=1\columnwidth]{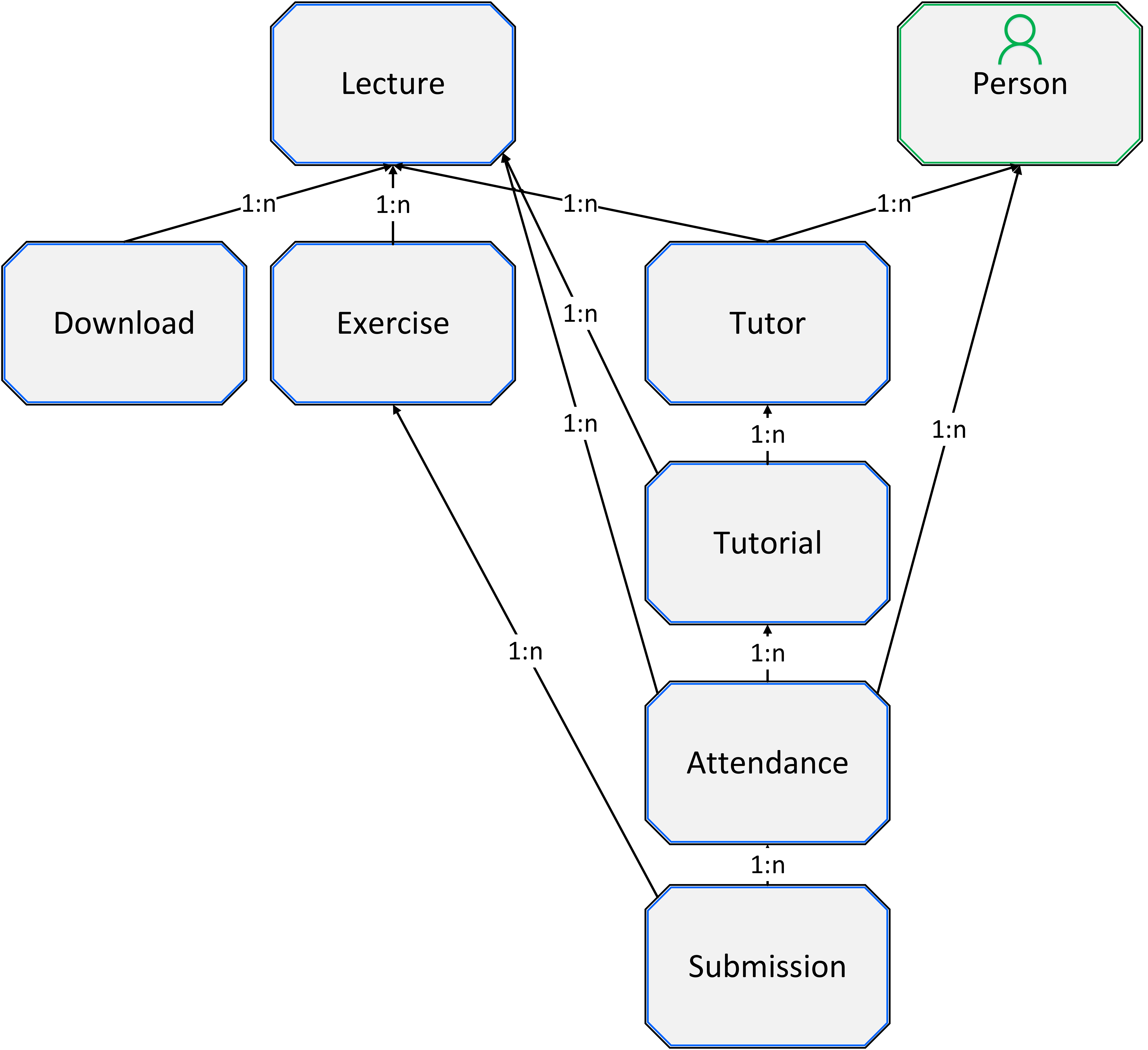}
\par\end{centering}
\caption{\label{fig:Phoodle-Relational-Process-Type}Relational Process Type
Structure of Phoodle}
\end{figure}

Figure \ref{fig:Phoodle-Relational-Process-Type} shows the relational
process type structure for Phoodle. The roles of student, supervisor
and tutor are realized via the relations between the $\mathit{Person}$
user type and other object types \citep{Andrews.2017}, e.g., a person
obtains the student role by creating an $\mathit{Attendance}$ for
a $\mathit{Lecture}$. Tutor and Supervisor roles are coupled to the
$\mathit{Tutor}$ and $\mathit{Lecture}$ object types, respectively.

Coordination processes are used to control the interactions $\mathit{Exercise}$
and $\mathit{Submission}$. Figure \ref{fig:Coordination-Process-for-Phoodle}
shows the lifecycle processes for $Exercise$ and $\mathit{Submission}$
as well as the coordination process. Once an $Exercise$ has been
created and published by a supervisor, students are allowed to create
and edit $\mathit{Submissions}$. Students must submit their solution
before a predefined deadline, but are permitted to alter and re-submit
their submission. Altering the submission is achieved with a backwards
transition from state $\mathit{Submit}$ to $\mathit{Edit}$ of the
$\mathit{Submission}$ object. 

State $\mathit{Submit}$ is designed to contain no activities, as
it is only used to indicate the status of the lifecycle and requires
no further functionality. However, this creates the following issue
in the $\mathit{Submission}$ process: Without a coordination process,
once state $\mathit{Submit}$ is activated, the $\mathit{Submission}$
lifecycle process immediately activates state $\mathit{Rate}$, where
a Tutor is supposed to grade the submission. This is unintended behavior,
as the student can no longer correct mistakes in his submission, as
there is no backwards transition from $\mathit{Rate}$ to $\mathit{Edit}$. 

In this case, the coordination process is used to correct this unintended
behavior. With the coordination process as shown in Figure \ref{fig:Coordination-Process-for-Phoodle},
a coordination constraint is introduced that prevents $\mathit{Submit}$
from activating. In consequence, when the student triggers the transition
from $\mathit{Edit}$ to $\mathit{Submit}$, $\mathit{Submit}$ is
marked as $\mathit{Pending}$ (cf. Table \ref{tab:State-Markings}).
States marked as $\mathit{Pending}$ permit activating backward transitions,
so the student may alter the submission. However, the $\mathit{Submission}$
may only go into state $\mathit{Rate}$ when the deadline for $\mathit{Submissions}$
to the $\mathit{Exercise}$ has been reached and the $\mathit{Exercise}$
is in state $Past\:Due$. As the state $\mathit{Submit}$ contains
no activities, state $\mathit{Rate}$ is activated immediately upon
the $\mathit{Exercise}$ reaching state $Past\:Due$. This is the
fully realized and intended behavior by Phoodle, which cannot be achieved
without a coordination process.

During the semester, the instance of the Phoodle business process
registered 136 distinct users (students, supervisors and tutors) and
logged over 40,000 interactions from these users. In total, the instance
comprised 2898 microservices representing 848 object instances, 1542
instances of relations between these objects, 5 coordination process
instances, and 503 uploaded files. 

In summary, process coordination is indispensable for the correct
functioning of the Phoodle business process. The Phoodle business
process furthermore showed the importance of considering complex relationships
between processes. Finally, the case study demonstrated the viability
of PHILharmonicFlows and object-aware business process management
in practice.

\section{\label{sec:Related-Work}Related Work}

The topic of coordinating interdependent processes is connected to
many different subjects. Primarily, data-centric approaches or other
approaches relying on multiple interdependent processes, summarized
in short as \emph{interaction-centric approaches,} are closely related
work. Activity-centric processes and choreographies are also relevant
due to their prevalence in both the BPM field and the industry. Overall,
as this paper is concerned with process coordination and its enactment,
preference is given to related work that describes interactions between
processes, covering \emph{the execution phase of the BPM lifecycle}
\citep{Dumas.2013}. Further emphasis is placed on the approach having
(formal) operational semantics.

\subsection{Interaction-centric Approaches}

Interaction-centric approaches are predominantly defined by the interactions
between different processes.

Proclets are small, lightweight processes that focus on interactions
between processes \citep{vanderAalst.2000b,vanderAalst.2001}. The
proclet approach was one of the first to abandon monolithic process
models in favor of small, interacting processes. Proclets are defined
using the well-known formalism of Petri nets, recognizing that instances
of proclets may need to communicate with more than one other proclet.
Therefore, the proclet approach supports one-to-many interactions
between proclets. For this purpose, the Petri net formalism is extended
with ports, enabling communication to other proclets. Ports are fully
integrated into the Petri net formalism, supporting the usual formal
analysis techniques known from standard Petri nets. Using ports, channels
connect to ports on other proclets, over which performatives may be
exchanged.

The actual communication between proclets over a channel is realized
by performatives, a special form of message. One of the major advantages
of the proclet approach is the support for the full range of formal
analysis techniques enabled by Petri nets. However, Petri nets can
only describe imperative processes, thus they are limited in their
support of flexible enactment of the processes.

Artifact-centric process management \citep{Nigam.2003} describes
business processes as interacting \emph{artifacts} with lifecycles\emph{.
}Central to this approach is the \emph{artifact}, which holds all
process-relevant information in an \emph{information model}. The artifact
lifecycle is specified using the Guard-Stage-Milestone (GSM) meta-model
\citep{Hull.2011b,Hull.2011}. An artifact may further interact with
other artifacts. However, GSM does not provide dedicated coordination
mechanisms or explicit artifact relations, as does object-aware process
management. Instead, GSM incorporates an arbitrary information model
and an expression framework with which artifact interactions may be
specified. While this, in theory, allows expressing any concept or
constraint, in practice much of the capabilities of artifact-centric
process management hinge on the power of the expression framework.
As a drawback, expressions might become very complex and must be supported
by a rule engine to realize the full potential of artifact-centric
process management. The concepts of the relational process structure
and the semantic relationships may be recreated with complicated expressions
to achieve at least the basic functionality of coordination processes.
While this is not impossible, it requires great effort on the side
of the modeler to achieve the same functionality as object-aware process
management provides out-of-the-box. Consequently, the fulfillment
of the challenges is possible in principle, but depends highly on
the used expression framework and the way the specific artifacts and
business processes are modeled.

Artifact-centric process management has been prototypically implemented
in the BizArtifact demo tool\footnote{https://sourceforge.net/projects/bizartifact/},
whose predecessors include Barcelona \citep{Heath.2013} and Siena
\citep{Cohn.2008}. Due to the complexity of an artifact-centric business
process, model verification \citep{Belardinelli.2012,BagheriHariri.2013}
constitutes an important aspect of artifact-centric process management.
Moreover, several variants of artifact-centric process management
exist. For artifact-centric process management and the modeling language
GSM, \citep{Damaggio.2013} have proved that incremental and fixpoint
semantics for GSM models are equivalent in their expressiveness. This
is an important stepping stone towards the full verification of GSM-based
artifact-centric process models. \citep{BagheriHariri.2013,Calvanese.2012}
investigate artifact-centric processes in regard to evolving databases
with the intent to verify process correctness, leading to the creation
of data-centric dynamic systems \citep{Russo.2013,Russo.2014,Montali.2014}.

Artifact-centric hubs are one of the first ideas to allow for collaboration
using artifacts \citep{Hull.2009}. However, the interactions take
place between process participants, not among the artifacts themselves.
Participants use artifacts to interact with one another, where an
artifact is similar to a bulletin board. \citep{Lohmann.2010b} reused
basic ideas that led to the creation of artifact-centric hubs, but
instead used these ideas for introducing an approach for artifact
choreographies. Process participants, called agents, use artifacts
and execute them, and artifacts have a specific location. By knowing
where artifacts are located and who is using them, a choreography
between these agents can be automatically generated. While both \citep{Hull.2009}
and \citep{Lohmann.2010b} have created approaches for managing interactions,
the interacting parties are not the artifacts themselves. Instead,
choreographies between participants are created, constituting a stepping
stone towards artifact-based interorganizational business processes.
Therefore, both approaches are not directly comparable to coordination
processes.

By contrast, \citep{Sun.2012} created an approach for providing declarative
choreographies for artifact-centric processes where artifacts actually
interact. The artifacts in this approach use a type-instance schema
as well. Declarative choreographies recognize the need for explicitly
knowing the relations between artifacts and their multiplicity. Consequently,
one-to-many relationships and many-to-many relationships are supported
by a concept called a correlation graph. The artifact instances are
coordinated using messages, which are exchanged based on the constraints
of the declarative choreography. These constraints are specified using
expressions, where the expressions require greater expressiveness
than the expressions used for semantic relationships, making expressions
for artifacts more complicated in comparison.

In \citep{Fahland.2011}, the need for supporting many-to-many relationships
when dealing with interactions between artifacts is recognized, i.e.,
Challenge 2 \emph{Complex Process Relations}. However, artifact lifecycles
are specified by using Petri nets, specifically proclets, instead
of GSM, with the intention of using the well-known formal properties
of Petri nets to verify an entire artifact-centric business process.
This work is expanded in \citep{Fahland.2019}, where many-to-many
interactions between processes are fully incorporated into the Petri
net descriptions of these processes. The interactions between different
Petri net processes are expressed in terms of correlation and cardinality
constraints, and full operational semantics is provided. This form
of description is accessible for formal reasoning and verification.
\citep{Fahland.2019} aims for notational simplicity by using as few
syntactical concepts as possible, i.e., Petri nets only. This stands
on the opposing site to coordination processes and object-aware process
management, aiming for high-level abstractions for different concepts
by using appropriate notations.

The coordination of large process structures with focus on the engineering
domain is considered in \citep{Muller.2007,Muller.2008}. The COREPRO
approach explicitly considers process relations with one-to-many cardinality
as well as dynamic changes at run-time, but transitive relations are
not considered. In comparison to COREPRO, semantic relationships correspond,
in principle, to external state transitions of a Lifecycle Coordination
Model. However, the external state transitions do not take the semantics
of the respective process interaction into account. 

\subsection{Activity-centric Approaches and Choreographies}

Activity-centric process management approaches mostly produce monolithic
models. Consequently, interactions between various processes mostly
occur only in an inter-organizational setting, where the business
processes of different companies must interact. One of the most popular
process modeling languages for activity-centric processes is described
in the BPMN 2.0 standard \citep{ObjectManagementGroup.2011}. The
BPMN 2.0 modeling language expresses process interactions as message
exchanges between processes, which are organized sequentially. Moreover,
BPMN 2.0 provides choreography diagrams as an abstractions over the
fleshed-out process diagrams in standard collaboration diagrams. Regarding
interactions, BPMN 2.0 choreographies possess the same capabilities
and expressiveness as collaboration diagrams, as both can seamlessly
be integrated with each other. Consequently, a discussion of choreography
diagrams is sufficient for assessing interactions in BPMN 2.0, a separate
discussion of collaboration diagrams is redundant .

BPMN 2.0 processes, as well as their message exchanges, use well-established
token semantics \citep{ObjectManagementGroup.2011}. Conceptually,
one or more tokens move through a BPMN 2.0 process diagram, indicating
which activities may be executed next. The flexibility limitations
of these semantics have been well established, as demonstrated with
declarative activity-centric languages \citep{Pesic.2006}. Moreover,
message exchanges in BPMN 2.0 are required to be synchronous, i.e.,
both processes must either be in a position to sent or receive a specific
message in order for a successful message exchange. Therefore, due
to the imperative nature of the BPMN 2.0 choreography diagram and
its synchronous message exchanges, there is less flexibility when
executing the interacting processes compared to a coordination process. 

BPMN 2.0 has become an industry standard and is the predominant modeling
language for processes in general, the following discusses each challenge
briefly. 
\begin{itemize}
\item Challenge 1 \emph{Asynchronous Concurrency} has only limited support
from BPMN. Concurrent processes only occur when modeled in \emph{pools},
which is mainly the case for inter-organizational settings. While
these processes may run asynchronously, this property is hampered
by the synchronous message exchanges between processes. Subprocesses
are executed completely synchonously to their parent process, and
the parent process is halted while the subprocesses are running.
\item BPMN 2.0 has means to describe one-to-many relationships between processes
with multi-instance pools and multi\textendash instance or loop activites.
BPMN 2.0 therefore partly fulfills Challenge 2 \emph{Complex Process
Relations.} However, cardinalities for the interacting processes cannot
be defined. Moreover, transitive relations and many-to-many relations
between processes are not supported. In general, explicit relations
between processes are not part of the concept of BPMN. The concept
of \emph{correlation keys} are used to identify groups of message
exchanges (``conversations'') and have therefor a very different
function as relations. Correlation keys are further incapable of providing
the same functionality as relations and are not defined at the model
level \citep{Meyer.2015}. Moreover, the support of multi-instance
pools and activities is lacking in BPMN 2.0 process engines \citep{Geiger.2018},
which is required for the complete fulfillment of Challenge 2 \emph{Complex
Process Relations}.
\item Challenge 3 \emph{Local Contexts} cannot be supported in BPMN 2.0
due to missing explicit relations and transitive relations.
\item BPMN 2.0 does not state any performance goals for the execution of
its processes nor for the engines that execute these processes. Moreover,
as BPMN 2.0 modelers do not consider vast structures of interrelated
processes \citep{zurMuehlen.2013}, support for Challenge 4 \emph{Immediate
Consistency} has understandably not been considered. It is also infeasible
to realize the requirements of Challenge 4 \emph{Immediate Consistency}
by using BPMN 2.0 data objects associated with activities, as neither
of these concepts is remotely suited for implementing consistency
\citep{ObjectManagementGroup.2011}. 
\item Regarding Challenge 5 \emph{Manageable Complexity}, as BPMN 2.0 does
not fulfill Challenges 1-4, it must be concluded that BPMN 2.0 does
not possess manageable complexity as defined by the challenge. However,
BPMN 2.0 has become an industry standard. As such, it is highly unlikely
that the complexity of BPMN 2.0 is prohibiting the use of the standard.
Otherwise, it would have hardly become an industry standard. Still,
process engines tend to struggle with implementing advanced concepts
of BPMN 2.0 \citep{Geiger.2018} and only a subset of modeling elements
is used in practice \citep{zurMuehlen.2013}.
\end{itemize}
iBPMN is an approach to specify process choreographies for BPMN 1.2
processes \citep{Decker.2011}. While BPMN 1.2 focused on interconnection
modeling, iBPMN introduced interaction modeling to process choreographies.
This allows avoiding many of the pitfalls of interconnection modeling.
The iBPMN approach has defined formal execution semantics based on
a Petri net variant called \emph{interaction Petri net}. As such,
iBPMN can leverage much of the analysis techniques of regular Petri
nets. iBPMN highly influenced the design of the BPMN 2.0 choreography
diagram. Consequently, iBPMN shares much of the same shortcomings
of BPMN 2.0 choreography diagrams as well. Namely, Challenges 1-3
are not or only rudimentarily supported.
\begin{table*}[t]
\begin{centering}
\caption{\label{tab:Comparison-of-Approaches}Comparison of Process Coordination
Approaches}
\par\end{centering}
\begin{centering}
{\scriptsize{}}%
\begin{tabular*}{1\textwidth}{@{\extracolsep{\fill}}c>{\centering}p{2.5cm}>{\centering}p{2.5cm}ccc>{\centering}p{2cm}}
\toprule 
{\scriptsize{}Comparison Criteria} & {\scriptsize{}Artifact-centric (GSM)} & {\scriptsize{}Artifact-centric (Proclet)} & {\scriptsize{}Proclets} & {\scriptsize{}BPMN} & {\scriptsize{}Corepro} & {\scriptsize{}Coordination Processes}\tabularnewline
\midrule
\midrule 
{\scriptsize{}Asynchronous Concurrency} & {\scriptsize{}(\CheckmarkBold )} &  &  & {\scriptsize{}(\CheckmarkBold )} &  & {\scriptsize{}\CheckmarkBold{}}\tabularnewline
\midrule 
{\scriptsize{}Complex Process Relations} & {\scriptsize{}(\CheckmarkBold )} & {\scriptsize{}\CheckmarkBold{}} &  & {\scriptsize{}(\CheckmarkBold )} & {\scriptsize{}(\CheckmarkBold )} & {\scriptsize{}\CheckmarkBold{}}\tabularnewline
\midrule 
{\scriptsize{}Local Contexts} & {\scriptsize{}(\CheckmarkBold )} &  &  &  &  & {\scriptsize{}\CheckmarkBold{}}\tabularnewline
\midrule 
{\scriptsize{}Immediate Consistency} & \textbf{\scriptsize{}?} & \textbf{\scriptsize{}?} & \textbf{\scriptsize{}?} &  & \textbf{\scriptsize{}?} & {\scriptsize{}\CheckmarkBold{}}\tabularnewline
\midrule 
{\scriptsize{}Manageable Complexity} & \textbf{\scriptsize{}?} & \textbf{\scriptsize{}?} & \textbf{\scriptsize{}?} & \textbf{\scriptsize{}?} & \textbf{\scriptsize{}?} & {\scriptsize{}\CheckmarkBold{}}\tabularnewline
\midrule 
{\scriptsize{}message-based} &  & {\scriptsize{}\CheckmarkBold{}} & {\scriptsize{}\CheckmarkBold{}} & {\scriptsize{}\CheckmarkBold{}} & {\scriptsize{}\CheckmarkBold{}} & \tabularnewline
\midrule 
{\scriptsize{}paradigm-agnostic} &  &  &  &  &  & {\scriptsize{}\CheckmarkBold{}}\tabularnewline
\bottomrule
\end{tabular*}{\scriptsize\par}
\par\end{centering}
\begin{centering}
{\scriptsize{}}%
\begin{tabular*}{1\textwidth}{@{\extracolsep{\fill}}ccc}
{\scriptsize{}\CheckmarkBold{} : Supported} & {\scriptsize{}(\CheckmarkBold ) : Partially supported} & \textbf{\scriptsize{}?}{\scriptsize{} : Unknown}\tabularnewline
\end{tabular*}{\scriptsize\par}
\par\end{centering}
\end{table*}

\citep{Meyer.2015,Meyer.2013} build upon BPMN 2.0 choreographies
to provide automated data exchange between process participants. Several
challenges are addressed, among them the data exchange between processes
in a simple one-to-many relationship. Neither transitive or many-to-many
relationships are considered. The approach elevates several properties
to the model level, such as message exchange formats and correlation
keys, and utilizes these to automatically derive SQL queries to populate
the content of messages at run-time.

As with artifact-centric process management, several articles are
concerned with the verification and correctness of choreographies
in an activity-centric setting. Specifically, the absence of deadlocks
is often topic of investigation \citep{Decker.2011,Gudemann.2013}.
Moreover, the issue of realizability of process choreographies is
investigated in \citep{Lohmann.2010,Lohmann.2011}. Realizability
investigates whether an implementation can produce every interaction
specified in the choreography. Partial and distributed realizability
are relaxed forms of this notion. In the general case, realizability
is not decidable even for one-to-one relationships between processes.
While coordination processes support many-to-many relationships and
should therefore be affected as well, coordination processes do not
adhere to procedural interactions, i.e., procedural message exchanges.
Coordination processes are specified on a greater level of abstraction,
making no claims on the sequence of interactions. In consequence,
not all of the basic assumptions made by \citep{Lohmann.2011} apply.
Therefore, the decidability results cannot be directly transferred
to coordination processes. Realizability of coordination processes
will be investigated in future work.

Declarative approaches to activity-centric process management use
constraints between activities to define a business process \citep{Pesic.2006,vanderAalst.2006}.
Several approaches for realizing a runtime exist, based on SAT-Solving
\citep{Ackermann.2018} or Dynamic Condition Response Graphs (DCR
Graphs) \citep{Hildebrandt.2010,Slaats.2013}. In context of DCR graphs,
concurrency and asynchronous process execution has been investigated
\citep{Debois.2015}.

\subsection{Business Process Architectures}

Business process architectures are concerned with interacting processes
as well \citep{EidSabbagh.2013,EidSabbagh.2013b}. Like coordination
processes, business process architectures have established that complex
relationships,e.g., many-to-many, between processes are common. Specifically,
at run-time, multiple process instances and complex interactions are
prevalent. However, business process architectures are concerned with
analyzing the interactions between the processes, whereas coordination
processes are concerned with enforcing constraints on these interactions. 

\subsection{Other Approaches}

Case Management/ Case Handling \citep{Reijers.2003,vanderAalst.2005}
and the standardized case management model and notation CMMN \citep{ObjectManagementGroup.2016}
have not been designed to allow for interactions between processes.
One fundamental property of case management is that everything relevant
to a case is subsumed under that case. In consequence, interactions
among cases are practically non-existent. Case Management, however,
enables flexible, data-centric process execution in the context of
the case. In fragment-based case management, different process fragments
may be flexibly composed to form a case \citep{Hewelt.2016}. As coordination
processes originated in a data-centric approach, case management is
considered related work.

Object-centric behavioral constraints (OCBC) \citep{Li.2017,Artale.2019,vanderAalst.2017,vanderAalst.2017b}
is a conformance checking approach tailored to real-world software
systems such as ERP and CRM systems. Object-centric behavioral constraint
has a formal model that integrates activities and data in a unified
manner. Activities are represented using a declarative language \citep{Pesic.2006}.
This representation is combined with a UML or ER data model. Between
data objects and data objects and activities and data objects, sophisticated
cardinalities and relationships (one-to-many, many-to-many) can be
defined. This model follows from the observation that real-world software
systems do not comply to the notion of process instance and that an
integrated Declare and UML model would fit better for mining these
systems. As OCBC rejects the notion of process instance, there are
consequently no interactions between process instances and therefore
no coordination is required or even possible. While interesting in
its own right as a fundamentally different data-centric approach,
the concepts as presented in this article are not compatible due to
rejecting the notion of process instances. Furthermore, if OCBC models
were to be executed, they would suffer from the same drawbacks as
declarative activity-centric models, namely the continuous solving
of the NP-hard constraint satisfaction problem. In consequence, OCBC
model execution performance would not be on par with classical imperative
activity-centric or data-centric business process management approaches.

Reo \citep{Reo.2020,Arbab.2002} is a channel-based exogenous coordination
language. Reo is designed to impose coordination pattern upon modular
components of complex systems by means of connectors. Connectors correspond
to coordinators. To preserve the modularity, Reo imposes the coordination
requirements on the modules from the outside, i.e., exogeneously.
This is similar to a coordination process and state-based views. Reo
provides a textual and visual syntax and a compiler that transforms
these representations into application code, e.g., in the java programming
language.

The BIP (Behavior, Interaction, Priority) language is used for modeling
heterogeneous real-time components \citep{Ananda.2006,BIPLanguage.2020}.
Components are formed out of the combination of behavior, interaction,
and priority layers. The layers imply a clear separation between the
different functionalities. Behavior is given as a Petri net, and interactions
are managed using connectors which link different components. Components
possess ports for receiving and sending data. This bears similarities
to the definition of proclets. 

Both Reo and the BIP language target component coordination at the
programming language level, as evidenced by the automatic compilation
of models to a target programming platform. Regarding the challenges,
as the focus of both approaches is different to coordination processes,
many of the prerequisites stated by the challenges do not directly
apply. One grand commonality between coordination processes and Reo/the
BIP languages is the focus on the run-time and the emphasis of the
performance of the resulting system.

Other data-centric approaches and their capabilities are out of scope
for this paper. A detailed assessment of the capabilities of data-centric
approaches may be found in \citep{Steinau.2018c}. 

\subsection{Summary}

Coordination processes support various features not covered by most
process coordination approaches, most notably the support of Challenge
1 \emph{Asynchronous Concurrency}. This gives coordination processes
a unique advantage. Table \ref{tab:Comparison-of-Approaches} provide
a comparison of interaction-based process management approaches.

Note that artifact-centric process management as a whole is paradigm-agnostic,
as artifacts can be specified with GSM, proclets, or any other language,
e.g., BPMN. The partial support for GSM-based artifacts in Table \ref{tab:Comparison-of-Approaches}
is due to the expression framework, with which the challenges might
be fulfilled. The possibility of fulfilling challenges is stated here
in principle and in regard to the expressiveness of the expression
framework. However, the challenges are not supported by dedicated
concepts of artifact-centric process management, as opposed to coordination
processes.

As can be seen from Table \ref{tab:Comparison-of-Approaches}, regarding
Challenges 1-5, coordination processes have a significant conceptual
advantage over the other approaches. Moreover, the operational semantics
of coordination process are not only theoretical work, but have been
actually implemented in a working object-aware process management
system. This is also an enormous practical advantage. The working
implementation in form of the PHILharmonicFlows prototype sets object-aware
process management apart from most other data-centric approaches.
Therefore, object-aware process management is among the most advanced
data-centric approaches in the BPM field.

\section{\label{sec:Conclusion}Summary and Outlook}

\inputencoding{latin9}A coordination process is an advanced concept
for coordinating a collection of individual processes. It provides
the superstructure to effectively employ relational process structures
as well as semantic relationships. A coordination process itself is
specified in a concise and comprehensive manner using coordination
steps, coordination transitions, and ports, abstracting from the complexity
of coordinating a multitude of interrelated processes. Coordination
processes allow for the automatic derivation of semantic relationships
from a coordination transition between two coordination steps. Complex
coordination constraints are expressed by combining multiple semantic
relationships using ports, and are configured using a comprehensive
context-aware expression framework.

Coordination processes fulfill five challenges in order to optimally
support interdependent process instances, e.g., by allowing for asynchronous
concurrency and complex process relations. At run-time, coordination
step instances represent individual process instances, which are organized
in coordination step containers. Coordination components represent
semantic relationships. They connect coordination step instances and
port instances, taking complex process relations into account. By
replicating coordination components for their instantiator entities,
local contexts emerge, allowing for individual coordination of individual
process instances. The result is a highly complex instance representation
of the coordination process. This representation comprises containers,
coordination step instances, port instances, and coordination component
instances that represents the structure of the dependencies between
processes. The instance representation is not static at run-time,
as new process instances might emerge and new semantic relationships
might be established. The proper handling of many-to-many and transitive
relationships as well as local contexts sets coordination processes
apart from approaches that only support less complex relationships,e.g.,
one-to-one and one-to many.

The actual enactment of this instance representation is accomplished
using markings. Each marking signifies a specific status of the execution
of processes, and each entity in a coordination process graph possesses
a marking. As the normal processes are executed, i.e., progress through
their states, the coordination process changes markings by defined
process rules. Markings determine whether or not coordination constraints
are fulfilled, which then allows activating a specific state of a
process. This enables superior flexibility in executing a data-centric
business process. Coordination processes are furthermore fully implemented
in PHILharmonicFlows, which implements the object-aware approach to
business process management.

Thereby, a sophisticated and fine-grained coordination of processes
a different complex relationships can be realized. Coordination processes
allow realizing very flexible business process executions, as the
restrictions placed on the constituting processes in complex relationships
are minimal. Processes that are coordinated are still able to run
concurrently, giving an edge in business process execution performance.
Coordination process only intervene when necessary, achieving a light-weight
coupling through state based-views. The abstraction of process by
state-based views further emphasizes this light-weight coupling by
allowing for the coordination of processes in any paradigm, not just
object-aware or data-centric processes. 

While coordination processes are already able to deal with a vast
number of coordination scenarios, there are still some areas left
for improvement. One challenge concerns the monitoring of a business
process which consists of multiple interacting, interdependent processes.
Coordination processes may be used to gain valuable insights into
the overall progress of the business process, as it may be used to
aggregate status information from the different constituting processes.
Moreover, currently coordination processes act passively by allowing
or disallowing the activation of states. There exist several scenarios
in which a more active role of the coordination process would be beneficial,
e.g., actively forcing the activation of a state instead of simply
allowing a user to activate the state. The operational semantics of
coordination processes may be expanded in this direction.

\bibliographystyle{spbasic}
\bibliography{references}

\end{document}